\documentclass[aps,pra,twocolumn,superscriptaddress,longbibliography,nofootinbib,10pt]{revtex4-2}
\usepackage{soul}
\usepackage{silence}
\usepackage{amsmath,amssymb,bm,graphicx,multirow,dcolumn,booktabs}
\usepackage{amsthm}
\usepackage{hyperref}
\usepackage{xcolor}

\newtheorem{proposition}{Proposition}

\newcommand{\dd}{\mathrm{d}}
\newcommand{\tr}{\mathrm{tr}}
\newcommand{\kb}{k_B}
\newcommand{\diag}{\mathrm{diag}}
\newcommand{\atanTwo}{\mathrm{atan2}}

\begin{document}

\title{Measurement-angle control of a continuously monitored quantum oscillator:\texorpdfstring{\\}{ } purification limits and optimal protocols}

\author{Ralph Sabbagh}
\affiliation{Department of Mechanical and Aerospace Engineering, University of California, Irvine, California 92697, USA}

\author{Olga Movilla Miangolarra}
\affiliation{Departamento de F\'{i}sica, Universidad de La Laguna, La Laguna 38203, Spain}
\affiliation{Instituto Universitario de Estudios Avanzados (IUdEA), Universidad de La Laguna, La Laguna 38203, Spain}

\author{Andrew N. Jordan}
\affiliation{Institute for Quantum Studies, Chapman University, Orange, California 92866, USA}
\affiliation{The Kennedy Chair in Physics, Chapman University, Orange, California 92866, USA}
\affiliation{Department of Physics and Astronomy, University of Rochester, Rochester, New York 14627, USA}

\author{Tryphon T. Georgiou}
\affiliation{Department of Mechanical and Aerospace Engineering, University of California, Irvine, California 92697, USA}

\begin{abstract}
We study the dynamic purification of a continuously monitored quantum harmonic oscillator coupled to a thermal bath, where the control is the quadrature measurement angle of up to two detectors with prescribed strengths and efficiencies. Within the Gaussian regime, the conditional state covariance obeys a deterministic Riccati equation, and we ask what purity can be reached and sustained by manipulating the angle in time, in analogy with rapid-purification protocols for qubits. We propose, analyze, and compare constant-rate, greedy, vibrational, and variational monitoring strategies on the basis of sustained purity starting from the thermal bath state, and quantify to what extent they are able to maintain higher purity levels compared to lab-fixed or phase-locked detectors. We highlight structural features, operational limits, and optimal protocols from each strategy and study how model parameters shape control performance. We observe that in the absence of measurement-frame anisotropy, the time-dependent strategies achieve higher but similar performance. Once this symmetry is broken, their performances diverge, either improving or worsening depending on the relative timescales of oscillator dynamics, thermalization, and measurement. 
\end{abstract}

\maketitle

\section{Introduction}\label{sec:intro}

The ability to purify quantum states lies at the heart of modern quantum science, as a means to overcome the inherent fragility of quantum systems as they interact with noisy environments. It is a prerequisite for high-fidelity computation, communication, and sensing, and has recently emerged as a salient thermodynamic resource in the context of continuously monitored quantum batteries \cite{mitchison2021charging,morrone2023daemonic,kua2025daemonic,cenedese2025boosting}. In essence, purification under continuous monitoring happens because the obtained measurement record carries information about the system, which in turn may be utilized to update or ``localize" the state via Bayesian inference~\cite{belavkin1987information,Wiseman_Milburn_2009,jacobs2006straightforward,jacobs2014quantum,jordan2024quantum,ALBARELLI2024129260}. The effectiveness of this process, which is determined in large part by the available detector specifications, can be further enhanced based on the choice of observables being monitored at different points in time. For instance, adaptively changing the measured spin component of a continuously monitored qubit can enhance its average purification rate~\cite{jacobs2003rapid}. Related enhancements have since been identified in higher-dimensional systems and for the simultaneous monitoring of complementary spin observables~\cite{combes2006rapid,combes2008rapid,ruskov2010three}. In general, whether one particular strategy outperforms another could also depend on the chosen figure of merit~\cite{wiseman2006reconsidering,jordan2006qubit}, as is common in quantum feedback control~\cite{zhang2017feedback}.

\begin{figure}[ht]
    \centering
    \includegraphics[width=\linewidth]{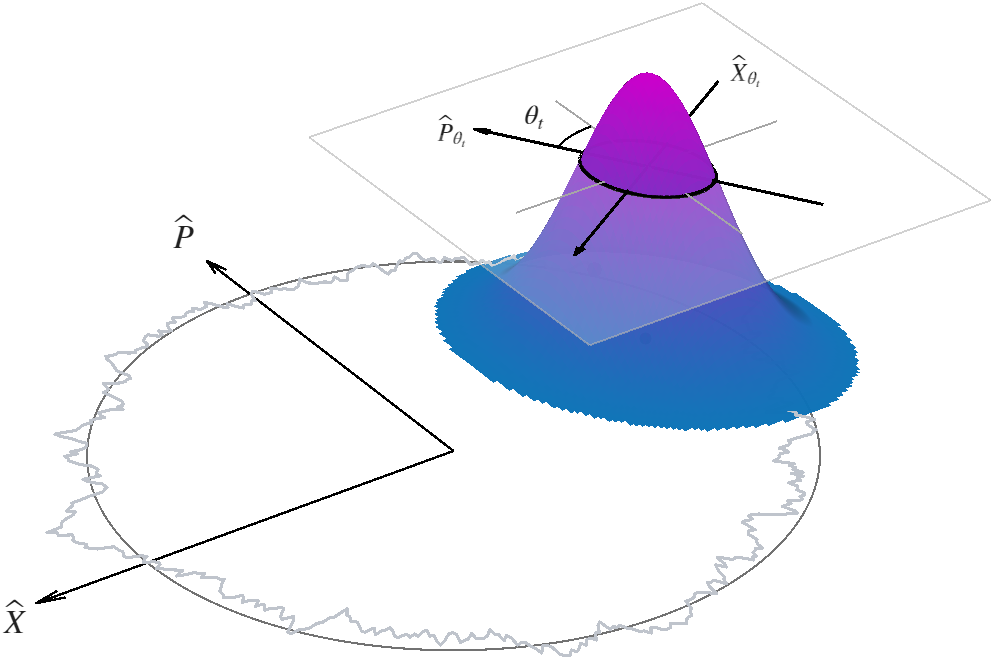}
    \caption{Phase-space schematic of a Gaussian state continuously monitored along two orthogonal quadratures, \(\hat X_{\theta_t}\) and \(\hat P_{\theta_t}\), whose angle \(\theta_t\) is the control. The measurement record drives random excursions of the conditional mean, while information gain and backaction reshape the conditional covariance deterministically. The black ellipse is a covariance contour. Its area is inversely proportional to the state purity.}
    \label{fig:F0}
\end{figure}

In continuous-variable systems, the interplay between information gain and noise becomes particularly transparent in the Gaussian regime. In this case, the conditional quantum state evolves under diffusive monitoring according to a quantum analogue of the classical Kalman--Bucy filter \cite[Chapter 8.1.3]{bechhoefer2021control}, in which the covariance matrix obeys a deterministic Riccati equation and only the first moments inherit the stochasticity of the measurement record~\cite{doherty1999feedback,Genoni02072016,serafini2017quantum}. Because the purity of a Gaussian state is fixed entirely by its covariance matrix~\cite{paris2003purity}, measurement-induced purification then reduces to a deterministic control problem. As a consequence, protocols that adapt the measurement to the evolving covariance can be designed in advance, without reference to any particular measurement record.

Earlier studies on measurement-based control in continuous-variable systems for quantum thermal machines and cooling applications have focused on either optimizing static measurement specifications, or designing feedback-actuated system  Hamiltonians~\cite{wiseman2001complete,wiseman2005optimal,mancini2007optimal,serafini2010determination,genoni2013optimal,genoni2014general,kua2025daemonic,rossi2018measurement,manikandan2022engine,bhandari2022boosted,karmakar2022stochastic,bhandari2023machines,manikandan2023cooling,sabbagh2025regulation}. 
Herein, we address a complementary situation in which the detectors are \textit{specified in advance} and only the orientation of the measured quadratures in phase space are accessible for control, as is the case in adaptive homodyne detection and backaction-evading measurements of mechanical motion~\cite{braginsky1980quantum,wiseman1995adaptive,armen2002adaptive,clerk2008backaction,hertzberg2010backaction,suh2014mechanically,shomroni2019optical,aspelmeyer2014cavity}. In this setting, we quantify the purity that can be reached and sustained in a thermally damped oscillator, and explain the dependence of purification performance on the system parameters.

To that end, we consider the harmonic oscillator\footnote{Here $\hat X=\sqrt{m\omega/\hbar}\,\hat x$ and $\hat P=\hat p/\sqrt{m\omega\hbar}$ are dimensionless position and momentum operators with $[\hat X,\hat P]=i$.}
\begin{align}
    \hat{H} = \frac{\hbar\omega}{2}(\hat{X}^2+\hat{P}^2),
\end{align}
monitored along two time-dependent quadratures
\begin{align}\label{rotatedquad}
    \begin{bmatrix}
        \hat{X}_{\theta_t}\\
        \hat{P}_{\theta_t}
    \end{bmatrix}
    :=
    \begin{bmatrix}
        \cos\theta_t&-\sin\theta_t\\
        \sin\theta_t&\phantom{-}\cos\theta_t
    \end{bmatrix}
    \begin{bmatrix}
        e^s &0\\
        0& e^{-s}
    \end{bmatrix}
    \begin{bmatrix}
        \hat{X}\\
        \hat{P}
    \end{bmatrix},
\end{align}
whose angle $\theta_t$ is the control. The detectors realize joint weak measurements of conjugate quadratures, in the spirit of Arthurs and Kelly~\cite{arthurs1965}, with specifications characterized by their measurement strengths \begin{align}
    k_1,k_2\geq 0,
\end{align}
and efficiencies \begin{align}
    0\leq \eta_1,\eta_2\leq 1.
\end{align}  
 The parameter $s$ quantifies anisotropy in the measurement relative to the natural setting where $\hat{X}$ and $\hat{P}$ are on equal footing. Specifically, when $s=0$ the measurement is symmetric, whereas $s\neq0$ provides the simplest breaking of this oscillator-measurement symmetry\footnote{
The parameter $s$ represents a source of anisotropy that squeezes the measurement frame relative to the natural phase space shape of the oscillator. It can arise, for instance, from a calibration mismatch which could happen if the detected quadratures are referenced to a mass $m_r$ and frequency $\omega_r$ that differ from those of the oscillator, in which case $s=\ln(m_r\omega_r/m\omega)/2$. The case $s=0$ is the idealized case, whereas $s\neq0$ provides the simplest breaking of this oscillator-measurement symmetry. Other sources of anisotropy, such as a relative squeeze between the detectors or an angular offset between their quadratures, may also be similarly considered.}  

Physically, the choice of measured quadrature matters because the measurement acquires information preferentially along that direction, tending to reduce its conditional variance while backaction affects the conjugate direction. A single-quadrature measurement can therefore squeeze the covariance, whereas monitoring complementary quadratures distributes these effects between the two directions.

Section~\ref{sec:model} develops the Gaussian conditional dynamics, identifies the separate contributions of Hamiltonian rotation, thermalization, information gain, and measurement backaction, and derives the corresponding purity dynamics. We first isolate those purification properties that are fixed by the detector and bath parameters and are independent of the control angle. Then, the remainder of the paper compares, for $s=0$ first, four increasingly general classes of measurement-angle control protocols:

\textit{i) Constant-rate protocols (Sec.~\ref{sec:constant}).}
Here, for
\begin{align}
    \theta_t=\Omega t+\theta_0,
\end{align}
we characterize in closed form the dependence of the sustained purity and relaxation rate on $\Omega$, and show, under natural assumptions, that they increase with the relative rotation $|\omega-\Omega|$. We then characterize the effectiveness of the fast-rate limit, i.e., when $|\Omega|\to\infty$. 

\textit{ii) Vibrational protocols (Sec.~\ref{sec:vibrational}).}
In this setting,
\begin{align}
    \theta_t=a\sin(\Omega t)+\theta_0,\,\,\Omega\gg 1,
\end{align}
representing high-frequency sinusoidal modulation. We characterize the resulting two-parameter family of effective measurement schemes and visualize the corresponding sustained purities, identifying the optimized strategy.

\textit{iii) Greedy protocols (Sec.~\ref{sec:greedy}).}
Here, the measurement angle is allowed to depend on the evolving covariance $\Sigma_t$,
\begin{align}
    \theta_t=\theta^\star(\Sigma_t),
\end{align}
and the protocol consists in selecting $\theta^\star$ to maximize the instantaneous purification rate. We derive the form of the optimal selection, which essentially aligns the measurement with the principal axes of the covariance. 

\textit{iv) Variational protocols (Sec.~\ref{sec:variational}).}
Here, we select $\theta_t$ that maximizes the average purity, 
\begin{align}
    -\int_0^T \ln\det\Sigma_t\,dt.
\end{align} 
We derive the form of the optimal control which turns out to be analogous to the greedy protocol, except for a dependence on the co-state of the optimization problem. 

Then, the impact of anisotropy, i.e., $s\neq 0$ on their performance is studied in Sec.~\ref{impactaniso}. It is observed to be either beneficial or detrimental depending on the interplay between the system timescales. Our main takeaway is that in the relevant parameter regimes these four control classes can theoretically sustain a higher average purity compared to lab-fixed or phase-locked detectors. The four strategies are observed to achieve a similar performance when $s=0$, and differ once the measurement frame becomes anisotropic, i.e., $s\neq 0$. For the representative anisotropic cases studied here, the best strategy was observed to be the variational one, followed closely by the vibrational and constant-rate strategies, and ending with the greedy strategy. Sec.~\ref{sec:outlook} discusses the relevant control timescales and directions of future work. Proofs and supporting derivations are collected in the Appendix.

\section{Conditional Gaussian Dynamics}\label{sec:model}

\subsection{It\^o stochastic master equation}

Let $W_t^A$ and $W_t^B$ be independent Wiener processes. Continuous monitoring of the conjugate quadratures $\hat X_{\theta_t}$ and $\hat P_{\theta_t}$ leads to a conditional state $\hat\rho_t$ that obeys the It\^{o} stochastic master equation \cite{belavkin1987information,jacobs2006straightforward,Wiseman_Milburn_2009,ALBARELLI2024129260}
\begin{align}
 \dd\hat\rho_t={}&\frac{1}{i\hbar}[\hat H,\hat\rho_t]\dd t
 +2\kappa\big[(\bar n+1)\mathcal D[\hat a]\hat\rho_t
 +\bar n\mathcal D[\hat a^\dagger]\hat\rho_t\big]\dd t \nonumber\\
 &+\frac{k_1}{4}\mathcal D[\hat{X}_{\theta_t}]\hat\rho_t\dd t
 +\frac{\sqrt{\eta_1k_1}}{2}\mathcal H[\hat{X}_{\theta_t}]\hat\rho_t\dd W_t^A \nonumber\\
 &+\frac{k_2}{4}\mathcal D[ \hat{P}_{\theta_t}]\hat\rho_t\dd t
 +\frac{\sqrt{\eta_2k_2}}{2}\mathcal H[ \hat{P}_{\theta_t}]\hat\rho_t\dd W_t^B,
 \label{eq:SME}
\end{align}
where
\begin{align}
\label{eq:superoperators}
\begin{cases}
 \mathcal D[\hat c]\hat\rho_t&=\hat c\hat\rho_t\hat c^\dagger-\frac12\{\hat c^\dagger\hat c,\hat\rho_t\},\\
 \mathcal H[\hat c]\hat\rho_t&=\hat c\hat\rho_t+\hat\rho_t\hat c^\dagger
 -\tr[(\hat c+\hat c^\dagger)\hat\rho_t]\hat\rho_t,
\end{cases}
\end{align}
are the Lindblad and measurement superoperators, and
\begin{align}\label{lowering}
\hat a=(\hat{X}+i \hat{P})/\sqrt{2}
\end{align}
is the annihilation operator. Here $k_1,k_2\ge0$ are the detector measurement strengths, $0\le\eta_1,\eta_2\le1$ are the detector efficiencies, $\kappa$ is the bath decay rate,
\begin{equation}
 \bar n=\frac{1}{\exp(\hbar\omega/\kb T)-1}
 \label{eq:nbar}
\end{equation}
is the bath occupation, and $\kb$ is the Boltzmann constant. Throughout, we assume that the stronger detector is also the most efficient, i.e.,
\begin{align}
    (k_1-k_2)(\eta_1-\eta_2)\geq0.
    \label{ass1}
\end{align}
The single-detector configuration is recovered in the limit as the strength of the weaker detector vanishes. The symbols $[~,~]$, $\{~,~\}$, and $\dagger$ denote the usual commutator, anti-commutator, and adjoint operations.

\subsection{Statistics and dynamics of Gaussian states}

A quantum state $\hat\rho_t$ is called \textit{Gaussian} if its quasi-characteristic function
\begin{align}
   f(\xi_x,\xi_p)=\tr (\hat{\rho}_te^{i\xi_x\hat{X}+i\xi_p\hat{P}}),\label{characfunc}
\end{align}
is the Fourier transform of a Gaussian with mean
\begin{align}
\mu_t=[\tr(\hat\rho_t\hat{X})~~\tr(\hat\rho_t\hat{P})]^\top,
\end{align}
and covariance
\begin{align}
    \Sigma_t = \frac{1}{2}\begin{bmatrix}
        2\tr(\hat\rho_t\hat{X}^2)&\tr(\hat\rho_t\{\hat{X},\hat{P}\})\\
        \tr(\hat\rho_t\{\hat{X},\hat{P}\})&2\tr(\hat\rho_t\hat{P}^2)
    \end{bmatrix}-\mu_t\mu_t^\top.
\end{align}
It is well-known that Gaussianity is preserved by \eqref{eq:SME} \cite{doherty1999feedback,Genoni02072016,serafini2017quantum}. The corresponding dynamics for $\mu_t$ and $\Sigma_t$ are
\begin{table}[t]
\caption{Summary of model parameters/matrices}
\label{tab:notation}
\renewcommand{\arraystretch}{1.15}
\setlength{\tabcolsep}{8pt}
\begin{ruledtabular}
\begin{tabular}{c l l}
& Symbol & Meaning \\
\midrule
&$m$, $\omega$ & oscillator parameters \\
\midrule
\multirow{4}{*}{\rotatebox{90}{bath}}
& $\kappa$ & bath decay rate \\
& $\bar{n}$ & bath occupation \\
& $\Sigma_{\rm th}$ & bath equilibrium covariance \\
\midrule
\multirow{9}{*}{\rotatebox{90}{measurement}}
& $k_1, k_2$ & detector strengths \\
& $\eta_1, \eta_2$ & detector efficiencies \\
& $s$ & anisotropy parameter \\
& $\theta_t$ & quadrature measurement angle \\
& $U_{\theta_t}$ & quadrature control matrix \\
& $R_{\theta_t}$ & $U_{\theta_t} R U_{\theta_t}^\top$, information matrix \\
& $K_{\theta_t}$ & $JU_{\theta_t} K U_{\theta_t}^\top J^\top $, backaction matrix
\end{tabular}
\end{ruledtabular}
\end{table}
\begin{subequations}
\label{eq:gaussianSDE}
\begin{align}
\hspace{-3mm} &\dd\mu_t=-(\omega J+\kappa I)\mu_t\dd t+\Sigma_tU_{\theta_t}R^{1/2}\dd W_t,\label{eq:meanSDE}\\
\hspace{-3mm}&\dot\Sigma_t=\omega[\Sigma_t,J]\hspace{-1mm}+2\kappa(\Sigma_{\rm th}\hspace{-0.5mm}-\hspace{-0.5mm}\Sigma_t)+K_{\theta_t}\hspace{-1mm}-\Sigma_tR_{\theta_t}\Sigma_t,\label{eq:RDE}
\end{align}
\end{subequations}
where $\dd W_t=[
     \dd W_t^A~~
     \dd W_t^B]^\top$,
\begin{align}
  \Sigma_{\rm th}
 &=
 (\bar{n}+0.5)I,&J&=\left[\begin{smallmatrix}0&-1\\1&\phantom{-}0\end{smallmatrix}\right],\\
 R_{\theta_t}&=U_{\theta_t} RU_{\theta_t}^\top,
 &R&=\diag(\eta_1k_1,\eta_2k_2), \\
 K_{\theta_t}&=JU_{\theta_t} U_{\theta_t}^\top J^{\top},
 &K&=\diag(k_1,k_2)/4,
\end{align}
and
\begin{align}
U_{\theta_t}:=\begin{bmatrix}
        e^s &0\\
        0& e^{-s}
    \end{bmatrix}\begin{bmatrix}
        \cos\theta_t&-\sin\theta_t\\
        \sin\theta_t&\phantom{-}\cos\theta_t
    \end{bmatrix}^\top .
\end{align}
The meaning of the above matrices and parameters is summarized in Table~\ref{tab:notation}. While \eqref{eq:meanSDE} is an It\^{o} SDE, \eqref{eq:RDE} is a deterministic Riccati ODE that is independent of the realized measurement record. Thus, the covariance trajectory is fully determined by the angle protocol, $\Sigma_0$, and the system parameters.

It is insightful to identify in \eqref{eq:gaussianSDE} the Hamiltonian, bath, and control-dependent measurement contributions:
\begin{align}
\text{Hamiltonian:}~~~&\begin{cases}
    \dd\mu_t^{\rm H}&:=-\omega J\mu_t\dd t\\
 \dot\Sigma_t^{\rm H}&:=\omega[\Sigma_t,J],
\end{cases}\label{eq:syscontrib}\\
\text{bath:}~~~&\begin{cases}
      \dd\mu_t^{\rm b}&:=-\kappa\mu_t\dd t\\
 \dot\Sigma_t^{\rm b}&:=2\kappa(\Sigma_{\rm th}-\Sigma_t),
\end{cases}\label{eq:bathcontrib}\\
\text{measurement:}~~~&\begin{cases}
    \dd\mu_{t,\theta_t}^{\rm m}&:=\Sigma_tU_{\theta_t}R^{1/2}\dd W_t\\
 \dot\Sigma_{t,\theta_t}^{\rm m}&:=K_{\theta_t}-\Sigma_tR_{\theta_t}\Sigma_t.
\end{cases}\label{eq:meascontrib}
\end{align}
Whereas the Hamiltonian terms in \eqref{eq:syscontrib} rotate the mean and covariance ellipse without changing its area, the bath terms in \eqref{eq:bathcontrib} pull the mean to the origin, driving $\Sigma_t$ to the bath equilibrium $\Sigma_{\rm th}$. At the same time, the measurement process \eqref{eq:meascontrib} contributes a negative information-gain term $-\Sigma_t R_{\theta_t}\Sigma_t$ and a positive backaction term $K_{\theta_t}$ in \eqref{eq:RDE}, while causing the mean to jitter in phase space.

The dependence of the covariance dynamics \eqref{eq:RDE} on $\theta_t$ enters solely through $R_{\theta_t},K_{\theta_t}$, which are quadratic in $U_{\theta_t}$. Consequently, the covariance dynamics are $\pi$ periodic in the angle. Finally, if the detectors are identical, equivalently $R,K\propto I$, then $R_{\theta_t}$ and $K_{\theta_t}$ no longer depend on the angle and control advantage is unavailable.

\subsection{Purity dynamics}\label{sec:puritydyn}

For Gaussian states, the purity $p_t:=\tr(\hat\rho_t^2)$ is completely determined by the covariance determinant \cite{paris2003purity},
\begin{equation}
 p_t=\frac{1}{2\sqrt{\det\Sigma_t}},
 \label{eq:purity}
\end{equation}
where the Robertson--Schr\"odinger uncertainty relation dictates that $\det\Sigma_t\geq1/4$, i.e., $p_t\leq1$.

For a $2\times2$ symmetric matrix, $\det(\Sigma_t) I = J\Sigma_t J^\top\Sigma_{t}$. Differentiating and taking the trace gives the corresponding purity dynamics
\begin{align}
\dot p_t={}&
 -\frac{p_t}{2}\left(\tr\!\left[\Sigma_t^{-1}\dot{\Sigma}_t^{\rm b}\right]+\tr\!\left[\Sigma_t^{-1}\dot\Sigma_{t,\theta_t}^{\rm m}\right]\right).
 \label{eq:detp}
\end{align}
As expected, the Hamiltonian contribution \eqref{eq:syscontrib} does not appear in \eqref{eq:detp}; only the heat bath and measurement contributions, \eqref{eq:bathcontrib} and \eqref{eq:meascontrib}, are relevant. The control enters through the second term, which may be written as
\begin{align}
    \tr\!\left[\Sigma_t^{-1}\dot\Sigma_{t,\theta_t}^{\rm m}\right]=\tr\left[\Sigma_tU_{\theta_t}\left[\begin{smallmatrix}
       k_1(p_t^2-\eta_1) &0\\
       0 &k_2(p_t^2-\eta_2)
    \end{smallmatrix}\right]U_{\theta_t}^\top\right].
    \label{eq:measterm}
\end{align}
The above readily shows how detector efficiencies can act as purity thresholds. If $p_t^2$ is less than $\eta_{\min}:=\min(\eta_1,\eta_2)$, the measurement will tend to increase purity irrespective of the control. Likewise, if $p_t^2$ is greater than $\eta_{\max}:=\max(\eta_1,\eta_2)$, the measurement will tend to decrease purity. In fact, in the absence of a heat bath, i.e., $\kappa=0$, one can show that the measurement will always \textit{trap} the system purity between $\sqrt{\eta_{\min}}$ and $\sqrt{\eta_{\max}}$ in finite time. We state this next; the proof is deferred to Appendix \ref{app:meas}.
\begin{proposition}
\label{prop:meas}
Assume that $K>0$ and $\kappa=0$. If $\eta_1\neq\eta_2$, then there is a finite time $\tau$, independent of the measurable angle protocol $\{\theta_t,t\geq0\}$, such that
\begin{align}\label{eq:finitetimeconfinement}
 p_t\in[\sqrt{\eta_{\min}},\sqrt{\eta_{\max}}],
\end{align}
for all $t\geq\tau$. For same-efficiency detectors, when $\eta_1=\eta_2=:\eta$, the same is true for any neighborhood of $\sqrt{\eta}$.
\end{proposition}

An analogous analysis holds at the other extreme where only the heat bath is present, i.e., $K=0$. In this case, \eqref{eq:RDE} is linear and the covariance reaches equilibrium $\Sigma_{\rm th}$. To make the notation more systematic, we denote the purity corresponding to $\Sigma_{\rm th}$ by $\sqrt{\eta_{\rm th}}:=(2\bar{n}+1)^{-1}$. A closer look at the heat bath contribution term in \eqref{eq:detp},
\begin{align}\label{conditth}
    \tr\left[\Sigma_t^{-1}\dot\Sigma_t^{\rm b}\right] &= 2\kappa\tr\left[\Sigma_t^{-1}\Sigma_{\rm th}-I\right]\nonumber\\
    &= 2\kappa\left[\frac{p_t}{\sqrt{\eta_{\rm th}}}\left(\sqrt{\chi_t}+\frac{1}{\sqrt{\chi_t}}\right)-2\right]
\end{align}
where $\chi_t$ is the condition number\footnote{The condition number of a matrix is the ratio between its maximum and minimum singular values.} of $\Sigma_t$, reveals a slightly more nuanced picture. When $p_t$ is greater than $\sqrt{\eta_{\rm th}}$, the heat bath will tend to decrease purity toward $\sqrt{\eta_{\rm th}}$. If $p_t$, however, is less than $\sqrt{\eta_{\rm th}}$, then the overall sign of the contribution at that instant depends on the shape mismatch between $\Sigma_t$ and $\Sigma_{\rm th}$ captured by $\chi_t$. Physically, even if the state has lower purity than the thermal equilibrium state, it may be strongly squeezed and {\em the bath may temporarily reduce the system purity further as it reshapes the ellipse toward the thermal covariance}. We summarize this next; the proof is deferred to Appendix~\ref{app:bath}.

\begin{proposition}\label{prop:bath}
    Assume that $\kappa>0$ and $K=0$. Then, the purity converges to $\sqrt{\eta_{\rm th}}$ following
\[
p_t=
\frac{\sqrt{\eta_{\rm th}}}
{\sqrt{
1+c_2e^{-2\kappa t}
+c_4e^{-4\kappa t}
}},
\]
where
\begin{align*}
    c_2 &= \frac{\sqrt{\eta_{\rm th}}}{p_0}\left(\sqrt{\chi_0}+\frac{1}{\sqrt{\chi_0}}\right)-2,\\
    c_4 &= \left(\frac{\sqrt{\eta_{\rm th}}}{p_0}\right)^2-c_2-1,
\end{align*}
and $\chi_0$ is the condition number of $\Sigma_0$.
\end{proposition}

\begin{figure}[tb]
\centering
    \includegraphics[width = \linewidth]{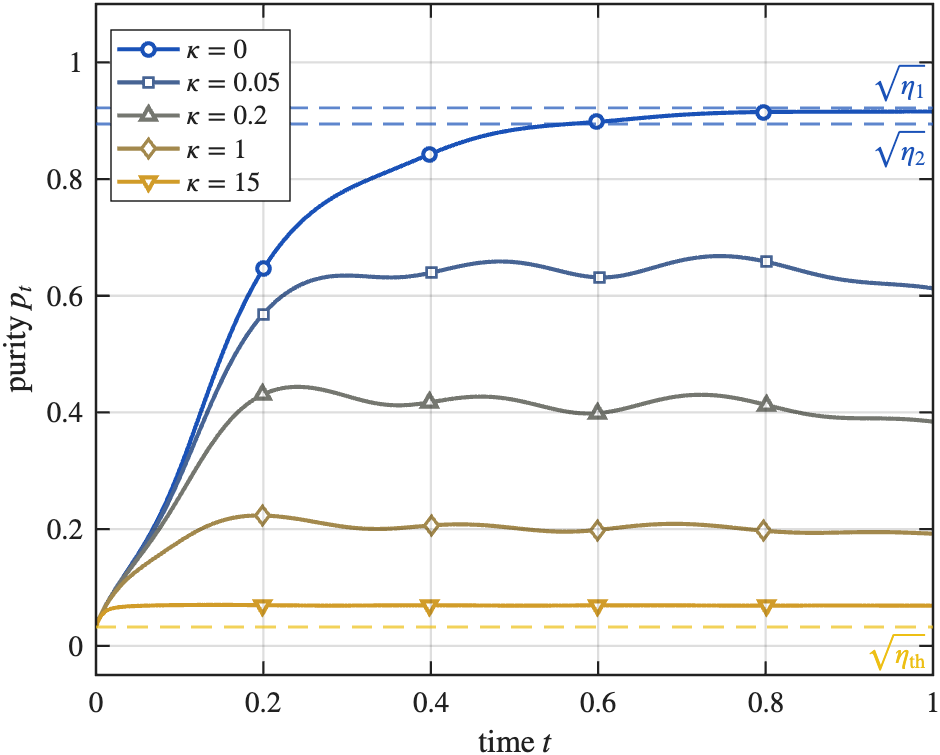}
 \caption{Purity under the angle protocol \(\theta_t=e^{-t}\cos(4\pi t)\) for several bath decay rates \(\kappa\) starting from the thermal state. As \(\kappa\) increases, the dynamics cross over from measurement-dominated behavior, with the purity constrained by the detector-efficiency interval \([\sqrt{\eta_2},\sqrt{\eta_1}]\), to bath-dominated behavior approaching the thermal purity \(\sqrt{\eta_{\rm th}}\). We take $\omega=1$, $\bar{n}=15$, $\eta_1=0.85$, $\eta_2=0.8$, $k_1=20$, and $k_2=1$.}
 \label{fig:F1}
\end{figure}

The upshot of Propositions \ref{prop:meas} and \ref{prop:bath} is that the long-time behavior of the purity is determined by the competition between the measurement process pulling the purity toward the interval $[\sqrt{\eta_{\min}},\sqrt{\eta_{\max}}]$, and thermalization pulling it toward $\sqrt{\eta_{\rm th}}$. We illustrate this in Fig.~\ref{fig:F1}.

\begin{figure*}[t]
    \centering
    \includegraphics[width=\linewidth]{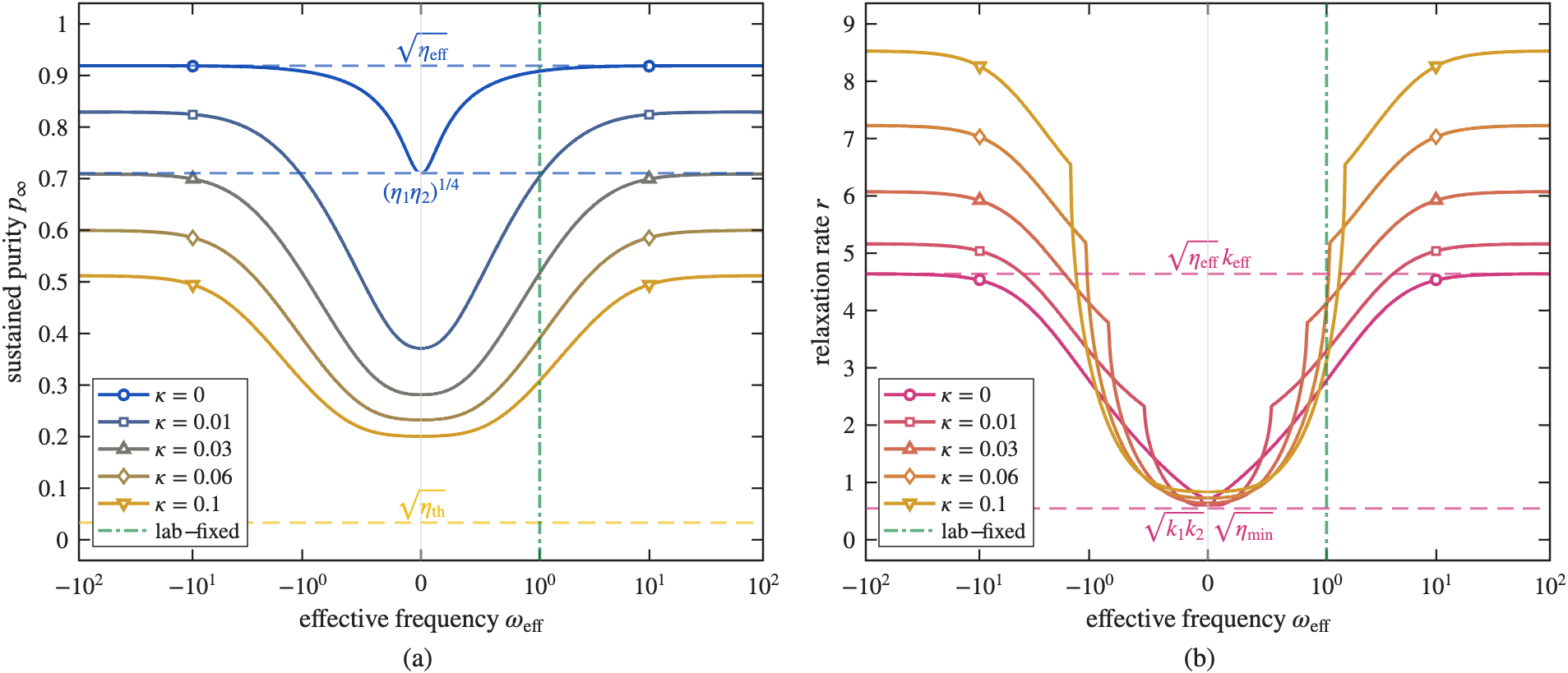}
    \caption{Left: steady-state purity \(p_\infty\) as a function of the effective detector frequency \(\omega_{\rm eff}=\omega-\Omega\) for detectors with \(k_1=10\), \(k_2=0.1\), \(\eta_1=0.85\), \(\eta_2=0.30\), and \(\bar n=14.5\), for several bath decay rates \(\kappa\). The purity is an even function of \(\omega_{\rm eff}\), is minimized at the co-rotating point \(\omega_{\rm eff}=0\), and approaches the fast-rotation plateau as \(|\omega_{\rm eff}|\) increases. For \(\kappa=0\), the dashed blue lines indicate the zero-frequency value \((\eta_1\eta_2)^{1/4}\) and the fast-rotation limit \(\sqrt{\eta_{\rm eff}}\). The dashed yellow line denotes the thermal purity \(\sqrt{\eta_{\rm th}}\). The green dash-dotted line marks the lab-fixed detector, \(\Omega=0\), corresponding here to \(\omega_{\rm eff}=\omega=1\). Right: relaxation rate \(r\) as a function of the effective detector frequency \(\omega_{\rm eff}=\omega-\Omega\) for the same detector parameters and bath decay rates \(\kappa\). The rate is even in \(\omega_{\rm eff}\), is smallest at the co-rotating point, and approaches its rotationally averaged value at large \(|\omega_{\rm eff}|\). The dashed pink lines show the \(\kappa=0\) limiting values \(\sqrt{k_1k_2}\sqrt{\eta_{\min}}\) at \(\omega_{\rm eff}=0\) and \(\sqrt{\eta_{\rm eff}}k_{\rm eff}\) in the fast-rotation limit. The green dash-dotted line marks the lab-fixed configuration \(\Omega=0\), for which \(\omega_{\rm eff}=\omega=1\).}
    \label{fig:F2}
\end{figure*}

\subsection{Dimensionless ratios}

It is natural to rewrite the bath terms in \eqref{eq:SME} as
\begin{align}
    k_{\rm th}\left(\left(\frac{1+\sqrt{\eta_{\rm th}}}{2}\right)\mathcal D[\hat a]\hat\rho_t
 +\left(\frac{1-\sqrt{\eta_{\rm th}}}{2}\right)\mathcal D[\hat a^\dagger]\hat\rho_t\right),
\end{align}
where
\begin{align}
   \eta_{\rm th}:=\frac{1}{(2\bar{n}+1)^2},
   \qquad
   k_{\rm th}:=\frac{2\kappa}{\sqrt{\eta_{\rm th}}},
\end{align}
can be interpreted as effective bath quality and strength, respectively. Moreover, for the two detectors, we define their averaged, ``effective" specifications 
\begin{align}\label{eq:effective-coef}
    k_{\rm eff}
    &:=
    \frac{k_1+k_2}{2},
    &
    \eta_{\rm eff}
    &:=
    \frac{\eta_1k_1+\eta_2k_2}{k_1+k_2}.
\end{align}
In doing so, one obtains the dimensionless ratios
\begin{align}\label{eq:ratios}
    \nu
    &:=
    \frac{k_{\rm eff}}{k_{\rm th}},
    &
    q
    &:=
    \frac{\eta_{\rm eff}}{\eta_{\rm th}},
\end{align}
which reflect effective measurement-to-bath strength and quality factors, respectively. Throughout, we assume 
\begin{align}
    q>1,
    \label{ass2}
\end{align}
so that the effective detector quality exceeds that of the thermal environment. 

\section{Constant rate protocols}\label{sec:constant}

In this section, we study protocols of the form
\begin{equation}\label{constantrate}
\theta_t=\Omega t+\theta_0,
\end{equation}
for $s=0$. In this case, the covariance in the detector frame
\begin{equation}\label{rotatedcov}
\widetilde{\Sigma}_t
:=U_{\theta_t}^{\top}\Sigma_tU_{\theta_t},
\qquad
\det\widetilde\Sigma_t=\det\Sigma_t,
\end{equation}
satisfies the Riccati differential equation
\begin{equation}\label{eq:rotatedRDE}
\dot{\widetilde{\Sigma}}_t
=
\omega_{\rm eff}[\widetilde{\Sigma}_t,J]
+2\kappa(\Sigma_{\rm th}-\widetilde{\Sigma}_t)
+JKJ^\top
-\widetilde{\Sigma}_tR\widetilde{\Sigma}_t,
\end{equation}
where
\begin{align}
    \omega_{\rm eff}:=\omega-\Omega,
\end{align}
is the effective rotation rate of the oscillator relative to the detectors, c.f. \eqref{eq:RDE}. Special values include
\begin{align}
    \omega_{\rm eff}=0,
\end{align}
for detectors co-rotating with the oscillator $(\Omega=\omega)$,
\begin{align}
    \omega_{\rm eff}=\omega,
\end{align}
for lab-fixed detectors $(\Omega=0)$, and
\begin{align}
    |\omega_{\rm eff}|\gg\omega,
\end{align}
when detector rotation dominates the relative motion. The dynamics \eqref{eq:rotatedRDE} are time invariant and depend on the protocol only through $\omega_{\rm eff}$. We solve them next and quantify purification advantage.

\subsection{Solution of the Riccati differential equation}

Since $R>0$ and
\begin{align}
    Q:=JKJ^\top+2\kappa\Sigma_{\rm th}>0,
\end{align}
standard Riccati theory \cite{anderson2007optimal,bechhoefer2021control} guarantees that $\widetilde\Sigma_t$ converges as $t\rightarrow\infty$ to the unique positive-definite solution $\widetilde\Sigma_\infty$ of
\begin{equation}
0=
\omega_{\rm eff}[\widetilde\Sigma_\infty,J]
-2\kappa\widetilde\Sigma_\infty
+Q
-\widetilde\Sigma_\infty R\widetilde\Sigma_\infty.
\label{eq:ARE}
\end{equation}
The single-detector configuration is recovered as a continuous limiting case. By \eqref{rotatedcov}, the covariance $\Sigma_t$ then approaches the dynamic steady state
\begin{align}
\|
\Sigma_t-
U_{\Omega t+\theta_0}
\widetilde\Sigma_\infty
U_{\Omega t+\theta_0}^{\top}
\|
\rightarrow0,
\end{align}
where $\|\cdot\|$ is any matrix norm. Moreover, the solution to \eqref{eq:rotatedRDE} is given by 
\begin{align}
\widetilde\Sigma_t
={}&
\widetilde\Sigma_\infty
+
e^{-\Gamma t}
\left[I+\Delta_0 G_t\right]^{-1}
\Delta_0 e^{-\Gamma^\top t},
\label{eq:transient}
\end{align}
where $\Delta_0:=\widetilde\Sigma_0-\widetilde\Sigma_\infty$ and
\begin{align}
\Gamma
&:=
\kappa I+\omega_{\rm eff}J+\widetilde\Sigma_\infty R,
\label{eq:Gamma}\\
G_t
&:=
\int_0^t
e^{-\Gamma^\top\tau}
R
e^{-\Gamma\tau}
\,\dd\tau .
\end{align}
This explicit transient solution is typically derived by defining the deviation from the steady state, $\Delta_t := \widetilde{\Sigma}_t - \widetilde{\Sigma}_\infty$, and then looking at the corresponding dynamics for its inverse $V_t = \Delta_t^{-1}$. This standard technique transforms the nonlinear Riccati equation into a linear differential equation which yields the solution. Finally, let $\sigma(\Gamma)$ denote the set of $\Gamma$ eigenvalues.
For any $\lambda\in\sigma(\Gamma)$ we have that $\Re(\lambda)>0,$ that is, the rate matrix $\Gamma$ has eigenvalues with strictly positive real parts. 

\subsection{Quantifying purification advantage}

The two metrics we consider are the \textit{steady-state purity}
\begin{align}
p_\infty
:=
\frac{1}{2\sqrt{\det\widetilde\Sigma_\infty}},
\end{align}
and the \textit{relaxation rate}
\begin{align}
r
:=
\min_{\lambda\in\sigma(\Gamma)}
2\Re(\lambda),
\end{align}
which characterize, respectively, the sustained purity and the rate at which it is approached. In what follows, we express them explicitly in terms of $\omega_{\rm eff}$ and the bath/detector parameters $\kappa$, $R$, and $Q$. The detailed proofs of these expressions can be found in Appendix \ref{app:closedform}.

Define
\begin{align}
\begin{cases}
S(\omega_{\rm eff})
:=
2\kappa^2+\tr(RQ)-2\omega_{\rm eff}^2,\\
P(\omega_{\rm eff})
:=
(\kappa^2+\omega_{\rm eff}^2)^2
+\kappa^2\tr(RQ)\\
\hspace{10mm}
+\omega_{\rm eff}^2\tr(RJQJ^\top)
+\det(RQ).
\end{cases}
\end{align}
Then
\begin{align}\label{rformula}
r(\omega_{\rm eff})
&=
\sqrt{S(\omega_{\rm eff})+2\sqrt P(\omega_{\rm eff})}
\nonumber\\&-
\sqrt{\max(S(\omega_{\rm eff})-2\sqrt P(\omega_{\rm eff}),0)}.
\end{align}
Likewise, define
\begin{align}
\begin{cases}
\tau(\omega_{\rm eff})
:=
\sqrt{S(\omega_{\rm eff})+2\sqrt P(\omega_{\rm eff})},\\
\alpha(\omega_{\rm eff})
:=
\tau(\omega_{\rm eff})^2+2\kappa\tau(\omega_{\rm eff})
-\frac12\tr R\tr Q,\\
\beta(\omega_{\rm eff})
:=
\tau(\omega_{\rm eff})^2+2\kappa\tau(\omega_{\rm eff})-\tr(RQ),\\
\zeta
:=
2(\tr Q)^2,\\
\delta
:=
8\det Q .
\end{cases}
\end{align}
Then
\begin{align}\label{pformula}
p_\infty(\omega_{\rm eff})
=
\sqrt{
\frac{
4\omega_{\rm eff}^2\alpha(\omega_{\rm eff})+\tau^2\beta(\omega_{\rm eff})
}{
4\omega_{\rm eff}^2\zeta+\tau^2\delta
}
}.
\end{align}
Note that both metrics, \eqref{rformula} and \eqref{pformula}, are even functions of $\omega_{\rm eff}$ and are independent of $\theta_0$. 

Under assumptions \eqref{ass1} and \eqref{ass2}, both $r$ and $p_\infty$ are increasing in $\omega_{\rm eff}^2$ (see Appendix \ref{app:monotone} for the detailed proof), and so the lowest values occur at $\omega_{\rm eff}=0$,
\begin{align}\label{bathat0}
\begin{cases}
r(0)
=
\tr G-\sqrt{(\tr G)^2-4\det G},
\\[1mm]
p_\infty(0)
=
\sqrt{
\cfrac{\det(\kappa I+G)}
{4\det Q}
},
\end{cases}
\end{align}
where $G:=\sqrt{\kappa^2I+RQ}$, and as $\omega_{\rm eff}^2\rightarrow\infty$ both approach their highest values given by the fast rate limit
\begin{align}\label{bathlargeomega}
\begin{cases}
r_{\rm fr}
=
2\kappa
\sqrt{1+q\nu(\nu+2)},
\\[1mm]
p_{\rm fr}
=
\left(\cfrac{
\sqrt{1+q\nu(\nu+2)}+1
}{
\nu+2
}\right)\sqrt{\eta_{\rm th}},
\end{cases}
\end{align}
with an $O(\omega_{\rm eff}^{-2})$ error. Here $\nu$ and $q$ are the dimensionless measurement-to-bath ratios defined in \eqref{eq:ratios}. Interestingly, the fast rate quantities \eqref{bathlargeomega} have no explicit dependence on the oscillator--bath coupling strength $\gamma:={\kappa}/{\omega}.$ 

Figure~\ref{fig:F2} illustrates these results for several bath decay rates. Both the sustained purity and relaxation rate are minimized at the co-rotating point \(\omega_{\rm eff}=0\) and increase toward their fast-rotation plateaus as \(|\omega_{\rm eff}|\) grows. The lab-fixed configuration \(\Omega=0\), corresponding to \(\omega_{\rm eff}=\omega\), lies between these two limits. As $\kappa\rightarrow0$, the endpoints reduce to combinations of arithmetic and geometric means
\begin{align}\label{at0}
\begin{cases}
p_\infty(0)
=
(\eta_1\eta_2)^{1/4},
&
p_{\rm fr}
=
\sqrt{\eta_{\rm eff}},
\\[1mm]
r(0)
=
\sqrt{k_1k_2}\sqrt{\eta_{\min}},
&
r_{\rm fr}
=
\sqrt{\eta_{\rm eff}}\,k_{\rm eff},
\end{cases}
\end{align}
with $k_{\rm eff}$ and $\eta_{\rm eff}$ as in \eqref{eq:effective-coef}, and as $\kappa \rightarrow \infty$
\begin{align}
\begin{cases}
    p_\infty(\omega_{\rm eff})
\rightarrow
\sqrt{\eta_{\rm th}},\\
r(\omega_{\rm eff})
\sim
2\kappa.
\end{cases} 
\end{align}

\begin{figure}
\centering
\includegraphics[width=\linewidth]{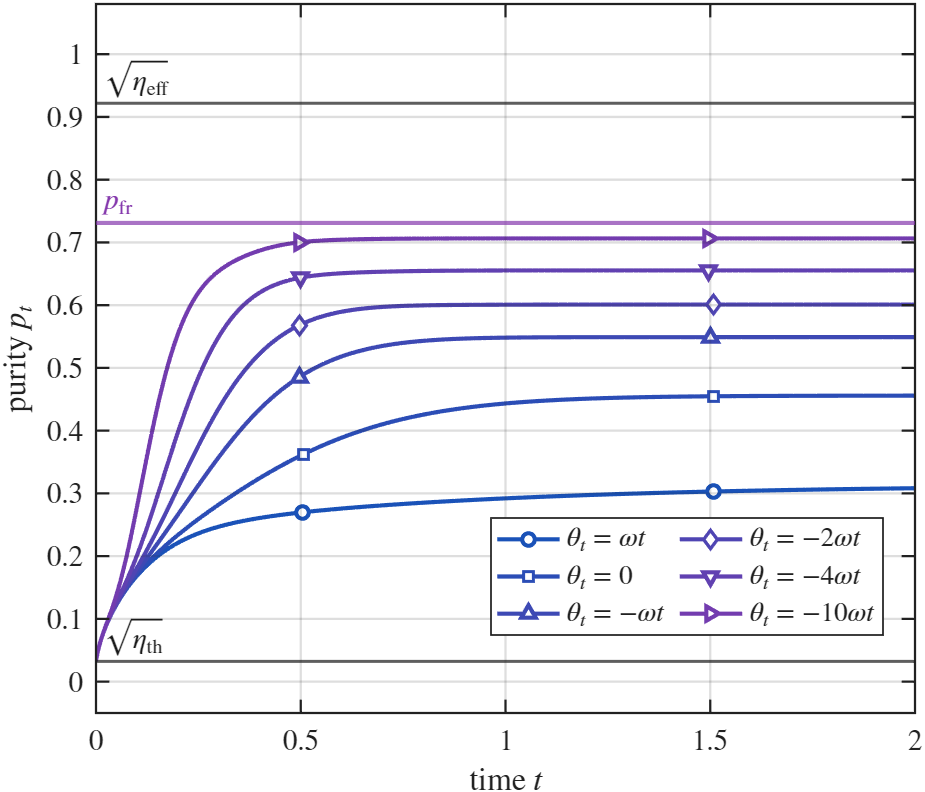}
\caption{Purity dynamics \(p_t\) for several constant-rate protocols starting from the thermal state \(\Sigma_{\rm th}\). Increasing the effective relative rotation rate accelerates convergence and raises the sustained purity. We take $\omega = 1$, \(k_1=10\), \(k_2=0.1\), \(\eta_1=0.85\), \(\eta_2=0.30\), and \(\bar n=14.5\).}
\label{fig:F3}
\end{figure}

Figure~\ref{fig:F3} shows corresponding transient responses starting from the thermal state. Increasing the relative rotation rate accelerates convergence and raises the sustained purity. For the chosen parameters,
\begin{align}
p_\infty(2\omega)\approx0.555,
\qquad
p_\infty(\omega)\approx0.462,
\end{align}
so a detector rotating opposite to the oscillator at the oscillator frequency already realizes a substantial fraction of the available improvement.

\subsection{The fast-rate limit}

We saw that the best sustained purity is achieved in the fast-rate limit \eqref{bathlargeomega}. The physical effect behind this is simply high-frequency  angular averaging. That is, as $|\Omega| \rightarrow\infty$, the Riccati dynamics \eqref{eq:RDE} become effectively  
\begin{align}\label{eq:avgRDE}
\dot{\bar\Sigma}_t
={}&
\omega[\bar\Sigma_t,J]
+2\kappa(\Sigma_{\rm th}-\bar\Sigma_t)
+\bar K-\bar\Sigma_t\bar R\bar\Sigma_t,
\end{align}
where $\bar{R}$ and {$\bar{K}$} are defined via
\begin{align}
\bar X
&:=
\frac{1}{T}\int_0^T X_{t}\,\dd t,~~T=2\pi/\Omega.
\end{align}
 This is the usual first-order high-frequency averaging approximation \cite{sanders2007averaging}. Utilizing the formula
\begin{align}
\bar M
:=
\frac{1}{2\pi}
\int_0^{2\pi}
e^{-\theta J}Me^{\theta J}\,\dd\theta
=
\frac{\tr M}{2}I,
\end{align}
the averaged measurement coefficients are simply
\begin{align}\label{eq:averaged-meas}
\bar R
&=
\eta_{\rm eff}k_{\rm eff}I,
&
\bar K
&=
\frac{k_{\rm eff}}{4}I.
\end{align}
Thus, rapid relative rotation realizes an effective {rotationally averaged} measurement with two identical detectors of strength $k_{\rm eff}$ and efficiency $\eta_{\rm eff}$, giving the purity $p_{\rm fr}$ in \eqref{bathlargeomega}.

\begin{figure*}[t]
\centering
\includegraphics[width=\textwidth]{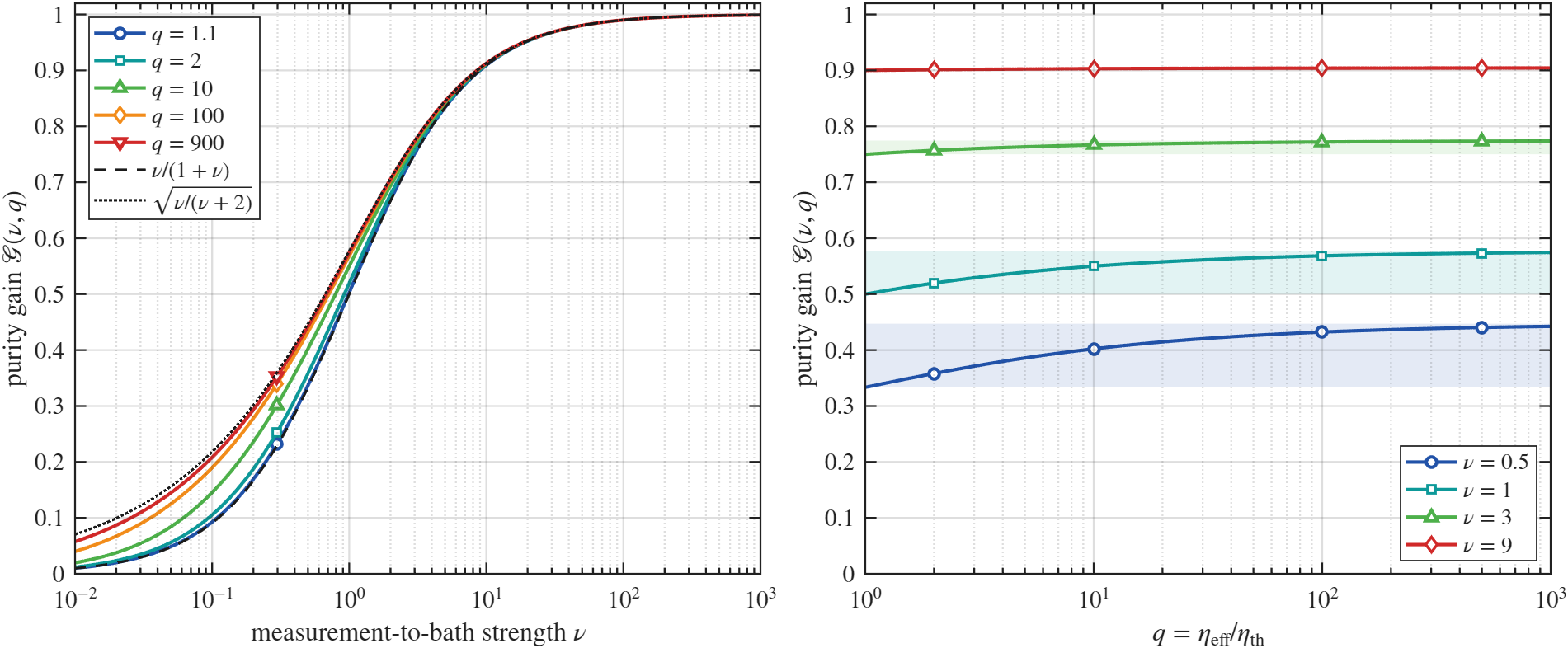}
\caption{Left: fast-rotation purity gain \(\mathcal G(\nu,q)\) versus the measurement-to-bath strength \(\nu\) for several detector-to-bath quality ratios \(q=\eta_{\rm eff}/\eta_{\rm th}\); dashed and dotted curves show the universal bounds \(\nu/(1+\nu)\) and \(\sqrt{\nu/(\nu+2)}\), respectively. Right: \(\mathcal G(\nu,q)\) versus \(q\) for several values of \(\nu\), with the corresponding \(\nu\)-dependent bounds shown as shaded bands. The measurement-to-bath strength \(\nu\) sets the scale over which improvements in detector quality translate into fast-rotation purity gain.}
\label{fig:gain}
\end{figure*}

It is not difficult to see that the fast-rate limit $p_{\rm fr}$ lies somewhere in the interval 
\begin{align}
  [\sqrt{\eta_{\rm th}},\sqrt{\eta_{\rm eff}}],
\end{align}
depending on how strong $\kappa$ is. Thus, a natural quantity capturing the effectiveness of the fast-rate limit is the fraction
\begin{align}
\mathcal G(\nu,q)
:=
\frac{
p_{\rm fr}-\sqrt{\eta_{\rm th}}
}{
\sqrt{\eta_{\rm eff}}-\sqrt{\eta_{\rm th}}
}
=
\frac{
\nu(\sqrt q+1)
}{
\sqrt{1+q\nu(\nu+2)}+\nu+1
}.
\label{eq:fractionalgain}
\end{align}
For fixed $\nu$, $\mathcal G$ is increasing in $q$ and obeys the universal bounds
\begin{align}
\frac{\nu}{1+\nu}
<
\mathcal G(\nu,q)
<
\sqrt{\frac{\nu}{\nu+2}},
\qquad q>1,
\label{eq:gainbound}
\end{align}
which are proven in Appendix \ref{app:gain}. Moreover, the upper and lower bounds are asymptotically sharp as $q\to\infty$ and $q\to 1$, respectively.
Thus, irrespective of the quality factor $q$, the strength ratio $\nu$ universally bounds the fraction of the available purity improvement reached by rapid rotation. For example, $\nu=1$, $3$, and $9$ correspond to lower bounds of $1/2$, $3/4$, and $9/10$, respectively. Figure~\ref{fig:gain} displays $\mathcal G(\nu,q)$ and these $\nu$-dependent bounds.

 We next consider vibrational protocols, whose additional angular degrees of freedom enlarge the set of effective measurements that can be realized. 

\section{Vibrational protocols}\label{sec:vibrational}

We now consider the open-loop sinusoidal, or \textit{vibrational}, protocols
\begin{align}\label{vibr}
\theta_t=a\sin(\Omega t)+\theta_0,
\end{align}
in the high-frequency regime, i.e. with $|\Omega|$ large compared to the relevant covariance evolution rates, for $s = 0$. The covariance dynamics \eqref{eq:RDE} are approximated by their averaged version
\begin{align}\label{eq:avgRDEvib}
\dot{\bar\Sigma}_t
={}&
\omega[\bar\Sigma_t,J]
+2\kappa(\Sigma_{\rm th}-\bar\Sigma_t)
+\bar K-\bar\Sigma_t\bar R\bar\Sigma_t,
\end{align}
where
\begin{align}\label{avgmeas}
\bar R
&:=
\frac{1}{T}\int_0^T R_{\theta_t}\,\dd t,
&
\bar K
&:=
\frac{1}{T}\int_0^T K_{\theta_t}\,\dd t,
\end{align}
and $T=2\pi/\Omega$. Again, this is the usual first-order high-frequency averaging approximation \cite{sanders2007averaging}.

\begin{figure}
\centering
\includegraphics[width=\linewidth]{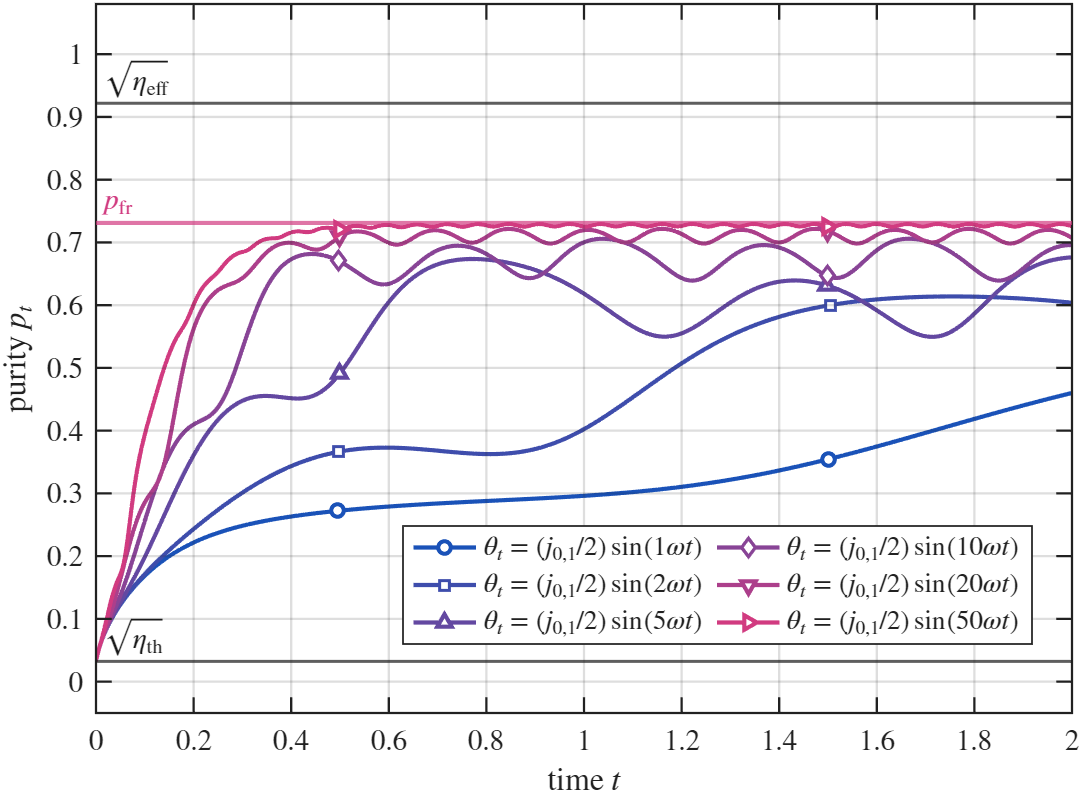}
\caption{Purity dynamics under the vibrational protocols \(\theta_t=(j_{0,1}/2)\sin(\Omega t)\) for several modulation frequencies \(\Omega\). As \(|\Omega|\) increases, the exact time-dependent dynamics approach the high-frequency averaged dynamics and the sustained purity converges to the fast-rate value \(p_{\rm fr}\). The parameters are \(\omega=1\), \(\kappa=0.05\), \(\eta_1=0.85\), \(\eta_2=0.80\), \(k_1=20\), \(k_2=0.1\), \(\bar n=15\), and \(s=0\), with the $\Sigma_0$ equal to the thermal state.}
\label{fig:F6}
\end{figure}

Because $R_{\theta_t}$ and $K_{\theta_t}$ depend quadratically on $U_{\theta_t}$, they contain only second angular harmonics,
\begin{align}\label{coeffsvib}
\begin{cases}
R_{\theta_t}
=
a_R+b_R\cos2{\theta_t}+c_R\sin2{\theta_t},
\\
K_{\theta_t}
=
a_K+b_K\cos2{\theta_t}+c_K\sin2{\theta_t},
\end{cases}
\end{align}
where
\begin{align}
\begin{cases}
a_R=(R_0+R_{\pi/2})/2=\eta_{\rm eff}k_{\rm eff}I,
\\
b_R=(R_0-R_{\pi/2})/2,
\\
c_R=R_{\pi/4}-a_R,
\end{cases}\end{align}
and
\begin{align}
\begin{cases}
a_K=(K_0+K_{\pi/2})/2=\frac{k_{\rm eff}}{4}I,
\\
b_K=(K_0-K_{\pi/2})/2,
\\
c_K=K_{\pi/4}-a_K.
\end{cases}
\end{align} Plugging \eqref{coeffsvib} in \eqref{avgmeas} and using the formula
\begin{align}
    \frac{1}{T}\int_0^T\cos2\theta_t dt= J_0(2a)\cos 2\theta_0,
\end{align}
yields 
\begin{align}\label{avging}
\begin{cases}
\bar R&=
a_R+J_0(2a)(b_R\cos2{\theta_0}+c_R\sin2{\theta_0}),
\\
&=
(1-J_0(2a))\eta_{\rm eff}k_{\rm eff}I
+
J_0(2a)R_{\theta_0},
\\
\bar K 
&=
a_K+J_0(2a)(b_K\cos2{\theta_0}+c_K\sin2{\theta_0}),\\
&=
(1-J_0(2a))\frac{k_{\rm eff}}{4}I
+
J_0(2a)K_{\theta_0},
\end{cases}
\end{align}
where $J_0$ is the Bessel function of the first kind. Observe that the vibration amplitude $a$ appearing in $J_0(2a)$ controls the degree of angular mixing. For $a=0$, the protocol is constant and no averaging occurs. At half of the first Bessel zero
\begin{align}
a=\frac{j_{0,1}}{2},
\qquad
j_{0,1}\simeq2.4048,
\end{align}
one has $J_0(2a)=0$, and
\begin{align}\label{avergd}
\bar R
=
\eta_{\rm eff}k_{\rm eff}I,~~~\bar K
=
\frac{k_{\rm eff}}{4}I.
\end{align}
These are precisely the fully averaged coefficients \eqref{eq:averaged-meas} obtained in the fast constant-rate limit. Because of this, the sustained steady-state purity in this case is again $p_{\rm fr}$, albeit using a vibrational protocol rather than a constant-rate one. In other words, the vibration protocol only requires access to a restricted angle range $\theta_t^*\in[-j_{0,1}/2,j_{0,1}/2]$ to obtain the same averaging effect as the fast-rate protocol, which sweeps over all possible angles.
Figure~\ref{fig:F6} illustrates the convergence of the exact vibrational dynamics toward this limit as the frequency $|\Omega|$ increases. 

\subsection{Characterizing all sustained purities}
We characterize next all sustained purities that can be obtained from the choices ($a$,$\theta_0$), which read naturally as radial coordinates. To this end, define 
\begin{align}
u
&=
J_0(2a)\cos2\theta_0,
&
v
&=
J_0(2a)\sin2\theta_0.
\label{eq:uvvibration}
\end{align}
Then equation~\eqref{avging} becomes
\begin{align}
(\bar R,\bar K)
=
(a_R,a_K)
+u(b_R,b_K)
+v(c_R,c_K).
\end{align}
By varying $a$ and $\theta_0$, the accessible averaged measurements fill the unit disk
\begin{align}
\mathcal D
:=
\{(u,v):u^2+v^2\leq1\}.
\end{align}
Its boundary $(J_0(2a)=1)$ corresponds to constant-angle monitoring, whereas its center {$(J_0(2a)=0)$},
\begin{align}
(u,v)=(0,0),
\end{align}
gives the fast-rate limit $p_{\rm fr}$. Since every point in $\mathcal D$ lies on a chord joining two boundary points, any $(\bar R,\bar K)$ can be written as
\begin{align}\label{cvxcomb}
(\bar R,\bar K)
=
\tau(R_{\phi_1},K_{\phi_1})
+
(1-\tau)(R_{\phi_2},K_{\phi_2}),
\quad
0\leq\tau\leq1.
\end{align}
Thus, high-frequency vibration may be viewed as engineering an effective mixture of constant-angle measurements with a continuously tunable angular bias. 

\subsection{Optimal vibrational strategy}
For each pair $(u,v)$ in the disk $\mathcal{D}$, let $p_\infty(u,v)$ denote the corresponding steady-state purity in the high-frequency limit. It is not difficult to show that because $s= 0$, $p_\infty(u,v)$ is radially symmetric, making the offset $\theta_0$ immaterial. 

The best sustained purity accessible within the disk is therefore
\begin{align}
p^\star
:=
\max_{(u,v)\in\mathcal D}
p_\infty(u,v),
\label{eq:hullceiling}
\end{align}
where, by construction, $p^\star\geq p_\infty(0,0)=p_{\rm fr}$. It turns out that the inequality saturates for $s = 0$, so that
\begin{align}
    p^\star= p_\infty(0,0)=p_{\rm fr}.
\end{align}
That is, the best vibrational strategy corresponds to the fast-rate limit $(u^\star,v^\star)=(0,0)$,
where
\begin{align}
\theta_t^\star
=
\frac{j_{0,1}}{2}\sin(\Omega t),
\end{align}
with \(|\Omega|\) sufficiently large. The reason for this is the same as the one observed in the constant-rate section: when one detector is decisively better than the other, i.e., better strength and efficiency \eqref{ass1}, it is always more favorable to average those higher detector specs over all angular directions in order to yield the best squeezing. Any off-center point, $(u,v)\neq 0$, reduces the extent of this mixing, thereby lowering the sustained purity. We illustrate this in the {left-hand side} of Figure~{\ref{fig:F7}}, where we plot the sustained purity $p_\infty(u,v)$ as a function of $(u,v)\in\mathcal{D}$ for different values of $\kappa$. In all three cases, the peak of the surface occurs at the center. A proof of this fact uses \eqref{cvxcomb} is given in Appendix~\ref{app:harmonics}.

\section{Greedy protocols}\label{sec:greedy}

We now consider the oscillator analogue of qubit rapid purification \cite{jacobs2003rapid}, obtained by selecting at each instant the measurement angle that maximizes the instantaneous purification rate \eqref{eq:detp}. By \eqref{eq:measterm}, the angle-dependent contribution is determined by
\begin{align}
\tr\!\left[\Sigma_t^{-1}\dot\Sigma^{\rm m}_{t,\theta}\right]
&=
\big\langle U_\theta^\top\Sigma_tU_\theta,\Lambda_t\big\rangle,
\label{eq:innerproduct}\\
\Lambda_t
&:=
\diag\!\left(
k_1(p_t^2-\eta_1),
k_2(p_t^2-\eta_2)
\right),
\nonumber
\end{align}
so that maximizing $\dot p_t$ is equivalent to minimizing the inner product in \eqref{eq:innerproduct}. We first determine this instantaneous optimum for $s=0$ and then discuss its implementation and long-time performance.

\subsection{Optimal instantaneous angle}

Since the dependence on $\theta$ is quadratic in $U_\theta$, the inner product contains only second angular harmonics,
\begin{align}
\big\langle U_\theta^\top\Sigma_tU_\theta,\Lambda_t\big\rangle
=
a_t+b_t\cos2\theta+c_t\sin2\theta,
\label{eq:greedyharmonic}
\end{align}
where, for $U_\theta=e^{-\theta J}$,
\begin{align}\label{eq:measharmonic}
\begin{cases}
a_t
=
\frac{1}{2}\tr\Lambda_t\,\tr \Sigma_t,
\\[1mm]
b_t
=
\frac{1}{2}
\tr(\Lambda_t\sigma_z)
\tr(\Sigma_t\sigma_z),
\\[1mm]
c_t
=
-\frac{1}{2}
\tr(\Lambda_t\sigma_z)
\tr(\Sigma_t\sigma_x),
\end{cases}
\end{align}
with
\begin{align}
\sigma_z=
\begin{bmatrix}
1&0\\
0&-1
\end{bmatrix},
\qquad
\sigma_x=
\begin{bmatrix}
0&1\\
1&0
\end{bmatrix}.
\end{align}
Whenever $b_t^2+c_t^2>0$, the minimizing angle is therefore
\begin{align}
\theta_t^\star
=
\frac{1}{2}\atanTwo(-c_t,-b_t)
\qquad(\mathrm{mod}\;\pi).
\label{eq:thetastar}
\end{align}
Here $\atanTwo(y,x)\in(-\pi,\pi]$ denotes the polar angle of the vector $(x,y)$, defined by
\begin{align}
\cos[\atanTwo(y,x)]
&=
\frac{x}{\sqrt{x^2+y^2}},
\nonumber\\
\sin[\atanTwo(y,x)]
&=
\frac{y}{\sqrt{x^2+y^2}}.
\end{align}
The modulo-$\pi$ identification follows because the measurement matrices are quadratic in $U_\theta$, so that $\theta$ and $\theta+\pi$ describe the same measurement. Thus, at the optimum \eqref{eq:thetastar},
\begin{align}
\min_\theta
\tr\!\left[
\Sigma_t^{-1}\dot\Sigma^{\rm m}_{t,\theta}
\right]
=
a_t-\sqrt{b_t^2+c_t^2},
\label{eq:greedyminimum}
\end{align}
giving the maximal instantaneous measurement-induced purification
\begin{align}\label{contgreedy}
\dot p_t^{\rm m}
=
\frac{p_t}{2}
(\sqrt{b_t^2+c_t^2}-a_t),
\end{align}

The geometric interpretation of this solution is most transparent in terms of the principal axes of $\Sigma_t$. To understand this, ignore first the common factor $\tr(\Lambda_t\sigma_z)$ in $b_t$ and $c_t$ and define the angle
\begin{align}
\phi_t
:=
\frac{1}{2}
\atanTwo\!\left(
-\tr(\Sigma_t\sigma_x),
\tr(\Sigma_t\sigma_z)
\right).
\end{align}
Then, $\phi_t$ as a control angle aligns the first detector axis with the major principal axis of $\Sigma_t$, while $\phi_t+\pi/2$ aligns it with the minor axis. Defining
\begin{align}
L_t
&:=
\tr(\Lambda_t\sigma_z)=
k_1(p_t^2-\eta_1)
-
k_2(p_t^2-\eta_2),
\end{align}
we then have 
\begin{align}
&\big\langle U_\theta^\top\Sigma_tU_\theta,\Lambda_t\big\rangle
=
a_t\nonumber\\&+
\frac{L_t}{2}
\sqrt{
\tr(\Sigma_t\sigma_z)^2+
\tr(\Sigma_t\sigma_x)^2
}
\cos\!\left[2(\theta-\phi_t)\right],
\end{align}
Hence,
\begin{align}
\theta_t^\star
=
\begin{cases}
\phi_t,&L_t<0,\\[1mm]
\phi_t+\frac{\pi}{2},&L_t>0,
\end{cases}
\qquad(\mathrm{mod}\;\pi).
\end{align}
Thus the greedy controller always assigns the major axis of $\Sigma_t$, namely the direction of largest uncertainty, to the detector having the smaller instantaneous coefficient $k_j(p_t^2-\eta_j)$. The sign of $L_t$ determines whether this is detector $1$ or detector $2$. Intuitively, the greedy protocol tries to equalize the two eigenvalues of \(\Sigma_t\). 
\subsection{Sample-and-hold implementation}

We implement the greedy protocol in a sample-and-hold form. Let $t_n=n\Delta t$. At each update time the minimizing angle is evaluated from the current covariance,
\begin{align}
\theta_n^\star
:=
\theta^\star(\Sigma_{t_n}),
\end{align}
and held fixed over the subsequent interval,
\begin{align}
\theta_t=\theta_n^\star,
\qquad
t_n\leq t<t_{n+1}.
\label{eq:samplehold}
\end{align}
The update period $\Delta t$ imposes a finite switching bandwidth, and the complete control sequence may in principle be computed offline since it does not require real-time access to the stochastic measurement record.

\begin{figure*}[t]
\centering
\includegraphics[width=\linewidth]{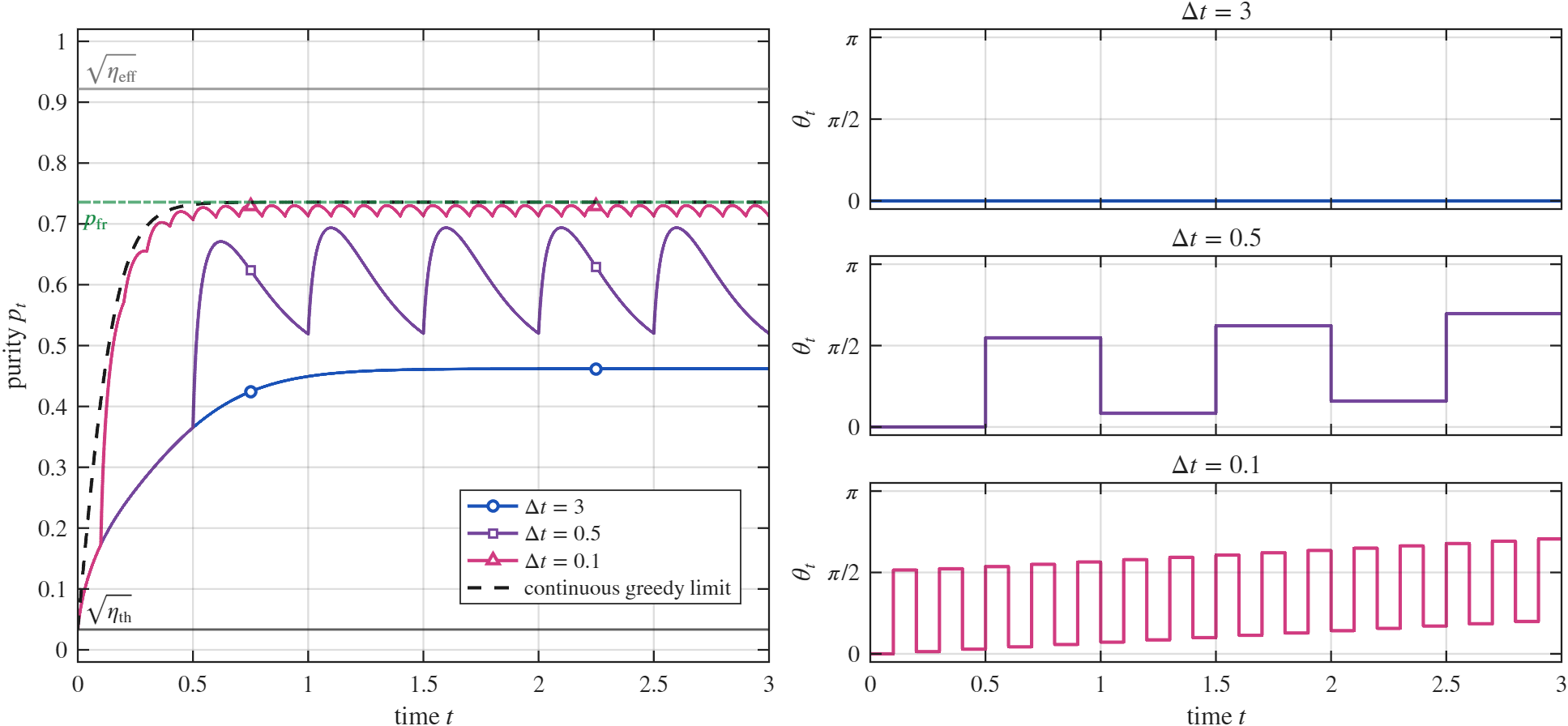}
\caption{Sample-and-hold implementation of the greedy protocol for the isotropic benchmark \(s=0\). Left: purity trajectories for update periods \(\Delta t=3\), \(0.5\), and \(0.1\), together with the continuous greedy limit and the reference levels \(\sqrt{\eta_{\rm th}}\), \(p_{\rm fr}\), and \(\sqrt{\eta_{\rm eff}}\). As the update period decreases, the sample-and-hold trajectories approach the continuous chattering limit and its sustained purity \(p_{\rm fr}\). Right: the corresponding held control angles \(\theta_t\), shown separately for each \(\Delta t\). Smaller update periods produce increasingly rapid switching between nearly orthogonal measurement orientations. The benchmark parameters are \(\omega=1\), \(\kappa=0.05\), \(\bar n=14.5\), \(\eta_1=0.85\), \(\eta_2=0.80\), \(k_1=20\), and \(k_2=0.1\).}
\label{fig:F9}
\end{figure*}

Figure~\ref{fig:F9} illustrates this implementation. As $\Delta t$ is reduced, the sample-and-hold purity approaches the continuous greedy trajectory \eqref{contgreedy}, while the corresponding controls switch increasingly rapidly between nearly orthogonal measurement orientations. The resulting rapid alternation gives equal weight to two orthogonal orientations and therefore reproduces the same effective measurement as uniform fast rotation. Consequently, the sustained greedy purity approaches $p_{\rm fr}$.

\subsection{Purification and system entropy rate}\label{sec:entropy}

The greedy protocol has a straightforward entropic  interpretation. For a single-mode Gaussian state, the von Neumann entropy is a function of the inverse purity,
\begin{align}
S(\hat\rho_t)
=
h(1/p_t),
\end{align}
where
\begin{align}
h(x)
:=
\frac{x+1}{2}\ln\frac{x+1}{2}
-
\frac{x-1}{2}\ln\frac{x-1}{2}.
\label{eq:vNentropy}
\end{align}
The function $h$ is strictly increasing for $x>1$ \cite{serafini2017quantum}, so 
\begin{align}
\frac{\dd S}{\dd t}
=
-\frac{h'(1/p_t)}{p_t^2}\dot p_t,
\qquad
\frac{h'(1/p_t)}{p_t^2}>0.
\end{align}
It follows immediately that
\begin{align}
\arg\min_\theta\frac{\dd S}{\dd t}
=
\arg\max_\theta\dot p_t.
\end{align}
Thus the greedy law may equivalently be viewed as the control that minimizes the instantaneous rate of change of the conditional system von Neumann entropy. A complete thermodynamic entropy-production balance, which may additionally contain environmental and information-flow contributions, is left for future study. Related formulations of entropy production under continuously monitored Gaussian dynamics, including treatments based on Wigner entropy, are discussed in Ref.~\cite{belenchia2020entropy}.
\section{Variational protocols}\label{sec:variational}

The preceding protocols restrict the measurement angle to specific, physically transparent families. In this section, we remove these restrictions and optimize the accumulated performance of the time-dependent control over a finite interval \(0\leq t\leq T\)
\begin{align}
\min_{\theta(\cdot)}
\mathcal J_T[\theta]
:=
\int_0^T
\ln\det\Sigma_t\,dt,
\label{eq:variationalproblem}
\end{align}
subject to
\begin{align}
\dot\Sigma_t
=
\omega[\Sigma_t,J]
+2\kappa(\Sigma_{\rm th}-\Sigma_t)
+K_{\theta_t}
-\Sigma_tR_{\theta_t}\Sigma_t.
\label{eq:variationaldynamics}
\end{align}
Since
\begin{align}
\ln\det\Sigma_t
=
-2\ln(2p_t),
\end{align}
minimizing \eqref{eq:variationalproblem} is equivalent to maximizing the time-averaged logarithmic purity. A convenient performance measure is therefore the geometric-mean purity
\begin{align}
p_{\rm geom}(T)
:=
\exp\left[
\frac{1}{T}
\int_0^T
\ln p_t\,dt
\right].
\label{eq:pgeom}
\end{align}

Using the double-angle decomposition of the measurement matrices introduced in
Eq.~\eqref{coeffsvib}, we reparameterize the physical control angle $\theta_t$ by
\begin{align}
u_t
:=
\begin{pmatrix}
\cos2\theta_t\\
\sin2\theta_t
\end{pmatrix},
\qquad
\|u_t\|=1.
\label{eq:variationalu}
\end{align}
The measurement matrices then take
the affine form
\begin{align}
R_u
&=
a_R+u_1b_R+u_2c_R,
\nonumber\\
K_u
&=
a_K+u_1b_K+u_2c_K.
\end{align}
The original optimization over $\theta_t$ may therefore be performed equivalently
over the unit-vector control $u_t$. Writing the covariance dynamics \eqref{eq:RDE} as
\begin{align}\label{state}
\dot\Sigma
=
A\Sigma+\Sigma A^\top
+2\kappa\Sigma_{\rm th}
+K_u-\Sigma R_u\Sigma,
\end{align}
with $A=-\omega J-\kappa I$, we introduce the symmetric costate matrix $\Pi_t$ and the Pontryagin Hamiltonian
\begin{align}
\mathcal H(\Sigma,\Pi,u)
&=
\ln\det\Sigma
\nonumber\\
&\quad+
\big\langle
\Pi,
A\Sigma+\Sigma A^\top
+2\kappa\Sigma_{\rm th}
+K_u-\Sigma R_u\Sigma
\big\rangle .
\label{eq:PHamiltonian}
\end{align}
For an optimal trajectory, the state evolves according to
Eq.~\eqref{eq:variationaldynamics}, the costate obeys
$\dot\Pi_t=-\partial\mathcal H/\partial\Sigma$, and the admissible control
$u_t\in S^1$ minimizes $\mathcal H$ pointwise in time. Since the objective contains
no terminal cost, the corresponding transversality condition is
\begin{align}
\Pi_T=0.
\label{eq:costateterminal}
\end{align}
Using
\begin{align}
\frac{\partial}{\partial\Sigma}
\ln\det\Sigma
=
\Sigma^{-1},
\end{align}
gives
\begin{align}
\dot\Pi_t
=
-\Sigma_t^{-1}
-A^\top\Pi_t-\Pi_tA
+\Pi_t\Sigma_tR_{u_t}
+R_{u_t}\Sigma_t\Pi_t.
\label{eq:costate}
\end{align}

The pointwise control minimization is particularly simple because the
double-angle parametrization makes the Hamiltonian affine in $u_t$.
Indeed, substituting the affine forms of \(R_u\) and \(K_u\) into
Eq.~\eqref{eq:PHamiltonian} separates the terms independent of the control from
those proportional to \(u_1\) and \(u_2\):
\begin{align}
\mathcal H_u
=
\mathcal A_t
+\mathcal B_tu_1
+\mathcal C_tu_2,
\end{align}
where
\begin{align}
\mathcal B_t
&=
\big\langle
\Pi_t,
{b_K}-\Sigma_tb_R\Sigma_t
\big\rangle,
\nonumber\\
\mathcal C_t
&=
\big\langle
\Pi_t,
{c_K}-\Sigma_tc_R\Sigma_t
\big\rangle.
\end{align}
At each time, we must therefore choose a unit vector $u_t$ that minimizes the
linear form $\mathcal B_tu_1+\mathcal C_tu_2$. The minimizing vector is given by 
\begin{align}
u_t^\star
=
-\frac{
(\mathcal B_t,\mathcal C_t)^\top
}{
\sqrt{\mathcal B_t^2+\mathcal C_t^2}
},
\label{eq:uPMP}
\end{align}
whenever $\mathcal B_t^2+\mathcal C_t^2>0$. If both coefficients vanish, the
Hamiltonian is instantaneously independent of the measurement angle. The physical
control is therefore
\begin{align}
\theta_t^\star
=
\frac12
\atanTwo(u_{2,t}^\star,u_{1,t}^\star)
\qquad(\mathrm{mod}\;\pi).
\end{align}
Equation~\eqref{eq:uPMP} also makes the relation to greedy control explicit.
If the costate is replaced by the instantaneous purity-gradient direction,
$\Pi_t\propto\Sigma_t^{-1}$, the same pointwise minimization reduces to the greedy
rule of Sec.~\ref{sec:greedy}. The distinction is that $\Pi_t$ in the variational
problem is determined by the backward equation \eqref{eq:costate} with
$\Pi_T=0$, and therefore encodes how a control choice made at time $t$ affects
the accumulated cost over the rest of the interval. 

The full finite-horizon comparison between all protocols, including the effect of \(s\neq0\), is
given in Sec.~\ref{impactaniso}.

\section{Impact of anisotropy}\label{impactaniso}
In this section, we study the impact of the anisotropy parameter $s$ on the performance of the four protocols. 

\subsection{Constant-rate protocols}\label{subsec:anirate}
For $s = 0$, we saw that the covariance in the fast-rate limit $|\Omega|\rightarrow\infty$ obeys the averaged dynamics 
\begin{align}
\dot{\bar\Sigma}_t
={}&
\omega[\bar\Sigma_t,J]
+2\kappa(\Sigma_{\rm th}-\bar\Sigma_t)+
{\bar K}
-
\bar\Sigma_t\bar R\bar\Sigma_t,
\label{eq:aavgRDE}
\end{align}  
where 
\begin{align}
\label{eq:avgFastMatrices}
\bar R
=
\eta_{\rm eff}k_{\rm eff},~~~~
{\bar K}
=
\frac{k_{\rm eff}}{4}I,
\end{align}
and yields the steady-state purity  
\begin{align}\label{avgformula}
p_{\rm fr}
=
\left(\cfrac{
\sqrt{1+q\nu(\nu+2)}+1
}{
\nu+2
}\right)\sqrt{\eta_{\rm th}}.
\end{align}
For $s\neq 0$, equations \eqref{eq:aavgRDE}, \eqref{eq:avgFastMatrices}, and \eqref{avgformula} become 
\begin{align}
\dot{\bar\Sigma}_t
&=
\omega[\bar\Sigma_t,J]
+2\kappa(\Sigma_{\rm th}-\bar\Sigma_t)+
{\bar K_s}
-
\bar\Sigma_t\bar R_s\bar\Sigma_t,
\label{eq:squeezedAvgRDE}\\
\label{eq:squeezedFastMatrices}
\bar R_s
&=
\eta_{\rm eff}k_{\rm eff}D_s^2,~~~~
{\bar K_s}=\frac{k_{\rm eff}}{4}D_s^{-2},\\
p_{\rm fr}(s)&=f(s,q,\nu,\gamma)\sqrt{\eta_{\rm th}},
\end{align}
where $D_s=\text{diag}(e^s,e^{-s})$ and \(f\) {is a function} we determine below. Notice that $p_{\rm fr}(s)$ now depends on the oscillator--bath coupling
$\gamma=\kappa/\omega$, which was noted absent when $s=0$. 

The formula for $f$ can be deduced from \eqref{pformula} as follows. Notice that \eqref{pformula} is the steady state purity associated with \eqref{eq:rotatedRDE}, which is the same as \eqref{eq:squeezedAvgRDE} modulo the replacement \begin{align}(\omega_{\rm eff},R,{K})\leftrightarrow  (\omega,\bar{R}_s,{\bar{K}_s}).\end{align}
Replacing these coefficients in \eqref{pformula} and simplifying yields 
\begin{equation}
f(s,q,\nu,\gamma)
=
\sqrt{\frac{
1+\sqrt{2\frac{d_s
+
\sqrt{\pi_s}-2}{\gamma^2}}+\frac{\sqrt{\pi_s}-1}{\gamma^2}
-\frac{2\nu^2q}{d_s
+
\sqrt{\pi_s}}s_s^2
}{(\nu+2c_s)^2+(\frac{2\nu^2}{d_s
+
\sqrt{\pi_s}}-4)s_s^2
}},
\label{eq:squeezedPhi}
\end{equation}
where 
\begin{align}
c_s &= \cosh(2s),~s_s=\sinh(2s),\\
d_s&=1+\gamma^2
(
1+q\nu(\nu+2c_s)
),\\
\pi_s
&=d_s^2+4\nu^2s_s^2\gamma^2q(1-\gamma^2q).
\end{align}
As a sanity check, when plugging $s =0$, we get 
\begin{align}
    c_0 &= 1,~~s_0 =0,\\
    d_0 &= 1+\gamma^2(1+q\nu(\nu+2)),\\
    \pi_0 &= d_0^2,
\end{align}
and formula \eqref{eq:squeezedPhi} simplifies to 
\begin{equation}
f(0,q,\nu,\gamma)
=
\frac{
\sqrt{\frac{{d_0}
-1}{\gamma^2}}+1
}{\nu+2
}= \frac{\sqrt{1+q\nu(\nu+2)}+1}{\nu+2},
\label{eq:simplify}
\end{equation}
which agrees with \eqref{avgformula}. 

\begin{figure*}
\centering
\includegraphics[width=\linewidth]{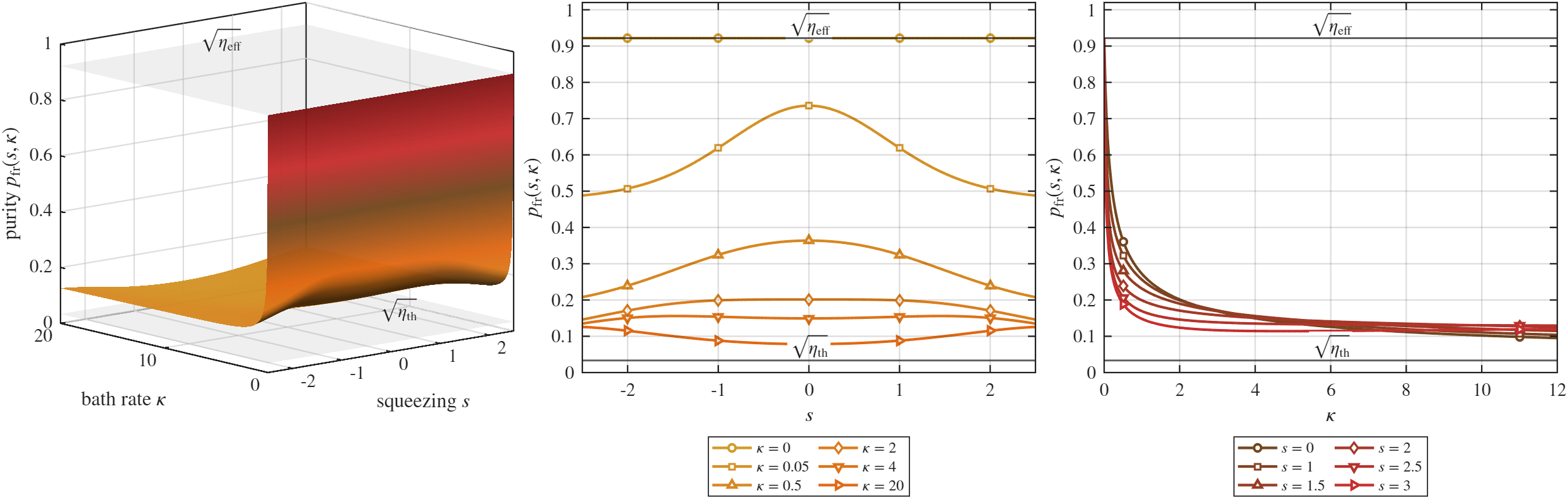}
\caption{Fast-rate sustained purity \(p_{\rm fr}(s,\kappa)\) in {an} anisotropic measurement frame along the fixed-detector parameter sweep used throughout the paper. Here \(q=\eta_{\rm eff}/\eta_{\rm th}\) is fixed, while increasing \(\kappa\) increases \(\gamma=\kappa/\omega\) and decreases \(\nu=k_{\rm eff}/k_{\rm th}\), with \(\gamma\nu\) fixed. At \(\kappa=0\), \(p_{\rm fr}=\sqrt{\eta_{\rm eff}}\) independently of \(s\). Along this trajectory, anisotropy first lowers the sustained purity, passes through a crossover, and eventually raises it. The translucent planes indicate \(\sqrt{\eta_{\rm eff}}\) and \(\sqrt{\eta_{\rm th}}\). We take \(\omega=1\), \(\eta_1=0.85\), \(\eta_2=0.80\), \(k_1=20\), \(k_2=0.1\), and \(\bar n=14.5\).}
\label{fig:F5}
\end{figure*}

If $\kappa =0$, one can show that 
\begin{align}
p_{\rm fr}(s)
=
\sqrt{\eta_{\rm eff}}.
\label{eq:squeezingNoBath}
\end{align}
To see this, take the limit as $\kappa\rightarrow 0$ of $f(s,q,\nu,\gamma)$, this causes $\gamma\rightarrow 0$ and $\nu\rightarrow \infty$ while 
\begin{align}\label{curves}
  q=
\frac{\eta_{\rm eff}}{\eta_{\rm th}},~~~  \gamma\nu =
\frac{\sqrt{\eta_{\rm th}}\,k_{\rm eff}}{2\omega},
\end{align}
are kept constant, and yields  
\begin{equation}
\lim_{\kappa\rightarrow0}f(s,q,\nu,\gamma)^2
=q.
\label{eq:squeezedPhi2}
\end{equation}
Therefore $\lim_{\kappa\rightarrow 0}p_{\rm fr}(s)={\sqrt q}\sqrt{\eta_{\rm th}}=\sqrt{\eta_{\rm eff}}$ \footnote{Another way of seeing this is to absorb the anisotropy into the covariance by defining $M_t:=D_s\bar\Sigma_t D_s$, and analyzing the resulting Riccati.}. This shows that anisotropy has no effect on the fast-rate purity without a competing heat bath. In fact, the dependence of $p_{\rm fr}(s)$ will stem from the competition between the heat bath trying to thermalize the state (which in our normalized setting has a radially symmetric shape), and the measurement trying to create a squeezed shape. The confluence of the corresponding timescales can actually affect whether this anisotropy is beneficial or detrimental to the sustained purity. Specifically, $p_{\rm fr}(s)$ can be either increasing, decreasing, or non-monotonic in $s$ depending on the location of the system in the dimensionless parameter space ($q$,$\nu$,$\gamma$). To illustrate this, we plot in Figure~\ref{fig:F5} the purity $p_{\rm fr}$ as a function of $s$ and $\kappa$, while keeping all other parameters fixed at select values. Indeed, looking at the center figure shows that the effect of $s$ is initially absent at $\kappa=0$, then degrades $p_{\rm fr}$ for small $\kappa$, passes through a crossover where the dependence is non-monotonic, and eventually enhances $p_{\rm fr}$ for large $\kappa$. 

\subsection{Vibrational protocols}

We saw in Section~\ref{sec:vibrational} that, for \(s=0\), high-frequency vibrational control provides no advantage over the fast-rate constant-rate limit, even though it has more degrees of freedom. When anisotropy is introduced, however, the vibrational control can be harnessed to further enhance or alleviate the impact of anisotropy compared to constant-rate protocols, depending on where the system is located in the $(q,\nu,\gamma)$ parameter space.
 
 To see this, note first that the center of the disk $(u,v)=(0,0)$ now corresponds to the averaged coefficients

\begin{align}\label{avergdsqueezed}
\bar R
= a_R=
\eta_{\rm eff}k_{\rm eff}D_s^2,~~~{\bar K}=a_K
=
\frac{k_{\rm eff}}{4}D_s^{-2}.
\end{align}
which are, as expected, the fully averaged coefficients \eqref{eq:squeezedFastMatrices} obtained in the fast constant-rate limit.

\begin{figure*}
\centering
\includegraphics[width=\linewidth]{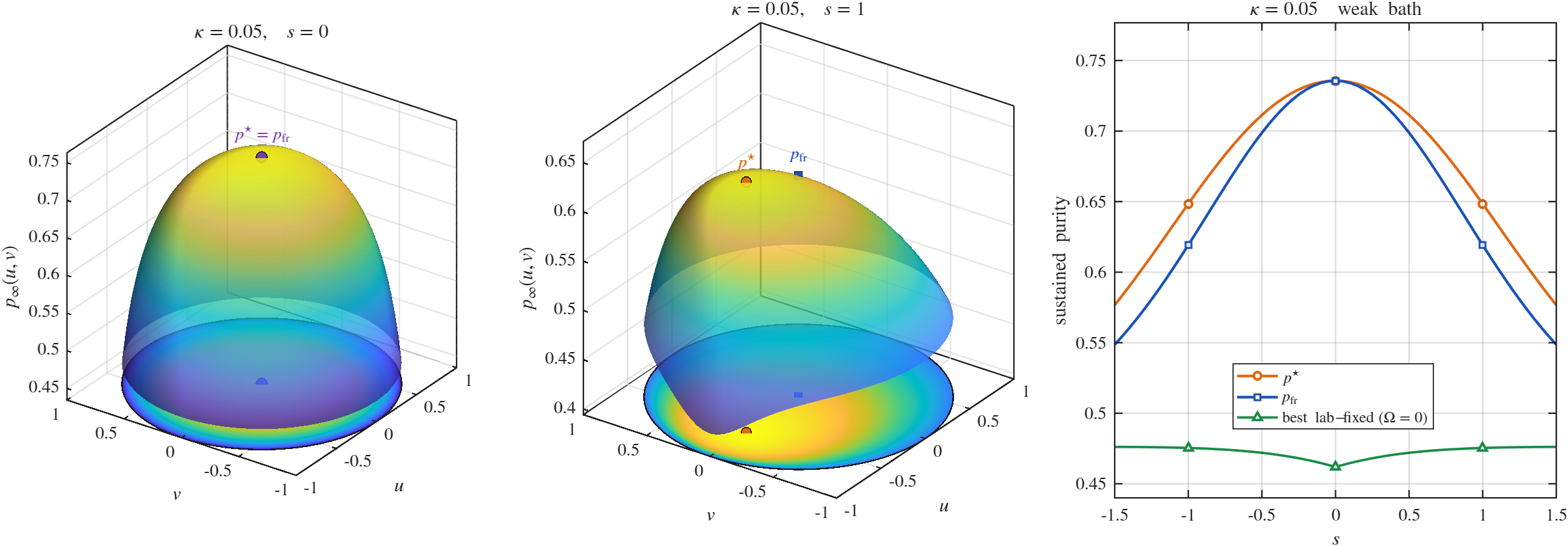}
\includegraphics[width=\linewidth]{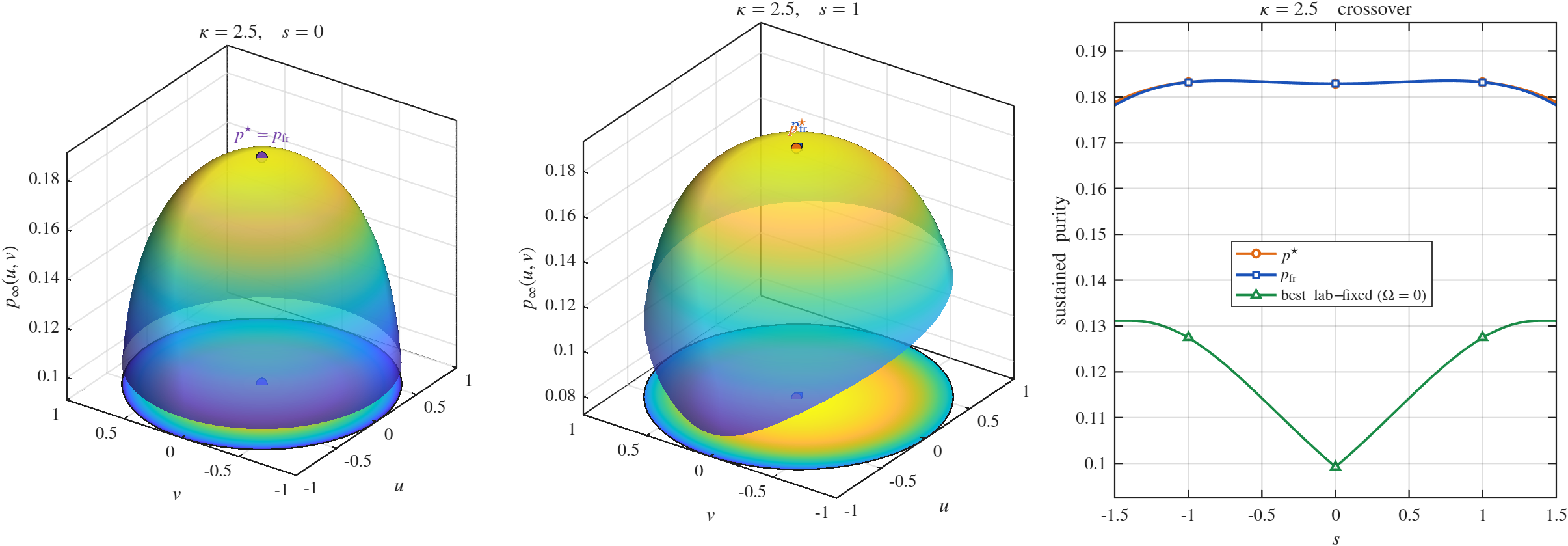}
\includegraphics[width=\linewidth]{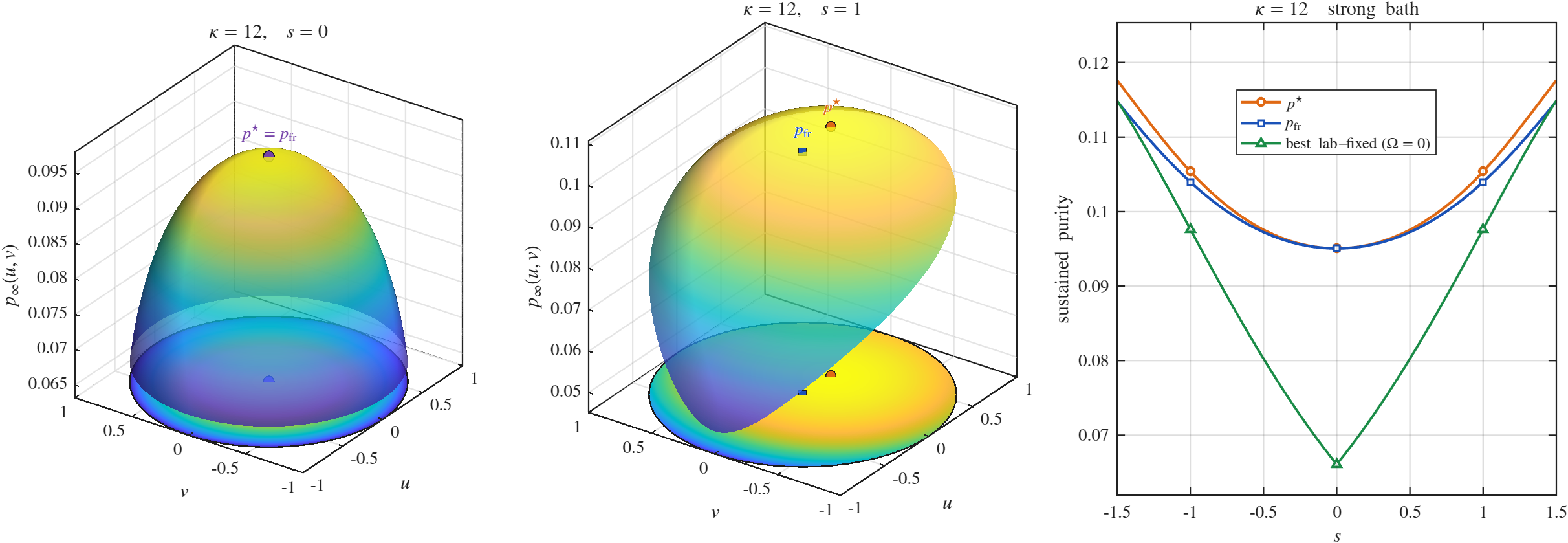}
\caption{Optimization over the accessible vibrational-measurement disk \(\mathcal D\) at three representative points along the fixed-detector \(\kappa\) sweep: \(\kappa=0.05\) (top), \(\kappa=2.5\) (middle), and \(\kappa=12\) (bottom). Along this sweep \(q=\eta_{\rm eff}/\eta_{\rm th}\) and \(\gamma\nu\) remain fixed, while increasing \(\kappa\) increases \(\gamma=\kappa/\omega\) and decreases \(\nu=k_{\rm eff}/k_{\rm th}\). In each row, the left and middle panels show \(p_\infty(u,v)\) for \(s=0\) and \(s=1\). The orange circle marks the optimized vibrational point \(p^\star\), while the blue square marks the fast-rate center \(p_{\rm fr}=p_\infty(0,0)\); the two coincide for \(s=0\). The right panels compare \(p^\star\), \(p_{\rm fr}\), and the best lab-fixed constant-angle purity as functions of \(s\). The preferred angular bias changes substantially along the trajectory, illustrating that the effect of anisotropy is set by the combined dimensionless ratios rather than by \(\kappa\) alone. The remaining parameters are \(\omega=1\), \(\eta_1=0.85\), \(\eta_2=0.80\), \(k_1=20\), \(k_2=0.1\), and \(\bar n=14.5\).}
\label{fig:F7}
\end{figure*}

Because $s\neq0$, the rotational symmetry of $p_\infty(u,v)$ observed in the $s=0$ case is now broken, and the center need no longer maximize the disk landscape
\begin{align}
    p^\star\neq p_\infty(0,0)=p_{\rm fr}.
\end{align}
This suggests that there may be a non-zero optimal strategy $(u^\star,v^\star)\neq(0,0)$ that can produce a higher steady-state purity than the one at the center of the disk. The point is that the anisotropy from the measurement can be further enhanced (if it raises the fast-rate purity) or countered (if it lowers the fast-rate purity) by maintaining a partial angular bias. We illustrate this in Figure ~\ref{fig:F7} for varying values of $\kappa$.

At $\kappa=0.05$, $s=1$ lowers the sustained purity, and 
\begin{align}
p_{\rm fr}
\simeq0.61923
\leq 
p^\star
&\simeq0.64847,
\end{align}
with optimum $(u^\star,v^\star)
\simeq
(-0.5924,0)$. Thus uniform angular averaging $(u,v)=(0,0)$ mixes more strongly than is optimal, so that retaining a partial angular bias actually helps recover a fraction of the anisotropy-induced loss. At $\kappa=2.5$, 
\begin{align}
p_{\rm fr}
\simeq0.183203\simeq
p^\star
\simeq0.183220,
\end{align}
with optimum $(u^\star,v^\star)
\simeq
(-0.0222,0)$. This point lies close to the center, meaning that uniform angular averaging is near-optimal in this region of the $(q,\nu,\gamma)$ parameter space. Finally, at $\kappa=12$, $s=1$ enhances the sustained purity, and
\begin{align}
p_{\rm fr}
\simeq0.10398\leq 
p^\star
\simeq0.10538,
\end{align}
with optimum $(u^\star,v^\star)
\simeq
(0.3198,0)$. Here, vibrational optimization provides a further improvement by retaining more of the favorable measurement anisotropy instead of averaging it uniformly. 

Interestingly, in all cases the {optimum moves} exclusively along the $v=0$ axis. This direction is associated specifically with the $s$ anisotropy. Movement along other directions in the disk is caused by other sources of anisotropy which may be added into the measurement model \eqref{rotatedquad}. 

\begin{figure*}
\centering
\includegraphics[width=\linewidth]{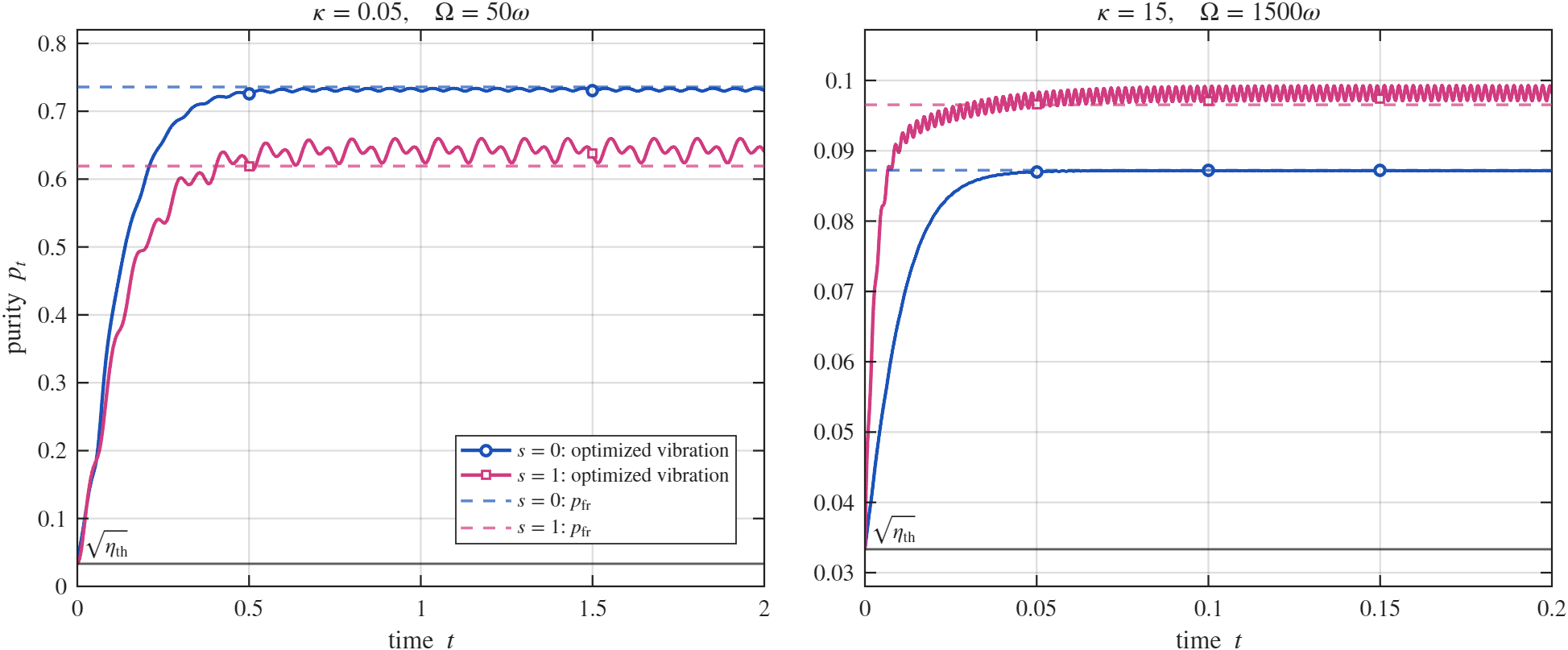}
\caption{Exact optimized-vibrational dynamics at two representative points along the fixed-detector \(\kappa\) sweep. Left: \(\kappa=0.05\), with \(\Omega=50\omega\). Increasing anisotropy from \(s=0\) to \(s=1\) lowers the absolute sustained purity, but the optimized vibrational trajectory remains above the corresponding fast-rate value \(p_{\rm fr}\). Right: \(\kappa=15\), with \(\Omega=1500\omega\). At this point the anisotropy raises the attainable purity, and optimized vibrational mixing provides an additional improvement over uniform fast-rate averaging. In each panel, solid curves show the exact optimized vibrational trajectories for \(s=0\) and \(s=1\), while dashed lines of matching color denote the corresponding fast-rate sustained purities. The black line marks the thermal purity \(\sqrt{\eta_{\rm th}}\). All trajectories start from \(\Sigma_0=\Sigma_{\rm th}\). The remaining parameters are \(\omega=1\), \(\eta_1=0.85\), \(\eta_2=0.80\), \(k_1=20\), \(k_2=0.1\), and \(\bar n=14.5\).}
\label{fig:F8}
\end{figure*}

Figure~\ref{fig:F8} illustrates the same comparison in the exact time-dependent dynamics. For $\kappa=0.05$, anisotropy lowers purity, and the optimized vibrational settles into a high-frequency oscillation whose cycle average remains above the fast-rate value. For $\kappa=15$, anisotropy increases purity, and the optimized vibration settles into a high-frequency oscillation whose cycle average again remains above the fast-rate value. {This shows that, compared with constant-rate control, vibrational modulation can supply additional control authority to suppress unfavorable anisotropy or retain and exploit it when it is beneficial.}

\subsection{Greedy protocols}
When $s\neq 0$, the analysis of Section \ref{sec:greedy} holds with $\Sigma_t$ replaced by its anisotropy-weighted version $M_t:=D_s\Sigma_tD_s$. This is because the measurement contribution \eqref{eq:innerproduct} retains a factor of $D_s$ on each side of $\Sigma_t$ coming from $U_\theta=D_se^{-\theta J}$. Thus, the same geometric interpretation holds in terms of the principal axes of $M_t$. That is, the greedy controller always assigns the major axis of $M_t$, which is the direction of the largest, now anisotropy-weighted covariance uncertainty, to the detector having the smaller instantaneous coefficient $k_j(p_t^2-\eta_j)$, and the sign of $L_t$ determines again whether this is detector $1$ or detector $2$. 

\begin{figure*}[t]
\centering
\includegraphics[width=\linewidth]{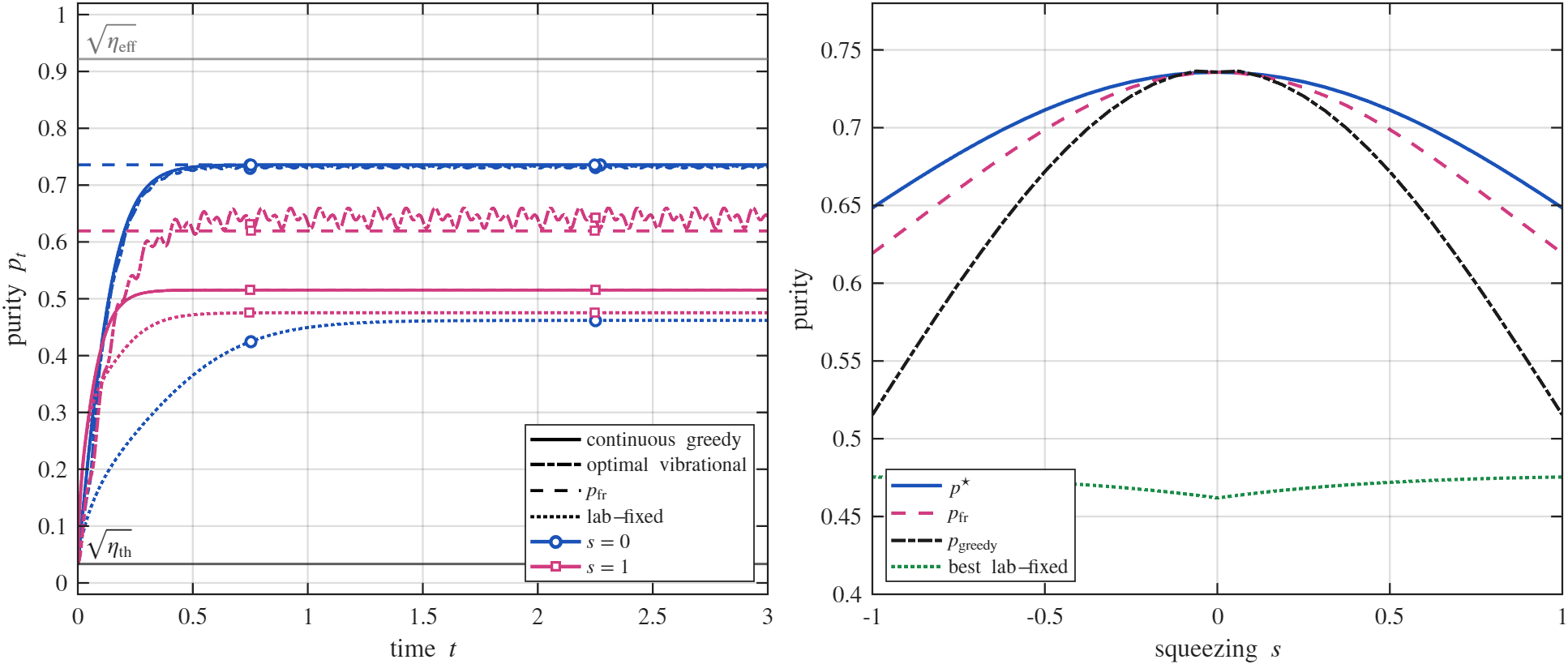}
\includegraphics[width=\linewidth]{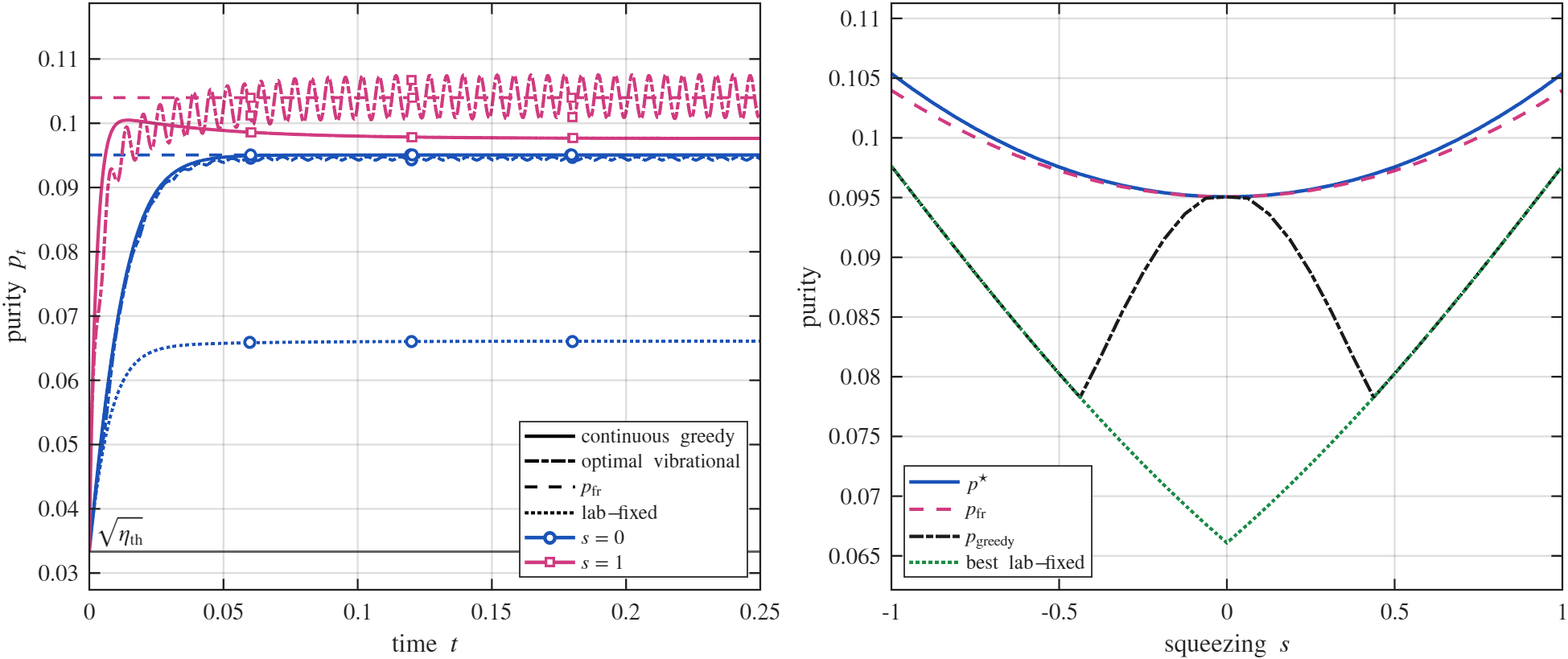}
\caption{Comparison of continuous greedy, optimized vibrational, fast-rate, and
lab-fixed control at two representative points along the fixed-detector
\(\kappa\) sweep. The upper pair corresponds to \(\kappa=0.05\), while the lower
pair corresponds to \(\kappa=12\); the remaining parameters are
\(\omega=1\), \(\bar n=14.5\), \(\eta_1=0.85\), \(\eta_2=0.80\),
\(k_1=20\), and \(k_2=0.1\). In each pair, the left panel shows purity dynamics
for \(s=0\) and \(s=1\), while the right panel compares \(p^\star\),
\(p_{\rm fr}\), \(p_{\rm greedy}\), and the best lab-fixed constant-angle purity
over \(|s|\leq1\). At \(s=0\), optimized vibration, fast rotation, and
continuous greedy control coincide. As \(|s|\) increases, the greedy solution
tends toward a nearly fixed measurement orientation and hence toward the
lab-fixed branch.}
\label{fig:F11}
\end{figure*}

The greedy protocol tries to equalize the two eigenvalues of \(M_t\), so that
$\Sigma_t$ tends towards $D_s^{-2}$. In this case, however, the oscillator rotation and thermal relaxation
continually perturb this preferred anisotropy-weighted covariance, and so the long-time
behavior differs from the fast-rate limit observed in the $s=0$ case. We compare in Fig.~\ref{fig:F11} the best lab-fixed\footnote{by ``best lab-fixed" we mean the constant-angle $\theta_t=\theta_0$ that gives the best sustained purity.}, fast-rate, vibrational, and greedy monitoring strategies,  for varying values of $\kappa$ and $s$. The upper and lower pairs of Fig.~\ref{fig:F11} correspond respectively to
$\kappa=0.05$ and $\kappa=12$. Within each pair, the left panel shows the
time-dependent purity for $s=0$ and $s=1$, while the right panel shows the
sustained purities as functions of $|s|$. At $s=0$, optimized vibration, fast
uniform rotation, and the continuous greedy controller
all realize the same rotationally averaged measurement and attain $p_{\rm fr}$.

At $\kappa=0.05$, anisotropy lowers the sustained fast-rate and vibrational
purities. For $s=1$,
\begin{align}
p_{\rm fr}\simeq0.6192\leq 
p^\star\simeq0.6485.
\end{align}
Thus anisotropy-adapted vibrational mixing improves on uniform fast rotation,
whereas the greedy protocol reaches a lower sustained purity. As $|s|$ increases, however,
the greedy branch bends toward the best lab-fixed result, indicating that the
preferred greedy angle is becoming increasingly stationary. At $\kappa=12$, the fast-rate and vibrational purities instead increase with
$|s|$. For $s=1$,
\begin{align}
p_{\rm fr}
\simeq0.10398\leq
p^\star
\simeq0.10538.
\end{align}
The greedy purity initially decreases with anisotropy and then rises again as
its preferred angle becomes nearly stationary.

{\subsection{Cross-protocol comparison}}
Thus far we have compared the three protocol classes: constant-rate, vibrational, and greedy. We now include the variational strategy for a complete comparison of all the classes, and report the performance metric $p_{\rm geom}(T)$ for each one. For the variational strategy, we numerically solve the coupled state--costate equations \eqref{state}-\eqref{eq:costate} by a forward--backward sweep.
Starting from an admissible control $u_t\in S^1$, the covariance is propagated
forward from its initial condition and the costate backward from $\Pi_T=0$.
Equation~\eqref{eq:uPMP} then gives an updated pointwise control, which is
under-relaxed with the previous iterate and normalized back onto $S^1$. The
procedure is repeated, retaining the iterate with the largest
$p_{\rm geom}(T)$.  

\begin{figure*}[t]
\centering
\includegraphics[width=\linewidth]{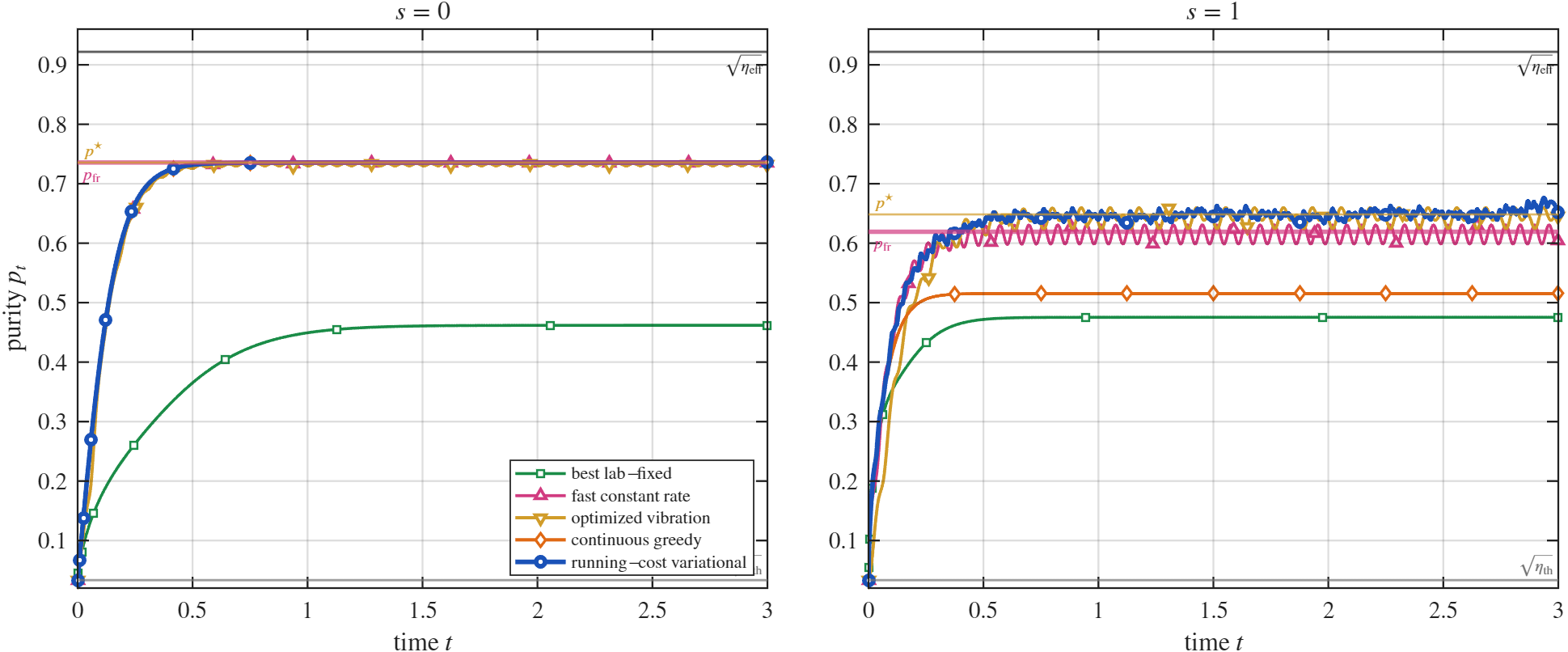}
\includegraphics[width=\linewidth]{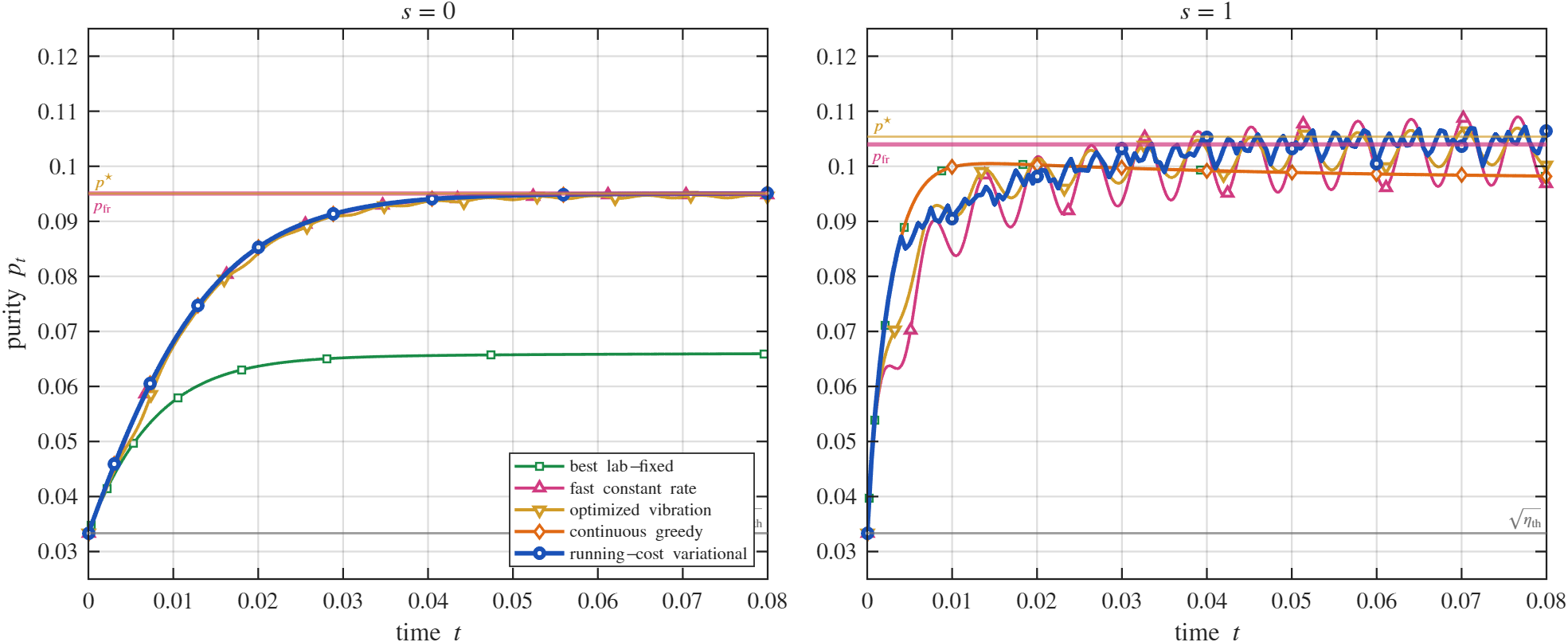}
\caption{Comparison of the best lab-fixed, fast constant-rate, optimized
vibrational, continuous greedy, and running-cost variational protocols over the
horizon \(T=3\). The upper row corresponds to \(\kappa=0.05\) and the lower row
to \(\kappa=12\); within each row, the left and right panels show \(s=0\) and
\(s=1\), respectively. The fast and vibrational protocols use
\(\Omega=50\omega\) for \(\kappa=0.05\) and \(\Omega=500\omega\) for
\(\kappa=12\). The lower row displays only \(0\leq t\leq0.08\) to resolve the
rapid oscillations, although all trajectories and accumulated-performance
measures are evaluated over the full interval \(0\leq t\leq3\). Horizontal
lines indicate \(\sqrt{\eta_{\rm th}}\), \(\sqrt{\eta_{\rm eff}}\),
\(p_{\rm fr}\), and \(p^\star\). All trajectories start from
\(\Sigma_0=\Sigma_{\rm th}\).}
\label{fig:F12}
\end{figure*}

Figure~\ref{fig:F12} and Table~\ref{tab:variationalcomparison}
compare the four protocols using the common finite-horizon measure
\(p_{\rm geom}(T)\). At \(s=0\), the time-dependent strategies have very similar
performance and approach the same rotationally averaged limit starting from the thermal initial condition
\(\Sigma_0=\Sigma_{\rm th}\). The reason for this is that complete angular averaging is a singular Pontryagin extremal of the relaxed problem. That is, if the unit-circle constraint on \(u_t\) is relaxed
to the unit disk, the center \(u_t=0\) preserves
\(\Sigma_t=\sigma_t I\) and \(\Pi_t=\pi_t I\), for which
\(\mathcal B_t=\mathcal C_t=0\) at all times. Thus physical
unit-circle controls approach this trajectory through increasingly rapid angular
modulation. The derivation is given in Appendix~\ref{app:variational-s0}. This
explains the numerical collapse of the variational solution toward rotational
averaging for \(s=0\), although it does not by itself establish global
finite-horizon optimality among all admissible physical controls.

Once \(s\neq0\),
their performances separate because they optimize different aspects of the
dynamics. For the representative cases shown here, the variational strategy
provides the strongest finite-horizon performance among the protocols
considered. The greedy control performs less well because pointwise
maximization of \(\dot p_t\) does not account for the future effect of the
chosen covariance trajectory. Overall, the comparison shows that
time-dependent angle control can substantially outperform a fixed measurement
orientation, while anisotropy distinguishes protocols that are nearly
equivalent when \(s=0\).

\begin{table*}
\centering
\caption{Geometric-mean purity \(p_{\rm geom}(T)\), defined in
Eq.~\eqref{eq:pgeom}, over \(0\leq t\leq T\) with \(T=3\), at the two
representative points \(\kappa=0.05\) and \(\kappa=12\) of the fixed-detector
sweep. The lab-fixed angle is optimized for the same running objective. The
fast constant-rate and vibrational protocols use \(\Omega=50\omega\) for
\(\kappa=0.05\) and \(\Omega=500\omega\) for \(\kappa=12\).}
\label{tab:variationalcomparison}
\begin{ruledtabular}
\begin{tabular}{lcccc}
Protocol
& \multicolumn{2}{c}{\(\kappa=0.05\)}
& \multicolumn{2}{c}{\(\kappa=12\)}
\\
& \(s=0\)
& \(s=1\)
& \(s=0\)
& \(s=1\)
\\
\hline
Best lab fixed
& \(0.4038\)
& \(0.4595\)
& \(0.0660\)
& \(0.0976\)
\\
Fast constant rate
& \(0.6895\)
& \(0.5912\)
& \(0.0946\)
& \(0.1031\)
\\
Optimized vibration
& \(0.6846\)
& \(0.5986\)
& \(0.0942\)
& \(0.1039\)
\\
Continuous greedy
& \(\mathbf{0.6929}\)
& \(0.5017\)
& \(\mathbf{0.0948}\)
& \(0.0976\)
\\
Running-cost variational
& \(\mathbf{0.6929}\)
& \(\mathbf{0.6201}\)
& \(\mathbf{0.0948}\)
& \(\mathbf{0.1051}\)
\end{tabular}
\end{ruledtabular}
\end{table*}
\vspace{3mm}

\subsection{Purification from a thermal state}
\label{sec:thermalprep}

\begin{figure}
    \centering
    \includegraphics[width=\linewidth]{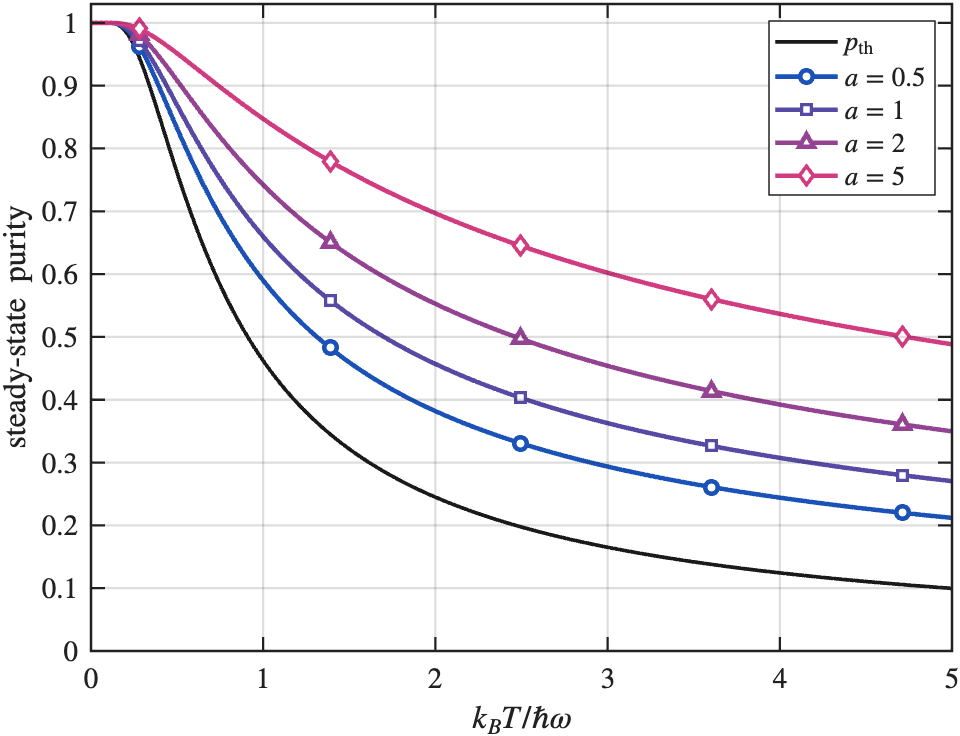}
    \caption{Fast-rate steady-state purity \(p_{\rm fr}(T)\) as a function of
    the dimensionless bath temperature \(k_B T/\hbar\omega\). The black curve
    shows the unmeasured thermal purity \(p_{\rm th}\), while the remaining
    curves show Eq.~\eqref{eq:pfrtemperature} for unit effective efficiency
    \(\eta_{\rm eff}=1\) and several fixed measurement-to-bath strength ratios
    \(a=k_{\rm eff}/(2\kappa)\). Increasing the measurement strength maintains
    a higher conditional steady-state purity as the bath temperature rises.}
    \label{fig:temperatureprep}
\end{figure}

Throughout this work we have taken the initial state to be the one in thermal equilibrium with the heat bath, i.e., $\Sigma_0=\Sigma_{\rm th}$. This represents a natural state-preparation setting, obtained by first allowing the oscillator
to equilibrate with the thermal environment and then turning on the
measurement, which has to compete with thermalization
to establish a new conditional steady state. It is therefore useful to ask how
the attainable purity changes with the bath temperature when the detector
resources are held fixed. To this end, we plot how the fast-rate limit $p_{\rm fr}$, which is observed as a common performance limit across all protocols for the $s=0$ case, changes as a function of the bath temperature. Let
\begin{align}
x
:=
2\bar n+1
=
\coth\!\left(
\frac{\hbar\omega}{2k_B T}
\right),
\end{align}
so that the unmeasured thermal purity is
\begin{align}
p_{\rm th}
=
\sqrt{\eta_{\rm th}}
=
\frac{1}{x}.
\end{align}
For fixed \(k_{\rm eff}\), \(\eta_{\rm eff}\), and \(\kappa\),
\begin{align}
\nu
&=
\frac{k_{\rm eff}}{2\kappa x},
&
q
&=
\eta_{\rm eff}x^2.
\end{align}
Thus varying the bath temperature traces a one-parameter path through the
\((q,\nu,\gamma)\) parameter space with
\begin{align}
\gamma
&=
\frac{\kappa}{\omega}
=
\mathrm{const.},
&
q\nu^2
&=
\frac{\eta_{\rm eff}k_{\rm eff}^2}{4\kappa^2}
=
\mathrm{const.}
\end{align}
Writing \(a:=k_{\rm eff}/(2\kappa)\), Eq.~\eqref{bathlargeomega}
becomes
\begin{align}
p_{\rm fr}(T)
=
\frac{
\sqrt{1+\eta_{\rm eff}a^2+2\eta_{\rm eff}ax}+1
}{
a+2x
}.
\label{eq:pfrtemperature}
\end{align}
Figure~\ref{fig:temperatureprep} plots the above function for
\(\eta_{\rm eff}=1\) and several fixed values of \(a\). Increasing the measurement strength raises the resulting steady-state purity and extends the range of temperatures over which substantial purification is maintained.

\section{Physical timescales and conclusions}\label{sec:outlook}

\emph{Physical timescales.}
Evidently, the relevant control timescales of the problem will roughly depend on how rapidly the conditional covariance evolves. For instance, in the constant-rate class, the exact solution suggests the rough characteristic scale 
\begin{align}
\omega_c
\sim
\max\!\left\{
\kappa,\,
\sqrt{\tr(RQ)},\,
\sqrt{\tr(RJQJ^\top)},\,
[\det(RQ)]^{1/4}
\right\},
\label{eq:omegac}
\end{align}
so that fast angular averaging requires control rates large compared with
\(\omega_c\). Related experimental ingredients, including phase-controlled
quadrature measurements, backaction-evading detection, and real-time
measurement-based control, have been demonstrated in cavity optomechanical and
electromechanical platforms
\cite{aspelmeyer2014cavity,hertzberg2010backaction,suh2014mechanically,
shomroni2019optical,wilson2015measurement,rossi2018measurement,
magrini2021real,tebbenjohanns2021quantum}.

\emph{Conclusions.}
We have studied how the orientation of continuously monitored quadratures can
be used as a control resource for purifying a thermally damped harmonic
oscillator when the underlying detector strengths or
efficiencies are fixed. The Gaussian setting makes this a deterministic covariance-control problem and isolates the competition among information gain, measurement backaction, Hamiltonian rotation, and thermalization. 

Four increasingly general control classes were considered, both in the isotropic setting $s=0$, and the anisotropic one $s\neq 0$. For the isotropic case
$s=0$, rotational symmetry strongly constrains the problem, and the different
time-dependent control strategies considered herein approach essentially the same
rotationally averaged behavior. This suggests that simply implementable protocols such as constant-rate or vibrational are near-optimal, and more complex optimizations are not necessary. Once $s\neq0$, however, this symmetry is broken and the
strategies separate. The anisotropy can either reduce or enhance purification,
depending on how the measurement-induced covariance deformation competes with
rotation and thermal relaxation.

A natural next step is to incorporate physical control constraints directly
into the optimization, including finite modulation bandwidth and latency. Because the covariance evolution is deterministic, the
measurement-angle protocols considered here can in principle be precomputed and
do not require real-time processing of the measurement record. Moreover, one
may consider more general sources of measurement anisotropy, such as relative
squeezing between detectors, nonorthogonal measurement axes, or
direction-dependent environmental noise, and investigate the
overall role of symmetries in control performance. Finally, multimode Gaussian
systems provide another natural setting in which several covariance directions
and measurement geometries {may compete} simultaneously.

Beyond the Gaussian regime, the conditional uncertainty itself can become
stochastic, so genuine feedback on the measurement record may offer further advantages. The same ideas may also be useful in monitored quantum
thermodynamics and quantum-battery settings, where purification and information
acquisition can act as operational resources
\cite{mitchison2021charging,morrone2023daemonic,kua2025daemonic,
cenedese2025boosting}. More general optimal-control formulations for
continuously monitored systems, including Pontryagin-based approaches
\cite{karmakar2025cdj}, provide a natural route for extending measurement-basis
control to these settings.

\begin{acknowledgments}
 RS and TG are supported by the National Science Foundation under grant ECCS-2347357, the Air Force Office of Scientific Research under FA9550-24-1-0278, and the Army Research Office under W911NF-22-1-0292. OMM was supported by the European Union's Horizon 2020 research and innovation programme under the Marie Skłodowska-Curie Grant Agreement No. 101151140, and acknowledges support from Ministerio de Ciencia e Innovaci\'on and European Union (FEDER) (PID2022-138269NB-I00). ANJ is supported by a grant from the John Templeton Foundation (Grant 63209) and the ARO under award number W911NF22-1-0258.
\end{acknowledgments}
\appendix*
\section{Proofs}\label{app:proofs}

This appendix collects the derivations used in the main text. Throughout,  $D_s=\diag(e^s,e^{-s})$ and
\begin{align*}
U_\theta=D_se^{-\theta J},~~\sigma_{\rm th}&=\bar n+\tfrac12
=\frac{1}{2\sqrt{\eta_{\rm th}}}.
\end{align*}
Moreover, for any symmetric invertible \(2\times2\) matrix \(X\),
\begin{align}
X^{-1}
=
\frac{JXJ^\top}{\det X},
\label{eq:app:inverse}
\end{align}
and for any traceless \(2\times2\) matrix \(X\),
\begin{align}
J^\top XJ=-X^\top .
\label{eq:app:traceless}
\end{align}

\subsection{Proposition~\ref{prop:meas}}
\label{app:meas}
Set \(\kappa=0\) and $\Lambda_t
:=
\diag\!\left(
k_1(p_t^2-\eta_1),
k_2(p_t^2-\eta_2)
\right)$. Then, equation~\eqref{eq:measterm} becomes
\begin{align}
\dot p_t
=
-\frac{p_t}{2}
\sum_{j=1}^2
(\Lambda_t)_{jj}
(U_{\theta_t}^\top\Sigma_tU_{\theta_t})_{jj}.
\label{eq:app:pdotsum}
\end{align}
Setting \(k_{\min}:=\min(k_1,k_2)\), we have
\begin{align}
\sum_{j=1}^2
k_j(U_\theta^\top\Sigma U_\theta)_{jj}
&\ge
k_{\min}\tr(\Sigma D_s^2)
\ge
\frac{e^{-2|s|}k_{\min}}{p},
\label{eq:app:boundA}
\end{align}
where we used
\(\tr\Sigma\ge2\sqrt{\det\Sigma}=1/p\). Moreover, let \(v\) be a unit eigenvector associated
with \(\lambda_{\min}(\Sigma_t)\). From
eq.~\eqref{eq:RDE} we have 
\begin{align*}
\frac{\dd}{\dd t}\lambda_{\min}
\ge
v^\top {K}_{\theta_t}v
-\lambda_{\min}^2v^\top R_{\theta_t}v\ge
n-\varrho\lambda_{\min}^2,
\end{align*}
where
\[
n:=\frac{e^{-2|s|}k_{\min}}{4},
\qquad
\varrho:=e^{2|s|}
\max(\eta_1k_1,\eta_2k_2).
\]
Set \(\ell:=\sqrt{n/(2\varrho)}\). Whenever
\(\lambda_{\min}<\ell\), its derivative is at least \(n/2\).
Consequently, after at most
\[
t_0:=\frac{2\ell}{n}
\]
one has \(\lambda_{\min}(\Sigma_t)\ge\ell\), and this lower bound is
preserved thereafter. Thus
\begin{align}
(U_{\theta_t}^\top\Sigma_tU_{\theta_t})_{jj}
\ge e^{-2|s|}\ell,
\qquad t\ge t_0.
\label{eq:app:boundB}
\end{align}

Suppose first that \(\eta_1\neq\eta_2\), and write
\(\Delta\eta:=\eta_{\max}-\eta_{\min}>0\). Let \(j_+\) and \(j_-\)
denote the detectors with efficiencies \(\eta_{\max}\) and
\(\eta_{\min}\), respectively. If \(p^2\le\eta_{\min}\), all diagonal
entries of \(\Lambda_t\) are nonpositive and
\((\Lambda_t)_{j_+j_+}\le-k_{j_+}\Delta\eta\). Equations
\eqref{eq:app:boundA}--\eqref{eq:app:boundB} therefore give
\begin{align*}
\dot p_t
\ge
\frac{e^{-2|s|}}{2}
\max\!\left\{
k_{\min}(\eta_{\min}-p_t^2),
k_{j_+}\ell\,\Delta\eta\,p_t
\right\}.
\end{align*}
The maximum on the right is bounded below by a positive constant on
\(0\le p\le\sqrt{\eta_{\min}}\), so the lower boundary is reached in
finite time. The vector field points inward there, so the purity
cannot subsequently cross below it.

Similarly, if \(p^2\ge\eta_{\max}\), then for \(t\ge t_0\),
\begin{align*}
\dot p_t
\le
-\frac12
\sqrt{\eta_{\max}}\,
e^{-2|s|}
k_{j_-}\ell\,\Delta\eta
<0.
\end{align*}
Hence the upper boundary is also reached in finite time and cannot
subsequently be crossed. This proves the finite-time confinement in
Eq.~\eqref{eq:finitetimeconfinement}.

If instead \(\eta_1=\eta_2=\eta\), Eq.~\eqref{eq:app:boundA} shows
that, for every \(\epsilon>0\),
\[
|\dot p_t|
\ge
\frac12e^{-2|s|}k_{\min}\epsilon
\qquad
\text{whenever }
|p_t^2-\eta|\ge\epsilon,
\]
with the sign directed toward \(\sqrt\eta\). Every neighborhood of
\(\sqrt\eta\) is therefore entered in finite time and is invariant
thereafter; equivalently, \(p_t\to\sqrt\eta\).
\hfill\(\blacksquare\)

\subsection{Proposition~\ref{prop:bath}}
\label{app:bath}

With \({K}=R=0\), Eq.~\eqref{eq:RDE} is linear. Since
\(\Sigma_{\rm th}=\sigma_{\rm th}I\) commutes with \(J\),
\begin{align*}
\Sigma_t
=
\epsilon_t e^{-\omega Jt}\Sigma_0e^{\omega Jt}
+
(1-\epsilon_t)\sigma_{\rm th}I,
\qquad
\epsilon_t:=e^{-2\kappa t}.
\end{align*}
If \(\lambda_1,\lambda_2\) are the eigenvalues of \(\Sigma_0\), the
eigenvalues of \(\Sigma_t\) are
\(\epsilon_t\lambda_i+(1-\epsilon_t)\sigma_{\rm th}\). Therefore
\begin{align*}
\frac{\det\Sigma_t}{\sigma_{\rm th}^2}
=
1+c_2\epsilon_t+c_4\epsilon_t^2,
\end{align*}
where
\begin{align*}
c_2
&=
\frac{\lambda_1+\lambda_2}{\sigma_{\rm th}}-2,\\
c_4
&=
\frac{\lambda_1\lambda_2}{\sigma_{\rm th}^2}-c_2-1.
\end{align*}
Using
\[
\lambda_1\lambda_2=\frac{1}{4p_0^2},
\qquad
\lambda_1+\lambda_2
=
\sqrt{\lambda_1\lambda_2}
\left(
\sqrt{\chi_0}+\frac{1}{\sqrt{\chi_0}}
\right)
\]
gives the coefficients stated in Proposition~\ref{prop:bath}. The
purity formula follows from \(p_t=1/(2\sqrt{\det\Sigma_t})\).
\hfill\(\blacksquare\)

\subsection{Steady-state purity and relaxation rate}
\label{app:closedform}

The formulas \eqref{rformula} and \eqref{pformula} follow from the
stationary Riccati equation
\begin{align}
\omega_{\rm eff}[\Sigma,J]
-2\kappa\Sigma
+Q-\Sigma R\Sigma
=0,
\label{eq:app:ARE}
\end{align}
where \(R=\diag(r_1,r_2)>0\) and
\(Q=\diag(q_1,q_2)>0\). To see this, define
\[
\Gamma
:=
\kappa I+\omega_{\rm eff}J+\Sigma R
\]
and the Hamiltonian matrix
\begin{align*}
\mathcal H
:=
\begin{bmatrix}
-\kappa I+\omega_{\rm eff}J & -R\\
-Q & \kappa I+\omega_{\rm eff}J
\end{bmatrix}.
\end{align*}
Equation~\eqref{eq:app:ARE} is equivalent to
\[
\mathcal H
\begin{bmatrix}I\\ \Sigma\end{bmatrix}
=
\begin{bmatrix}I\\ \Sigma\end{bmatrix}
(-\Gamma^\top).
\]
Since \(\mathcal H\) is Hamiltonian, its eigenvalues occur in
opposite pairs \cite{anderson2007optimal}. If
\(\lambda_{1,2}\) are the eigenvalues of \(\Gamma\), then
\[
\sigma(\mathcal H)
=
\{\lambda_1,\lambda_2,-\lambda_1,-\lambda_2\}.
\]
Direct evaluation gives
\begin{align*}
\frac12\tr\mathcal H^2
&=
2\kappa^2-2\omega_{\rm eff}^2+\tr(RQ)
=S,\\
\det\mathcal H
&=
(\kappa^2+\omega_{\rm eff}^2)^2
+\kappa^2\tr(RQ)\\
&\quad
+\omega_{\rm eff}^2\tr(RJQJ^\top)
+\det(RQ)
=P.
\end{align*}
Because \(\Re\lambda_i>0\),
\(\lambda_1\lambda_2=\sqrt P\), and hence
\begin{align*}
(\lambda_1+\lambda_2)^2
&=
S+2\sqrt P,\\
(\lambda_1-\lambda_2)^2
&=
S-2\sqrt P.
\end{align*}
This gives Eq.~\eqref{rformula}. In particular,
\begin{align}
\tau
:=
\tr\Gamma
=
2\kappa+\tr(\Sigma R)
=
\sqrt{S+2\sqrt P}.
\label{eq:app:tau}
\end{align}

For the purity, let \(Y:=\Sigma^{-1}\), so
\(\det Y=4p_\infty^2\). Multiplying Eq.~\eqref{eq:app:ARE} by \(Y\)
and taking the trace gives
\begin{align}
\tr(YQ)=\tau+2\kappa.
\label{eq:app:trYQ}
\end{align}
Multiplying on both sides by \(Y\) gives
\[
YQY
=
R+2\kappa Y+\omega_{\rm eff}[Y,J].
\]
On the other hand, the Cayley--Hamilton identity for \(YQ\) gives
\[
YQY
=
(\tau+2\kappa)Y
-\det Y\,\operatorname{adj}Q.
\]
Thus
\begin{align}
\tau Y-\omega_{\rm eff}[Y,J]
=
F,
\qquad
F:=R+\det Y\,\operatorname{adj}Q.
\label{eq:app:linear}
\end{align}
Write any symmetric matrix as
\(X=\tfrac12(\tr X)I+X_0\), with \(X_0\) traceless. Since a
traceless symmetric \(2\times2\) matrix anticommutes with \(J\),
Eq.~\eqref{eq:app:linear} gives
\begin{align*}
\tr Y
&=
\frac{\tr F}{\tau},\\
Y_0
&=
\frac{
\tau F_0+2\omega_{\rm eff}F_0J
}{
\tau^2+4\omega_{\rm eff}^2
}.
\end{align*}
Substitution into Eq.~\eqref{eq:app:trYQ}, together with
\[
F_0=R_0-\det Y\,Q_0,
\qquad
\tr(Q_0^2)
=
\frac12(\tr Q)^2-2\det Q,
\]
yields
\begin{align*}
4\det Y
\left[
\omega_{\rm eff}^2(\tr Q)^2+\tau^2\det Q
\right]
=
8\omega_{\rm eff}^2\alpha+2\tau^2\beta.
\end{align*}
Since \(p_\infty^2=\det Y/4\), this is Eq.~\eqref{pformula}.

The derivation only uses that \(R\) and \(Q\) are diagonal. It
therefore also applies to the averaged dynamics for {$s\neq0$}
\eqref{eq:squeezedAvgRDE} after the substitutions
\[
\omega_{\rm eff}\mapsto\omega,
\qquad
R\mapsto\bar R_s,
\qquad
Q\mapsto {\bar K_s}+2\kappa\Sigma_{\rm th},
\]
which gives Eq.~\eqref{eq:squeezedPhi2}.
\hfill\(\blacksquare\)

\subsection{Monotonicity and limiting values}
\label{app:monotone}

Set \(w:=\omega_{\rm eff}^2\). Under the ordering assumption
\eqref{ass1}, exchange detector labels if necessary so that
\[
k_1\ge k_2,
\qquad
\eta_1\ge\eta_2.
\]
If the detectors are identical, \(R,{K}\propto I\) and both metrics are
independent of \(w\). We therefore consider the nonidentical case.
With
\begin{align*}
r_1&=\eta_1k_1,
&
r_2&=\eta_2k_2,\\
q_1&=\frac{k_2}{4}+m,
&
q_2&=\frac{k_1}{4}+m,
&
m&:=2\kappa\sigma_{\rm th},
\end{align*}
we have \(r_1>r_2\) and \(q_2\ge q_1\).

For the relaxation rate, let
\[
t_1:=\tr(RQ),
\qquad
t_2:=\tr(RJQJ^\top).
\]
Differentiation gives
\begin{align*}
\frac{\dd(S\pm2\sqrt P)}{\dd w}
=
-2
\pm
\frac{2(\kappa^2+w)+t_2}{\sqrt P}.
\end{align*}
The lower-sign expression is negative. The upper-sign expression is
nonnegative because
\begin{align*}
&[2(\kappa^2+w)+t_2]^2-4P
=
4\kappa^2(t_2-t_1)
+t_2^2-4\det(RQ)\\
&=
4\kappa^2(r_1-r_2)(q_2-q_1)
+(r_1q_2-r_2q_1)^2
\ge0.
\end{align*}
Hence \(\tau=\sqrt{S+2\sqrt P}\) is nondecreasing while
\(S-2\sqrt P\) is nonincreasing, and Eq.~\eqref{rformula} shows that
\(r\) is nondecreasing in \(w\). Under the standing quality
assumption \eqref{ass2}, the nonidentical physical detector pair has
\(r_1q_2-r_2q_1\neq0\), so the increase is strict.

For the purity, eliminating \(w\) between
\((\tau^2-S)^2=4P\) and Eq.~\eqref{pformula} gives
\begin{align}
p_\infty^2
=
\frac{
(r_1-r_2)(\tau^2+2\kappa\tau)
-\xi\,\tr R
}{
4\left[
\xi\,\tr Q
+(q_2-q_1)(\tau^2-2\kappa\tau)
\right]
},
\label{eq:app:ptau}
\end{align}
where
\[
\xi:=q_1r_1-q_2r_2,
\qquad
\zeta:=q_2r_1-q_1r_2.
\]
Differentiating $p_\infty$ directly gives
\begin{align*}
\frac{\dd p_\infty^2}{\dd\tau}
=
\frac{16f(\tau)}{\mathcal D(\tau)^2},
\qquad
f(\tau)
=
\kappa\xi^2+\tau\xi\zeta-\kappa g\tau^2,
\end{align*}
where \(\mathcal D(\tau)\) is the denominator of
Eq.~\eqref{eq:app:ptau} and
\[
g:=(q_2-q_1)(r_1-r_2).
\]
For the ordered nonidentical detector pair,
\(\xi>0\), \(\zeta>0\), and \(g\ge0\). 

If \(\kappa=0\), then
\(f(\tau)=\tau\xi\zeta>0\), so \(p_\infty\) increases strictly with
\(\tau\), and hence with \(w\).

Now suppose \(\kappa>0\). The function \(f\) is concave and
\(f(0)>0\). As \(w\to\infty\),
\[
\tau\longrightarrow
T:=\sqrt{4\kappa^2+\Pi},
\qquad
\Pi:=\tr R\,\tr Q.
\]
It is therefore enough to show \(f(T)>0\). Using
\(g=(\zeta^2-\xi^2)/\Pi\),
\begin{align*}
(T^2-4\kappa^2)f(T)
=
\big[
(T^2-2\kappa^2)\xi-\kappa T\zeta
\big]
\big[
2\kappa\xi+T\zeta
\big].
\end{align*}
Thus \(f(T)>0\) follows from
\begin{align}
\frac{\zeta}{\xi}
<
\frac{T^2-2\kappa^2}{\kappa T}.
\label{eq:app:ratio}
\end{align}
To verify this bound, write
\[
a:=\frac{k_1}{4},
\qquad
b:=\frac{k_2}{4},
\qquad
A:=\eta_1a,
\qquad
B:=\eta_2b.
\]
Then
\begin{align*}
\xi
&=
4\left[
ab(\eta_1-\eta_2)+m(A-B)
\right],\\
\zeta-\xi
&=
4(a-b)(A+B).
\end{align*}
Since
\(A-B\ge(a-b)\eta_1\) and
\(A+B\le(a+b)\eta_1\),
\begin{align*}
\frac{\zeta}{\xi}
\le
1+\rho,
\qquad
\rho:=\frac{a+b}{m}.
\end{align*}
Moreover,
\[
\rho=\nu,
\qquad
\frac{\Pi}{\kappa^2}
=
4q\rho(\rho+2).
\]
Because \(q>1\), the right-hand side of
Eq.~\eqref{eq:app:ratio} is larger than its value at \(q=1\),
\[
\frac{2\rho^2+4\rho+1}{\rho+1}
>
1+\rho.
\]
Thus \(f(T)>0\), and \(p_\infty\) is strictly increasing in \(w\).
This proves the monotonicity stated in the main text.
\hfill\(\blacksquare\)

At \(w=0\), Eq.~\eqref{eq:app:ARE} decouples:
\[
q_i-2\kappa\sigma_i-r_i\sigma_i^2=0.
\]
The positive roots are
\[
\sigma_i
=
\frac{q_i}{\kappa+g_i},
\qquad
g_i:=\sqrt{\kappa^2+r_iq_i},
\]
which gives Eq.~\eqref{bathat0}. As \(w\to\infty\),
\[
\sqrt P
=
w+\kappa^2+\frac{t_2}{2}+O(w^{-1}),
\]
so
\[
\tau
\longrightarrow
T=\sqrt{4\kappa^2+\tr R\,\tr Q},
\qquad
S-2\sqrt P\longrightarrow-\infty.
\]
Hence \(r\to T\). Evaluating Eq.~\eqref{eq:app:ptau} at
\(\tau=T\) gives
\begin{align*}
p_{\rm fr}^2
=
\frac{\tr R\,(T+2\kappa)}
{4\tr Q\,(T-2\kappa)}
=
\frac{(T+2\kappa)^2}{4(\tr Q)^2},
\end{align*}
which is Eq.~\eqref{bathlargeomega}. Since
\(\tau-T=O(w^{-1})\), both limiting corrections are
\(O(\omega_{\rm eff}^{-2})\).

\subsection{Fast-rate purity and gain bounds}
\label{app:gain}

For \(\kappa>0\),
\[
\tr R=2\eta_{\rm eff}k_{\rm eff},
\qquad
\tr Q=\frac{k_{\rm eff}}{2}+4\kappa\sigma_{\rm th}.
\]
Since
\[
k_{\rm eff}
=
4\kappa\sigma_{\rm th}\nu,
\qquad
4\sigma_{\rm th}^2\eta_{\rm eff}=q,
\]
we obtain
\begin{align*}
\tr Q
&=
2\kappa\sigma_{\rm th}(\nu+2),\\
\tr R\,\tr Q
&=
4\kappa^2q\nu(\nu+2).
\end{align*}
Substitution into Eq.~\eqref{bathlargeomega} yields the expression for
\(p_{\rm fr}\) used in Eq.~\eqref{eq:fractionalgain}. Rationalizing
that equation gives
\[
\mathcal G(\nu,q)
=
\frac{
\nu(\sqrt q+1)
}{
\sqrt{1+q\nu(\nu+2)}+\nu+1
}.
\]

For the bounds, set
\[
x:=\sqrt q,
\qquad
c:=\nu(\nu+2),
\qquad
h(x)
:=
\frac{1+x}{\sqrt{1+cx^2}+\nu+1},
\]
so that \(\mathcal G=\nu h\). Up to a positive factor,
\[
h'(x)
\sim
1-cx+(\nu+1)\sqrt{1+cx^2}.
\]
This is positive for \(x>0\). If \(cx>1\), then the above is positive because 
\[
(\nu+1)^2(1+cx^2)-(cx-1)^2
=
c(1+x)^2>0.
\]
For \(cx\le1\) positivity is immediate. Thus
\(\mathcal G\) increases strictly with \(q\). Finally,
\[
\lim_{q\to1^+}\mathcal G
=
\frac{\nu}{1+\nu},
\qquad
\lim_{q\to\infty}\mathcal G
=
\sqrt{\frac{\nu}{\nu+2}},
\]
which proves Eq.~\eqref{eq:gainbound} and the sharpness of both
bounds.
\hfill\(\blacksquare\)

\subsection{Optimal vibrational protocol}
\label{app:harmonics}

We explain herein why the \(s=0\) vibrational optimum stated in
Sec.~\ref{sec:vibrational} coincides with the fast rate limit
\begin{align}
    p^\star=p_{\rm fr}.
\end{align}
Since
\(p_\infty(u,v)\) is rotationally symmetric, i.e., depends only on
\(\rho:=\sqrt{u^2+v^2}\), we can take \((u,v)=(\rho,0)\) without loss of generality. Now define
\[
\tau:=\frac{1+\rho}{2}.
\]
The corresponding averaged measurement is the mixture of two
orthogonal constant-angle measurements with weights
\(\tau\) and \(1-\tau\). It is equivalent to an {effective} detector
pair with
\begin{align*}
k_1'
&=
\tau k_1+(1-\tau)k_2,\\
k_2'
&=
\tau k_2+(1-\tau)k_1,
\end{align*}
and
\begin{align*}
\eta_1'k_1'
&=
\tau\eta_1k_1+(1-\tau)\eta_2k_2,\\
\eta_2'k_2'
&=
\tau\eta_2k_2+(1-\tau)\eta_1k_1.
\end{align*}
This transformation preserves \(k_{\rm eff}\) and
\(\eta_{\rm eff}\), and therefore preserves \(q\) and \(\nu\).
Moreover, for positive detector strengths,
\begin{align*}
(k_1'-k_2')(\eta_1'-\eta_2')
=
\frac{
\rho^2(k_1-k_2)k_1k_2(\eta_1-\eta_2)
}{
k_1'k_2'
}
\ge0,
\end{align*}
so the effective pair satisfies the same detector ordering. The monotonicity result
of Appendix~\ref{app:monotone} therefore implies
\[
p_\infty(u,v)
\le
p_{\rm fr}
=
p_\infty(0,0).
\]
Thus the disk center is a global maximizer for \(s=0\), as stated in
the main text.

{
\subsection{The \texorpdfstring{$s=0$}{s=0} variational limit}
\label{app:variational-s0}

We justify the variational interpretation of the fast-rate limit used in
Sec.~\ref{sec:variational}. If we relax the instantaneous constraint
\(\|u_t\|=1\) to the disk \(\|u_t\|\leq1\). Then for \(s=0\), the center
\(u_t=0\) gives complete angular averaging, with
\begin{align}
a_R
&=
\eta_{\rm eff}k_{\rm eff}I,
&
a_K
&=
\frac{k_{\rm eff}}{4}I.
\end{align}
The remaining coefficient matrices \(b_R,c_R,b_K,c_K\) are traceless. Thus, starting from \(\Sigma_0=\Sigma_{\rm th}\), the relaxed control \(u_t=0\)
preserves
\begin{align}
\Sigma_t=\sigma_t I,
\end{align}
where
\begin{align}
\dot\sigma_t
=
2\kappa(\sigma_{\rm th}-\sigma_t)
+\frac{k_{\rm eff}}{4}
-\eta_{\rm eff}k_{\rm eff}\sigma_t^2.
\label{eq:app:variationalSigma}
\end{align}
The terminal condition \(\Pi_T=0\) and the costate equation likewise preserve
\begin{align}
\Pi_t=\pi_t I,
\end{align}
with
\begin{align}
\dot\pi_t
=
-\frac{1}{\sigma_t}
+
2\left(
\kappa+\eta_{\rm eff}k_{\rm eff}\sigma_t
\right)\pi_t.
\label{eq:app:variationalPi}
\end{align}
Along this trajectory,
\begin{align}
\mathcal B_t
&=
\pi_t\tr\!\left(b_K-\sigma_t^2b_R\right)=0,
&
\mathcal C_t
&=
\pi_t\tr\!\left(c_K-\sigma_t^2c_R\right)=0,
\end{align}
because the corresponding coefficient matrices are traceless. Thus the
Pontryagin Hamiltonian is independent of the relaxed control at \(u_t=0\), and
the completely averaged trajectory is a singular extremal of the relaxed
problem.

The point is that \(u_t=0\) is not itself a physical measurement angle, since physical
controls satisfy \(\|u_t\|=1\), but it is approached, by increasingly
rapid angular modulation whose second harmonic averages to zero. The fast-rate
limit is therefore the physical high-frequency realization of this relaxed
trajectory. This necessary-condition argument explains the observed numerical
variational solution at \(s=0\), but of course does not by itself prove global
finite-horizon optimality over all admissible physical controls.
}

\bibliography{refs}

\begin{thebibliography}{51}%
\makeatletter
\providecommand \@ifxundefined [1]{%
 \@ifx{#1\undefined}
}%
\providecommand \@ifnum [1]{%
 \ifnum #1\expandafter \@firstoftwo
 \else \expandafter \@secondoftwo
 \fi
}%
\providecommand \@ifx [1]{%
 \ifx #1\expandafter \@firstoftwo
 \else \expandafter \@secondoftwo
 \fi
}%
\providecommand \natexlab [1]{#1}%
\providecommand \enquote  [1]{``#1''}%
\providecommand \bibnamefont  [1]{#1}%
\providecommand \bibfnamefont [1]{#1}%
\providecommand \citenamefont [1]{#1}%
\providecommand \href@noop [0]{\@secondoftwo}%
\providecommand \href [0]{\begingroup \@sanitize@url \@href}%
\providecommand \@href[1]{\@@startlink{#1}\@@href}%
\providecommand \@@href[1]{\endgroup#1\@@endlink}%
\providecommand \@sanitize@url [0]{\catcode `\\12\catcode `\$12\catcode
  `\&12\catcode `\#12\catcode `\^12\catcode `\_12\catcode `\%12\relax}%
\providecommand \@@startlink[1]{}%
\providecommand \@@endlink[0]{}%
\providecommand \url  [0]{\begingroup\@sanitize@url \@url }%
\providecommand \@url [1]{\endgroup\@href {#1}{\urlprefix }}%
\providecommand \urlprefix  [0]{URL }%
\providecommand \Eprint [0]{\href }%
\providecommand \doibase [0]{https://doi.org/}%
\providecommand \selectlanguage [0]{\@gobble}%
\providecommand \bibinfo  [0]{\@secondoftwo}%
\providecommand \bibfield  [0]{\@secondoftwo}%
\providecommand \translation [1]{[#1]}%
\providecommand \BibitemOpen [0]{}%
\providecommand \bibitemStop [0]{}%
\providecommand \bibitemNoStop [0]{.\EOS\space}%
\providecommand \EOS [0]{\spacefactor3000\relax}%
\providecommand \BibitemShut  [1]{\csname bibitem#1\endcsname}%
\let\auto@bib@innerbib\@empty
\bibitem [{\citenamefont {Mitchison}\ \emph {et~al.}(2021)\citenamefont
  {Mitchison}, \citenamefont {Goold},\ and\ \citenamefont
  {Prior}}]{mitchison2021charging}%
  \BibitemOpen
  \bibfield  {author} {\bibinfo {author} {\bibfnamefont {M.~T.}\ \bibnamefont
  {Mitchison}}, \bibinfo {author} {\bibfnamefont {J.}~\bibnamefont {Goold}},\
  and\ \bibinfo {author} {\bibfnamefont {J.}~\bibnamefont {Prior}},\ }\href
  {https://doi.org/10.22331/q-2021-07-13-500} {\bibfield  {journal} {\bibinfo
  {journal} {Quantum}\ }\textbf {\bibinfo {volume} {5}},\ \bibinfo {pages}
  {500} (\bibinfo {year} {2021})}\BibitemShut {NoStop}%
\bibitem [{\citenamefont {Morrone}\ \emph {et~al.}(2023)\citenamefont
  {Morrone}, \citenamefont {Rossi},\ and\ \citenamefont
  {Genoni}}]{morrone2023daemonic}%
  \BibitemOpen
  \bibfield  {author} {\bibinfo {author} {\bibfnamefont {D.}~\bibnamefont
  {Morrone}}, \bibinfo {author} {\bibfnamefont {M.~A.~C.}\ \bibnamefont
  {Rossi}},\ and\ \bibinfo {author} {\bibfnamefont {M.~G.}\ \bibnamefont
  {Genoni}},\ }\href {https://doi.org/10.1103/PhysRevApplied.20.044073}
  {\bibfield  {journal} {\bibinfo  {journal} {Physical Review Applied}\
  }\textbf {\bibinfo {volume} {20}},\ \bibinfo {pages} {044073} (\bibinfo
  {year} {2023})}\BibitemShut {NoStop}%
\bibitem [{\citenamefont {Kua}\ \emph {et~al.}(2025)\citenamefont {Kua},
  \citenamefont {Serafini},\ and\ \citenamefont {Genoni}}]{kua2025daemonic}%
  \BibitemOpen
  \bibfield  {author} {\bibinfo {author} {\bibfnamefont {K.~H.}\ \bibnamefont
  {Kua}}, \bibinfo {author} {\bibfnamefont {A.}~\bibnamefont {Serafini}},\ and\
  \bibinfo {author} {\bibfnamefont {M.~G.}\ \bibnamefont {Genoni}},\ }\href
  {https://doi.org/10.1088/2058-9565/ae2200} {\bibfield  {journal} {\bibinfo
  {journal} {Quantum Science and Technology}\ }\textbf {\bibinfo {volume}
  {11}},\ \bibinfo {pages} {015014} (\bibinfo {year} {2025})}\BibitemShut
  {NoStop}%
\bibitem [{\citenamefont {Cenedese}\ \emph {et~al.}(2026)\citenamefont
  {Cenedese}, \citenamefont {Benenti}, \citenamefont {Ferraro},\ and\
  \citenamefont {Genoni}}]{cenedese2025boosting}%
  \BibitemOpen
  \bibfield  {author} {\bibinfo {author} {\bibfnamefont {G.}~\bibnamefont
  {Cenedese}}, \bibinfo {author} {\bibfnamefont {G.}~\bibnamefont {Benenti}},
  \bibinfo {author} {\bibfnamefont {D.}~\bibnamefont {Ferraro}},\ and\ \bibinfo
  {author} {\bibfnamefont {M.~G.}\ \bibnamefont {Genoni}},\ }\href
  {https://doi.org/10.1103/f591-hx6x} {\bibfield  {journal} {\bibinfo
  {journal} {Physical Review Letters}\ }\textbf {\bibinfo {volume} {137}},\
  \bibinfo {pages} {050406} (\bibinfo {year} {2026})}\BibitemShut {NoStop}%
\bibitem [{\citenamefont {Belavkin}(1989)}]{belavkin1987information}%
  \BibitemOpen
  \bibfield  {author} {\bibinfo {author} {\bibfnamefont {V.~P.}\ \bibnamefont
  {Belavkin}},\ }in\ \href@noop {} {\emph {\bibinfo {booktitle} {Modeling and
  Control of Systems}}},\ \bibinfo {series} {Lecture Notes in Control and
  Information Sciences}, Vol.\ \bibinfo {volume} {121},\ \bibinfo {editor}
  {edited by\ \bibinfo {editor} {\bibfnamefont {A.}~\bibnamefont
  {Blaqui{\`e}re}}}\ (\bibinfo  {publisher} {Springer},\ \bibinfo {address}
  {Berlin},\ \bibinfo {year} {1989})\ pp.\ \bibinfo {pages}
  {245--265}\BibitemShut {NoStop}%
\bibitem [{\citenamefont {Wiseman}\ and\ \citenamefont
  {Milburn}(2009)}]{Wiseman_Milburn_2009}%
  \BibitemOpen
  \bibfield  {author} {\bibinfo {author} {\bibfnamefont {H.~M.}\ \bibnamefont
  {Wiseman}}\ and\ \bibinfo {author} {\bibfnamefont {G.~J.}\ \bibnamefont
  {Milburn}},\ }\href@noop {} {\emph {\bibinfo {title} {Quantum Measurement and
  Control}}}\ (\bibinfo  {publisher} {Cambridge University Press},\ \bibinfo
  {address} {Cambridge},\ \bibinfo {year} {2009})\BibitemShut {NoStop}%
\bibitem [{\citenamefont {Jacobs}\ and\ \citenamefont
  {Steck}(2006)}]{jacobs2006straightforward}%
  \BibitemOpen
  \bibfield  {author} {\bibinfo {author} {\bibfnamefont {K.}~\bibnamefont
  {Jacobs}}\ and\ \bibinfo {author} {\bibfnamefont {D.~A.}\ \bibnamefont
  {Steck}},\ }\href {https://doi.org/10.1080/00107510601101934} {\bibfield
  {journal} {\bibinfo  {journal} {Contemporary Physics}\ }\textbf {\bibinfo
  {volume} {47}},\ \bibinfo {pages} {279} (\bibinfo {year} {2006})}\BibitemShut
  {NoStop}%
\bibitem [{\citenamefont {Jacobs}(2014)}]{jacobs2014quantum}%
  \BibitemOpen
  \bibfield  {author} {\bibinfo {author} {\bibfnamefont {K.}~\bibnamefont
  {Jacobs}},\ }\href@noop {} {\emph {\bibinfo {title} {Quantum Measurement
  Theory and its Applications}}}\ (\bibinfo  {publisher} {Cambridge University
  Press},\ \bibinfo {address} {Cambridge},\ \bibinfo {year} {2014})\BibitemShut
  {NoStop}%
\bibitem [{\citenamefont {Jordan}\ and\ \citenamefont
  {Siddiqi}(2024)}]{jordan2024quantum}%
  \BibitemOpen
  \bibfield  {author} {\bibinfo {author} {\bibfnamefont {A.~N.}\ \bibnamefont
  {Jordan}}\ and\ \bibinfo {author} {\bibfnamefont {I.~A.}\ \bibnamefont
  {Siddiqi}},\ }\href@noop {} {\emph {\bibinfo {title} {Quantum Measurement:
  Theory and Practice}}}\ (\bibinfo  {publisher} {Cambridge University Press},\
  \bibinfo {address} {Cambridge},\ \bibinfo {year} {2024})\BibitemShut
  {NoStop}%
\bibitem [{\citenamefont {Albarelli}\ and\ \citenamefont
  {Genoni}(2024)}]{ALBARELLI2024129260}%
  \BibitemOpen
  \bibfield  {author} {\bibinfo {author} {\bibfnamefont {F.}~\bibnamefont
  {Albarelli}}\ and\ \bibinfo {author} {\bibfnamefont {M.~G.}\ \bibnamefont
  {Genoni}},\ }\href {https://doi.org/10.1016/j.physleta.2023.129260}
  {\bibfield  {journal} {\bibinfo  {journal} {Physics Letters A}\ }\textbf
  {\bibinfo {volume} {494}},\ \bibinfo {pages} {129260} (\bibinfo {year}
  {2024})}\BibitemShut {NoStop}%
\bibitem [{\citenamefont {Jacobs}(2003)}]{jacobs2003rapid}%
  \BibitemOpen
  \bibfield  {author} {\bibinfo {author} {\bibfnamefont {K.}~\bibnamefont
  {Jacobs}},\ }\href {https://doi.org/10.1103/PhysRevA.67.030301} {\bibfield
  {journal} {\bibinfo  {journal} {Physical Review A}\ }\textbf {\bibinfo
  {volume} {67}},\ \bibinfo {pages} {030301(R)} (\bibinfo {year}
  {2003})}\BibitemShut {NoStop}%
\bibitem [{\citenamefont {Combes}\ and\ \citenamefont
  {Jacobs}(2006)}]{combes2006rapid}%
  \BibitemOpen
  \bibfield  {author} {\bibinfo {author} {\bibfnamefont {J.}~\bibnamefont
  {Combes}}\ and\ \bibinfo {author} {\bibfnamefont {K.}~\bibnamefont
  {Jacobs}},\ }\href {https://doi.org/10.1103/PhysRevLett.96.010504} {\bibfield
   {journal} {\bibinfo  {journal} {Physical Review Letters}\ }\textbf {\bibinfo
  {volume} {96}},\ \bibinfo {pages} {010504} (\bibinfo {year}
  {2006})}\BibitemShut {NoStop}%
\bibitem [{\citenamefont {Combes}\ \emph {et~al.}(2008)\citenamefont {Combes},
  \citenamefont {Wiseman},\ and\ \citenamefont {Jacobs}}]{combes2008rapid}%
  \BibitemOpen
  \bibfield  {author} {\bibinfo {author} {\bibfnamefont {J.}~\bibnamefont
  {Combes}}, \bibinfo {author} {\bibfnamefont {H.~M.}\ \bibnamefont
  {Wiseman}},\ and\ \bibinfo {author} {\bibfnamefont {K.}~\bibnamefont
  {Jacobs}},\ }\href {https://doi.org/10.1103/PhysRevLett.100.160503}
  {\bibfield  {journal} {\bibinfo  {journal} {Physical Review Letters}\
  }\textbf {\bibinfo {volume} {100}},\ \bibinfo {pages} {160503} (\bibinfo
  {year} {2008})}\BibitemShut {NoStop}%
\bibitem [{\citenamefont {Ruskov}\ \emph {et~al.}(2010)\citenamefont {Ruskov},
  \citenamefont {Korotkov},\ and\ \citenamefont {M{\o}lmer}}]{ruskov2010three}%
  \BibitemOpen
  \bibfield  {author} {\bibinfo {author} {\bibfnamefont {R.}~\bibnamefont
  {Ruskov}}, \bibinfo {author} {\bibfnamefont {A.~N.}\ \bibnamefont
  {Korotkov}},\ and\ \bibinfo {author} {\bibfnamefont {K.}~\bibnamefont
  {M{\o}lmer}},\ }\href {https://doi.org/10.1103/PhysRevLett.105.100506}
  {\bibfield  {journal} {\bibinfo  {journal} {Physical Review Letters}\
  }\textbf {\bibinfo {volume} {105}},\ \bibinfo {pages} {100506} (\bibinfo
  {year} {2010})}\BibitemShut {NoStop}%
\bibitem [{\citenamefont {Wiseman}\ and\ \citenamefont
  {Ralph}(2006)}]{wiseman2006reconsidering}%
  \BibitemOpen
  \bibfield  {author} {\bibinfo {author} {\bibfnamefont {H.~M.}\ \bibnamefont
  {Wiseman}}\ and\ \bibinfo {author} {\bibfnamefont {J.~F.}\ \bibnamefont
  {Ralph}},\ }\href {https://doi.org/10.1088/1367-2630/8/6/090} {\bibfield
  {journal} {\bibinfo  {journal} {New Journal of Physics}\ }\textbf {\bibinfo
  {volume} {8}},\ \bibinfo {pages} {90} (\bibinfo {year} {2006})}\BibitemShut
  {NoStop}%
\bibitem [{\citenamefont {Jordan}\ and\ \citenamefont
  {Korotkov}(2006)}]{jordan2006qubit}%
  \BibitemOpen
  \bibfield  {author} {\bibinfo {author} {\bibfnamefont {A.~N.}\ \bibnamefont
  {Jordan}}\ and\ \bibinfo {author} {\bibfnamefont {A.~N.}\ \bibnamefont
  {Korotkov}},\ }\href {https://doi.org/10.1103/PhysRevB.74.085307} {\bibfield
  {journal} {\bibinfo  {journal} {Physical Review B}\ }\textbf {\bibinfo
  {volume} {74}},\ \bibinfo {pages} {085307} (\bibinfo {year}
  {2006})}\BibitemShut {NoStop}%
\bibitem [{\citenamefont {Zhang}\ \emph {et~al.}(2017)\citenamefont {Zhang},
  \citenamefont {Liu}, \citenamefont {Wu}, \citenamefont {Jacobs},\ and\
  \citenamefont {Nori}}]{zhang2017feedback}%
  \BibitemOpen
  \bibfield  {author} {\bibinfo {author} {\bibfnamefont {J.}~\bibnamefont
  {Zhang}}, \bibinfo {author} {\bibfnamefont {Y.-x.}\ \bibnamefont {Liu}},
  \bibinfo {author} {\bibfnamefont {R.-B.}\ \bibnamefont {Wu}}, \bibinfo
  {author} {\bibfnamefont {K.}~\bibnamefont {Jacobs}},\ and\ \bibinfo {author}
  {\bibfnamefont {F.}~\bibnamefont {Nori}},\ }\href
  {https://doi.org/10.1016/j.physrep.2017.02.003} {\bibfield  {journal}
  {\bibinfo  {journal} {Physics Reports}\ }\textbf {\bibinfo {volume} {679}},\
  \bibinfo {pages} {1} (\bibinfo {year} {2017})}\BibitemShut {NoStop}%
\bibitem [{\citenamefont {Bechhoefer}(2021)}]{bechhoefer2021control}%
  \BibitemOpen
  \bibfield  {author} {\bibinfo {author} {\bibfnamefont {J.}~\bibnamefont
  {Bechhoefer}},\ }\href@noop {} {\emph {\bibinfo {title} {Control Theory for
  Physicists}}}\ (\bibinfo  {publisher} {Cambridge University Press},\ \bibinfo
  {address} {Cambridge},\ \bibinfo {year} {2021})\BibitemShut {NoStop}%
\bibitem [{\citenamefont {Doherty}\ and\ \citenamefont
  {Jacobs}(1999)}]{doherty1999feedback}%
  \BibitemOpen
  \bibfield  {author} {\bibinfo {author} {\bibfnamefont {A.~C.}\ \bibnamefont
  {Doherty}}\ and\ \bibinfo {author} {\bibfnamefont {K.}~\bibnamefont
  {Jacobs}},\ }\href {https://doi.org/10.1103/PhysRevA.60.2700} {\bibfield
  {journal} {\bibinfo  {journal} {Physical Review A}\ }\textbf {\bibinfo
  {volume} {60}},\ \bibinfo {pages} {2700} (\bibinfo {year}
  {1999})}\BibitemShut {NoStop}%
\bibitem [{\citenamefont {Genoni}\ \emph {et~al.}(2016)\citenamefont {Genoni},
  \citenamefont {Lami},\ and\ \citenamefont {Serafini}}]{Genoni02072016}%
  \BibitemOpen
  \bibfield  {author} {\bibinfo {author} {\bibfnamefont {M.~G.}\ \bibnamefont
  {Genoni}}, \bibinfo {author} {\bibfnamefont {L.}~\bibnamefont {Lami}},\ and\
  \bibinfo {author} {\bibfnamefont {A.}~\bibnamefont {Serafini}},\ }\href
  {https://doi.org/10.1080/00107514.2015.1125624} {\bibfield  {journal}
  {\bibinfo  {journal} {Contemporary Physics}\ }\textbf {\bibinfo {volume}
  {57}},\ \bibinfo {pages} {331} (\bibinfo {year} {2016})}\BibitemShut
  {NoStop}%
\bibitem [{\citenamefont {Serafini}(2017)}]{serafini2017quantum}%
  \BibitemOpen
  \bibfield  {author} {\bibinfo {author} {\bibfnamefont {A.}~\bibnamefont
  {Serafini}},\ }\href@noop {} {\emph {\bibinfo {title} {Quantum Continuous
  Variables: A Primer of Theoretical Methods}}}\ (\bibinfo  {publisher} {CRC
  Press},\ \bibinfo {address} {Boca Raton},\ \bibinfo {year}
  {2017})\BibitemShut {NoStop}%
\bibitem [{\citenamefont {Paris}\ \emph {et~al.}(2003)\citenamefont {Paris},
  \citenamefont {Illuminati}, \citenamefont {Serafini},\ and\ \citenamefont
  {De~Siena}}]{paris2003purity}%
  \BibitemOpen
  \bibfield  {author} {\bibinfo {author} {\bibfnamefont {M.~G.~A.}\
  \bibnamefont {Paris}}, \bibinfo {author} {\bibfnamefont {F.}~\bibnamefont
  {Illuminati}}, \bibinfo {author} {\bibfnamefont {A.}~\bibnamefont
  {Serafini}},\ and\ \bibinfo {author} {\bibfnamefont {S.}~\bibnamefont
  {De~Siena}},\ }\href {https://doi.org/10.1103/PhysRevA.68.012314} {\bibfield
  {journal} {\bibinfo  {journal} {Physical Review A}\ }\textbf {\bibinfo
  {volume} {68}},\ \bibinfo {pages} {012314} (\bibinfo {year}
  {2003})}\BibitemShut {NoStop}%
\bibitem [{\citenamefont {Wiseman}\ and\ \citenamefont
  {Di{\'o}si}(2001)}]{wiseman2001complete}%
  \BibitemOpen
  \bibfield  {author} {\bibinfo {author} {\bibfnamefont {H.~M.}\ \bibnamefont
  {Wiseman}}\ and\ \bibinfo {author} {\bibfnamefont {L.}~\bibnamefont
  {Di{\'o}si}},\ }\href@noop {} {\bibfield  {journal} {\bibinfo  {journal}
  {Chemical Physics}\ }\textbf {\bibinfo {volume} {268}},\ \bibinfo {pages}
  {91} (\bibinfo {year} {2001})}\BibitemShut {NoStop}%
\bibitem [{\citenamefont {Wiseman}\ and\ \citenamefont
  {Doherty}(2005)}]{wiseman2005optimal}%
  \BibitemOpen
  \bibfield  {author} {\bibinfo {author} {\bibfnamefont {H.~M.}\ \bibnamefont
  {Wiseman}}\ and\ \bibinfo {author} {\bibfnamefont {A.~C.}\ \bibnamefont
  {Doherty}},\ }\href {https://doi.org/10.1103/PhysRevLett.94.070405}
  {\bibfield  {journal} {\bibinfo  {journal} {Physical Review Letters}\
  }\textbf {\bibinfo {volume} {94}},\ \bibinfo {pages} {070405} (\bibinfo
  {year} {2005})}\BibitemShut {NoStop}%
\bibitem [{\citenamefont {Mancini}\ and\ \citenamefont
  {Wiseman}(2007)}]{mancini2007optimal}%
  \BibitemOpen
  \bibfield  {author} {\bibinfo {author} {\bibfnamefont {S.}~\bibnamefont
  {Mancini}}\ and\ \bibinfo {author} {\bibfnamefont {H.~M.}\ \bibnamefont
  {Wiseman}},\ }\href {https://doi.org/10.1103/PhysRevA.75.012330} {\bibfield
  {journal} {\bibinfo  {journal} {Physical Review A}\ }\textbf {\bibinfo
  {volume} {75}},\ \bibinfo {pages} {012330} (\bibinfo {year}
  {2007})}\BibitemShut {NoStop}%
\bibitem [{\citenamefont {Serafini}\ and\ \citenamefont
  {Mancini}(2010)}]{serafini2010determination}%
  \BibitemOpen
  \bibfield  {author} {\bibinfo {author} {\bibfnamefont {A.}~\bibnamefont
  {Serafini}}\ and\ \bibinfo {author} {\bibfnamefont {S.}~\bibnamefont
  {Mancini}},\ }\href {https://doi.org/10.1103/PhysRevLett.104.220501}
  {\bibfield  {journal} {\bibinfo  {journal} {Physical Review Letters}\
  }\textbf {\bibinfo {volume} {104}},\ \bibinfo {pages} {220501} (\bibinfo
  {year} {2010})}\BibitemShut {NoStop}%
\bibitem [{\citenamefont {Genoni}\ \emph {et~al.}(2013)\citenamefont {Genoni},
  \citenamefont {Mancini},\ and\ \citenamefont {Serafini}}]{genoni2013optimal}%
  \BibitemOpen
  \bibfield  {author} {\bibinfo {author} {\bibfnamefont {M.~G.}\ \bibnamefont
  {Genoni}}, \bibinfo {author} {\bibfnamefont {S.}~\bibnamefont {Mancini}},\
  and\ \bibinfo {author} {\bibfnamefont {A.}~\bibnamefont {Serafini}},\ }\href
  {https://doi.org/10.1103/PhysRevA.87.042333} {\bibfield  {journal} {\bibinfo
  {journal} {Physical Review A}\ }\textbf {\bibinfo {volume} {87}},\ \bibinfo
  {pages} {042333} (\bibinfo {year} {2013})}\BibitemShut {NoStop}%
\bibitem [{\citenamefont {Genoni}\ \emph {et~al.}(2014)\citenamefont {Genoni},
  \citenamefont {Mancini},\ and\ \citenamefont {Serafini}}]{genoni2014general}%
  \BibitemOpen
  \bibfield  {author} {\bibinfo {author} {\bibfnamefont {M.~G.}\ \bibnamefont
  {Genoni}}, \bibinfo {author} {\bibfnamefont {S.}~\bibnamefont {Mancini}},\
  and\ \bibinfo {author} {\bibfnamefont {A.}~\bibnamefont {Serafini}},\ }\href
  {https://doi.org/10.1134/S1061920814030054} {\bibfield  {journal} {\bibinfo
  {journal} {Russian Journal of Mathematical Physics}\ }\textbf {\bibinfo
  {volume} {21}},\ \bibinfo {pages} {329} (\bibinfo {year} {2014})}\BibitemShut
  {NoStop}%
\bibitem [{\citenamefont {Rossi}\ \emph {et~al.}(2018)\citenamefont {Rossi},
  \citenamefont {Mason}, \citenamefont {Chen}, \citenamefont {Tsaturyan},\ and\
  \citenamefont {Schliesser}}]{rossi2018measurement}%
  \BibitemOpen
  \bibfield  {author} {\bibinfo {author} {\bibfnamefont {M.}~\bibnamefont
  {Rossi}}, \bibinfo {author} {\bibfnamefont {D.}~\bibnamefont {Mason}},
  \bibinfo {author} {\bibfnamefont {J.}~\bibnamefont {Chen}}, \bibinfo {author}
  {\bibfnamefont {Y.}~\bibnamefont {Tsaturyan}},\ and\ \bibinfo {author}
  {\bibfnamefont {A.}~\bibnamefont {Schliesser}},\ }\href
  {https://doi.org/10.1038/s41586-018-0643-8} {\bibfield  {journal} {\bibinfo
  {journal} {Nature}\ }\textbf {\bibinfo {volume} {563}},\ \bibinfo {pages}
  {53} (\bibinfo {year} {2018})}\BibitemShut {NoStop}%
\bibitem [{\citenamefont {Manikandan}\ \emph {et~al.}(2022)\citenamefont
  {Manikandan}, \citenamefont {Elouard}, \citenamefont {Murch}, \citenamefont
  {Auff{\`e}ves},\ and\ \citenamefont {Jordan}}]{manikandan2022engine}%
  \BibitemOpen
  \bibfield  {author} {\bibinfo {author} {\bibfnamefont {S.~K.}\ \bibnamefont
  {Manikandan}}, \bibinfo {author} {\bibfnamefont {C.}~\bibnamefont {Elouard}},
  \bibinfo {author} {\bibfnamefont {K.~W.}\ \bibnamefont {Murch}}, \bibinfo
  {author} {\bibfnamefont {A.}~\bibnamefont {Auff{\`e}ves}},\ and\ \bibinfo
  {author} {\bibfnamefont {A.~N.}\ \bibnamefont {Jordan}},\ }\href
  {https://doi.org/10.1103/PhysRevE.105.044137} {\bibfield  {journal} {\bibinfo
   {journal} {Physical Review E}\ }\textbf {\bibinfo {volume} {105}},\ \bibinfo
  {pages} {044137} (\bibinfo {year} {2022})}\BibitemShut {NoStop}%
\bibitem [{\citenamefont {Bhandari}\ and\ \citenamefont
  {Jordan}(2022)}]{bhandari2022boosted}%
  \BibitemOpen
  \bibfield  {author} {\bibinfo {author} {\bibfnamefont {B.}~\bibnamefont
  {Bhandari}}\ and\ \bibinfo {author} {\bibfnamefont {A.~N.}\ \bibnamefont
  {Jordan}},\ }\href {https://doi.org/10.1103/PhysRevResearch.4.033103}
  {\bibfield  {journal} {\bibinfo  {journal} {Physical Review Research}\
  }\textbf {\bibinfo {volume} {4}},\ \bibinfo {pages} {033103} (\bibinfo {year}
  {2022})}\BibitemShut {NoStop}%
\bibitem [{\citenamefont {Karmakar}\ \emph {et~al.}(2022)\citenamefont
  {Karmakar}, \citenamefont {Lewalle},\ and\ \citenamefont
  {Jordan}}]{karmakar2022stochastic}%
  \BibitemOpen
  \bibfield  {author} {\bibinfo {author} {\bibfnamefont {T.}~\bibnamefont
  {Karmakar}}, \bibinfo {author} {\bibfnamefont {P.}~\bibnamefont {Lewalle}},\
  and\ \bibinfo {author} {\bibfnamefont {A.~N.}\ \bibnamefont {Jordan}},\
  }\href {https://doi.org/10.1103/PRXQuantum.3.010327} {\bibfield  {journal}
  {\bibinfo  {journal} {PRX Quantum}\ }\textbf {\bibinfo {volume} {3}},\
  \bibinfo {pages} {010327} (\bibinfo {year} {2022})}\BibitemShut {NoStop}%
\bibitem [{\citenamefont {Bhandari}\ \emph {et~al.}(2023)\citenamefont
  {Bhandari}, \citenamefont {Czupryniak}, \citenamefont {Erdman},\ and\
  \citenamefont {Jordan}}]{bhandari2023machines}%
  \BibitemOpen
  \bibfield  {author} {\bibinfo {author} {\bibfnamefont {B.}~\bibnamefont
  {Bhandari}}, \bibinfo {author} {\bibfnamefont {R.}~\bibnamefont
  {Czupryniak}}, \bibinfo {author} {\bibfnamefont {P.~A.}\ \bibnamefont
  {Erdman}},\ and\ \bibinfo {author} {\bibfnamefont {A.~N.}\ \bibnamefont
  {Jordan}},\ }\href {https://doi.org/10.3390/e25020204} {\bibfield  {journal}
  {\bibinfo  {journal} {Entropy}\ }\textbf {\bibinfo {volume} {25}},\ \bibinfo
  {pages} {204} (\bibinfo {year} {2023})}\BibitemShut {NoStop}%
\bibitem [{\citenamefont {Manikandan}\ and\ \citenamefont
  {Qvarfort}(2023)}]{manikandan2023cooling}%
  \BibitemOpen
  \bibfield  {author} {\bibinfo {author} {\bibfnamefont {S.~K.}\ \bibnamefont
  {Manikandan}}\ and\ \bibinfo {author} {\bibfnamefont {S.}~\bibnamefont
  {Qvarfort}},\ }\href {https://doi.org/10.1103/PhysRevA.107.023516} {\bibfield
   {journal} {\bibinfo  {journal} {Physical Review A}\ }\textbf {\bibinfo
  {volume} {107}},\ \bibinfo {pages} {023516} (\bibinfo {year}
  {2023})}\BibitemShut {NoStop}%
\bibitem [{\citenamefont {Sabbagh}\ \emph {et~al.}(2025)\citenamefont
  {Sabbagh}, \citenamefont {Miangolarra},\ and\ \citenamefont
  {Georgiou}}]{sabbagh2025regulation}%
  \BibitemOpen
  \bibfield  {author} {\bibinfo {author} {\bibfnamefont {R.}~\bibnamefont
  {Sabbagh}}, \bibinfo {author} {\bibfnamefont {O.~M.}\ \bibnamefont
  {Miangolarra}},\ and\ \bibinfo {author} {\bibfnamefont {T.~T.}\ \bibnamefont
  {Georgiou}},\ }in\ \href@noop {} {\emph {\bibinfo {booktitle} {2025 IEEE 64th
  Conference on Decision and Control (CDC)}}}\ (\bibinfo {organization}
  {IEEE},\ \bibinfo {year} {2025})\ pp.\ \bibinfo {pages}
  {7635--7640}\BibitemShut {NoStop}%
\bibitem [{\citenamefont {Braginsky}\ \emph {et~al.}(1980)\citenamefont
  {Braginsky}, \citenamefont {Vorontsov},\ and\ \citenamefont
  {Thorne}}]{braginsky1980quantum}%
  \BibitemOpen
  \bibfield  {author} {\bibinfo {author} {\bibfnamefont {V.~B.}\ \bibnamefont
  {Braginsky}}, \bibinfo {author} {\bibfnamefont {Y.~I.}\ \bibnamefont
  {Vorontsov}},\ and\ \bibinfo {author} {\bibfnamefont {K.~S.}\ \bibnamefont
  {Thorne}},\ }\href {https://doi.org/10.1126/science.209.4456.547} {\bibfield
  {journal} {\bibinfo  {journal} {Science}\ }\textbf {\bibinfo {volume}
  {209}},\ \bibinfo {pages} {547} (\bibinfo {year} {1980})}\BibitemShut
  {NoStop}%
\bibitem [{\citenamefont {Wiseman}(1995)}]{wiseman1995adaptive}%
  \BibitemOpen
  \bibfield  {author} {\bibinfo {author} {\bibfnamefont {H.~M.}\ \bibnamefont
  {Wiseman}},\ }\href {https://doi.org/10.1103/PhysRevLett.75.4587} {\bibfield
  {journal} {\bibinfo  {journal} {Physical Review Letters}\ }\textbf {\bibinfo
  {volume} {75}},\ \bibinfo {pages} {4587} (\bibinfo {year}
  {1995})}\BibitemShut {NoStop}%
\bibitem [{\citenamefont {Armen}\ \emph {et~al.}(2002)\citenamefont {Armen},
  \citenamefont {Au}, \citenamefont {Stockton}, \citenamefont {Doherty},\ and\
  \citenamefont {Mabuchi}}]{armen2002adaptive}%
  \BibitemOpen
  \bibfield  {author} {\bibinfo {author} {\bibfnamefont {M.~A.}\ \bibnamefont
  {Armen}}, \bibinfo {author} {\bibfnamefont {J.~K.}\ \bibnamefont {Au}},
  \bibinfo {author} {\bibfnamefont {J.~K.}\ \bibnamefont {Stockton}}, \bibinfo
  {author} {\bibfnamefont {A.~C.}\ \bibnamefont {Doherty}},\ and\ \bibinfo
  {author} {\bibfnamefont {H.}~\bibnamefont {Mabuchi}},\ }\href
  {https://doi.org/10.1103/PhysRevLett.89.133602} {\bibfield  {journal}
  {\bibinfo  {journal} {Physical Review Letters}\ }\textbf {\bibinfo {volume}
  {89}},\ \bibinfo {pages} {133602} (\bibinfo {year} {2002})}\BibitemShut
  {NoStop}%
\bibitem [{\citenamefont {Clerk}\ \emph {et~al.}(2008)\citenamefont {Clerk},
  \citenamefont {Marquardt},\ and\ \citenamefont
  {Jacobs}}]{clerk2008backaction}%
  \BibitemOpen
  \bibfield  {author} {\bibinfo {author} {\bibfnamefont {A.~A.}\ \bibnamefont
  {Clerk}}, \bibinfo {author} {\bibfnamefont {F.}~\bibnamefont {Marquardt}},\
  and\ \bibinfo {author} {\bibfnamefont {K.}~\bibnamefont {Jacobs}},\ }\href
  {https://doi.org/10.1088/1367-2630/10/9/095010} {\bibfield  {journal}
  {\bibinfo  {journal} {New Journal of Physics}\ }\textbf {\bibinfo {volume}
  {10}},\ \bibinfo {pages} {095010} (\bibinfo {year} {2008})}\BibitemShut
  {NoStop}%
\bibitem [{\citenamefont {Hertzberg}\ \emph {et~al.}(2010)\citenamefont
  {Hertzberg}, \citenamefont {Rocheleau}, \citenamefont {Ndukum}, \citenamefont
  {Savva}, \citenamefont {Clerk},\ and\ \citenamefont
  {Schwab}}]{hertzberg2010backaction}%
  \BibitemOpen
  \bibfield  {author} {\bibinfo {author} {\bibfnamefont {J.~B.}\ \bibnamefont
  {Hertzberg}}, \bibinfo {author} {\bibfnamefont {T.}~\bibnamefont
  {Rocheleau}}, \bibinfo {author} {\bibfnamefont {T.}~\bibnamefont {Ndukum}},
  \bibinfo {author} {\bibfnamefont {M.}~\bibnamefont {Savva}}, \bibinfo
  {author} {\bibfnamefont {A.~A.}\ \bibnamefont {Clerk}},\ and\ \bibinfo
  {author} {\bibfnamefont {K.~C.}\ \bibnamefont {Schwab}},\ }\href
  {https://doi.org/10.1038/nphys1479} {\bibfield  {journal} {\bibinfo
  {journal} {Nature Physics}\ }\textbf {\bibinfo {volume} {6}},\ \bibinfo
  {pages} {213} (\bibinfo {year} {2010})}\BibitemShut {NoStop}%
\bibitem [{\citenamefont {Suh}\ \emph {et~al.}(2014)\citenamefont {Suh},
  \citenamefont {Weinstein}, \citenamefont {Lei}, \citenamefont {Wollman},
  \citenamefont {Steinke}, \citenamefont {Meystre}, \citenamefont {Clerk},\
  and\ \citenamefont {Schwab}}]{suh2014mechanically}%
  \BibitemOpen
  \bibfield  {author} {\bibinfo {author} {\bibfnamefont {J.}~\bibnamefont
  {Suh}}, \bibinfo {author} {\bibfnamefont {A.~J.}\ \bibnamefont {Weinstein}},
  \bibinfo {author} {\bibfnamefont {C.~U.}\ \bibnamefont {Lei}}, \bibinfo
  {author} {\bibfnamefont {E.~E.}\ \bibnamefont {Wollman}}, \bibinfo {author}
  {\bibfnamefont {S.~K.}\ \bibnamefont {Steinke}}, \bibinfo {author}
  {\bibfnamefont {P.}~\bibnamefont {Meystre}}, \bibinfo {author} {\bibfnamefont
  {A.~A.}\ \bibnamefont {Clerk}},\ and\ \bibinfo {author} {\bibfnamefont
  {K.~C.}\ \bibnamefont {Schwab}},\ }\href
  {https://doi.org/10.1126/science.1253258} {\bibfield  {journal} {\bibinfo
  {journal} {Science}\ }\textbf {\bibinfo {volume} {344}},\ \bibinfo {pages}
  {1262} (\bibinfo {year} {2014})}\BibitemShut {NoStop}%
\bibitem [{\citenamefont {Shomroni}\ \emph {et~al.}(2019)\citenamefont
  {Shomroni}, \citenamefont {Qiu}, \citenamefont {Malz}, \citenamefont
  {Nunnenkamp},\ and\ \citenamefont {Kippenberg}}]{shomroni2019optical}%
  \BibitemOpen
  \bibfield  {author} {\bibinfo {author} {\bibfnamefont {I.}~\bibnamefont
  {Shomroni}}, \bibinfo {author} {\bibfnamefont {L.}~\bibnamefont {Qiu}},
  \bibinfo {author} {\bibfnamefont {D.}~\bibnamefont {Malz}}, \bibinfo {author}
  {\bibfnamefont {A.}~\bibnamefont {Nunnenkamp}},\ and\ \bibinfo {author}
  {\bibfnamefont {T.~J.}\ \bibnamefont {Kippenberg}},\ }\href
  {https://doi.org/10.1038/s41467-019-10024-3} {\bibfield  {journal} {\bibinfo
  {journal} {Nature Communications}\ }\textbf {\bibinfo {volume} {10}},\
  \bibinfo {pages} {2086} (\bibinfo {year} {2019})}\BibitemShut {NoStop}%
\bibitem [{\citenamefont {Aspelmeyer}\ \emph {et~al.}(2014)\citenamefont
  {Aspelmeyer}, \citenamefont {Kippenberg},\ and\ \citenamefont
  {Marquardt}}]{aspelmeyer2014cavity}%
  \BibitemOpen
  \bibfield  {author} {\bibinfo {author} {\bibfnamefont {M.}~\bibnamefont
  {Aspelmeyer}}, \bibinfo {author} {\bibfnamefont {T.~J.}\ \bibnamefont
  {Kippenberg}},\ and\ \bibinfo {author} {\bibfnamefont {F.}~\bibnamefont
  {Marquardt}},\ }\href {https://doi.org/10.1103/RevModPhys.86.1391} {\bibfield
   {journal} {\bibinfo  {journal} {Reviews of Modern Physics}\ }\textbf
  {\bibinfo {volume} {86}},\ \bibinfo {pages} {1391} (\bibinfo {year}
  {2014})}\BibitemShut {NoStop}%
\bibitem [{\citenamefont {Arthurs}\ and\ \citenamefont
  {Kelly}(1965)}]{arthurs1965}%
  \BibitemOpen
  \bibfield  {author} {\bibinfo {author} {\bibfnamefont {E.}~\bibnamefont
  {Arthurs}}\ and\ \bibinfo {author} {\bibfnamefont {J.~L.}\ \bibnamefont
  {Kelly}},\ }\href {https://doi.org/10.1002/j.1538-7305.1965.tb01684.x}
  {\bibfield  {journal} {\bibinfo  {journal} {Bell System Technical Journal}\
  }\textbf {\bibinfo {volume} {44}},\ \bibinfo {pages} {725} (\bibinfo {year}
  {1965})}\BibitemShut {NoStop}%
\bibitem [{\citenamefont {Anderson}\ and\ \citenamefont
  {Moore}(2007)}]{anderson2007optimal}%
  \BibitemOpen
  \bibfield  {author} {\bibinfo {author} {\bibfnamefont {B.~D.~O.}\
  \bibnamefont {Anderson}}\ and\ \bibinfo {author} {\bibfnamefont {J.~B.}\
  \bibnamefont {Moore}},\ }\href@noop {} {\emph {\bibinfo {title} {Optimal
  Control: Linear Quadratic Methods}}}\ (\bibinfo  {publisher} {Dover},\
  \bibinfo {address} {Mineola, NY},\ \bibinfo {year} {2007})\BibitemShut
  {NoStop}%
\bibitem [{\citenamefont {Sanders}\ \emph {et~al.}(2007)\citenamefont
  {Sanders}, \citenamefont {Verhulst},\ and\ \citenamefont
  {Murdock}}]{sanders2007averaging}%
  \BibitemOpen
  \bibfield  {author} {\bibinfo {author} {\bibfnamefont {J.~A.}\ \bibnamefont
  {Sanders}}, \bibinfo {author} {\bibfnamefont {F.}~\bibnamefont {Verhulst}},\
  and\ \bibinfo {author} {\bibfnamefont {J.}~\bibnamefont {Murdock}},\
  }\href@noop {} {\emph {\bibinfo {title} {Averaging Methods in Nonlinear
  Dynamical Systems}}},\ \bibinfo {edition} {2nd}\ ed.\ (\bibinfo  {publisher}
  {Springer},\ \bibinfo {address} {New York},\ \bibinfo {year}
  {2007})\BibitemShut {NoStop}%
\bibitem [{\citenamefont {Belenchia}\ \emph {et~al.}(2020)\citenamefont
  {Belenchia}, \citenamefont {Mancino}, \citenamefont {Landi},\ and\
  \citenamefont {Paternostro}}]{belenchia2020entropy}%
  \BibitemOpen
  \bibfield  {author} {\bibinfo {author} {\bibfnamefont {A.}~\bibnamefont
  {Belenchia}}, \bibinfo {author} {\bibfnamefont {L.}~\bibnamefont {Mancino}},
  \bibinfo {author} {\bibfnamefont {G.~T.}\ \bibnamefont {Landi}},\ and\
  \bibinfo {author} {\bibfnamefont {M.}~\bibnamefont {Paternostro}},\ }\href
  {https://doi.org/10.1038/s41534-020-00334-6} {\bibfield  {journal} {\bibinfo
  {journal} {npj Quantum Information}\ }\textbf {\bibinfo {volume} {6}},\
  \bibinfo {pages} {97} (\bibinfo {year} {2020})}\BibitemShut {NoStop}%
\bibitem [{\citenamefont {Wilson}\ \emph {et~al.}(2015)\citenamefont {Wilson},
  \citenamefont {Sudhir}, \citenamefont {Piro}, \citenamefont {Schilling},
  \citenamefont {Ghadimi},\ and\ \citenamefont
  {Kippenberg}}]{wilson2015measurement}%
  \BibitemOpen
  \bibfield  {author} {\bibinfo {author} {\bibfnamefont {D.~J.}\ \bibnamefont
  {Wilson}}, \bibinfo {author} {\bibfnamefont {V.}~\bibnamefont {Sudhir}},
  \bibinfo {author} {\bibfnamefont {N.}~\bibnamefont {Piro}}, \bibinfo {author}
  {\bibfnamefont {R.}~\bibnamefont {Schilling}}, \bibinfo {author}
  {\bibfnamefont {A.~H.}\ \bibnamefont {Ghadimi}},\ and\ \bibinfo {author}
  {\bibfnamefont {T.~J.}\ \bibnamefont {Kippenberg}},\ }\href
  {https://doi.org/10.1038/nature14672} {\bibfield  {journal} {\bibinfo
  {journal} {Nature}\ }\textbf {\bibinfo {volume} {524}},\ \bibinfo {pages}
  {325} (\bibinfo {year} {2015})}\BibitemShut {NoStop}%
\bibitem [{\citenamefont {Magrini}\ \emph {et~al.}(2021)\citenamefont
  {Magrini}, \citenamefont {Rosenzweig}, \citenamefont {Bach}, \citenamefont
  {Deutschmann-Olek}, \citenamefont {Hofer}, \citenamefont {Hong},
  \citenamefont {Kiesel}, \citenamefont {Kugi},\ and\ \citenamefont
  {Aspelmeyer}}]{magrini2021real}%
  \BibitemOpen
  \bibfield  {author} {\bibinfo {author} {\bibfnamefont {L.}~\bibnamefont
  {Magrini}}, \bibinfo {author} {\bibfnamefont {P.}~\bibnamefont {Rosenzweig}},
  \bibinfo {author} {\bibfnamefont {C.}~\bibnamefont {Bach}}, \bibinfo {author}
  {\bibfnamefont {A.}~\bibnamefont {Deutschmann-Olek}}, \bibinfo {author}
  {\bibfnamefont {S.~G.}\ \bibnamefont {Hofer}}, \bibinfo {author}
  {\bibfnamefont {S.}~\bibnamefont {Hong}}, \bibinfo {author} {\bibfnamefont
  {N.}~\bibnamefont {Kiesel}}, \bibinfo {author} {\bibfnamefont
  {A.}~\bibnamefont {Kugi}},\ and\ \bibinfo {author} {\bibfnamefont
  {M.}~\bibnamefont {Aspelmeyer}},\ }\href
  {https://doi.org/10.1038/s41586-021-03602-3} {\bibfield  {journal} {\bibinfo
  {journal} {Nature}\ }\textbf {\bibinfo {volume} {595}},\ \bibinfo {pages}
  {373} (\bibinfo {year} {2021})}\BibitemShut {NoStop}%
\bibitem [{\citenamefont {Tebbenjohanns}\ \emph {et~al.}(2021)\citenamefont
  {Tebbenjohanns}, \citenamefont {Mattana}, \citenamefont {Rossi},
  \citenamefont {Frimmer},\ and\ \citenamefont
  {Novotny}}]{tebbenjohanns2021quantum}%
  \BibitemOpen
  \bibfield  {author} {\bibinfo {author} {\bibfnamefont {F.}~\bibnamefont
  {Tebbenjohanns}}, \bibinfo {author} {\bibfnamefont {M.~L.}\ \bibnamefont
  {Mattana}}, \bibinfo {author} {\bibfnamefont {M.}~\bibnamefont {Rossi}},
  \bibinfo {author} {\bibfnamefont {M.}~\bibnamefont {Frimmer}},\ and\ \bibinfo
  {author} {\bibfnamefont {L.}~\bibnamefont {Novotny}},\ }\href
  {https://doi.org/10.1038/s41586-021-03617-w} {\bibfield  {journal} {\bibinfo
  {journal} {Nature}\ }\textbf {\bibinfo {volume} {595}},\ \bibinfo {pages}
  {378} (\bibinfo {year} {2021})}\BibitemShut {NoStop}%
\bibitem [{\citenamefont {Karmakar}\ and\ \citenamefont
  {Jordan}(2026)}]{karmakar2025cdj}%
  \BibitemOpen
  \bibfield  {author} {\bibinfo {author} {\bibfnamefont {T.}~\bibnamefont
  {Karmakar}}\ and\ \bibinfo {author} {\bibfnamefont {A.~N.}\ \bibnamefont
  {Jordan}},\ }\href {https://doi.org/10.22331/q-2026-03-24-2043} {\bibfield
  {journal} {\bibinfo  {journal} {Quantum}\ }\textbf {\bibinfo {volume} {10}},\
  \bibinfo {pages} {2043} (\bibinfo {year} {2026})}\BibitemShut {NoStop}%
\end{thebibliography}%
\bibliographystyle{apsrev4-2}

\end{document}